\documentclass[11pt,a4paper,fleqn]{article}
\makeatletter
\def~{\ifmmode\;\else\penalty\@M\ \fi}
\def\@setmcodes#1#2#3{{\count0=#1 \count1=#3
  \loop \global\mathcode\count0=\count1 \ifnum \count0<#2
  \advance\count0 by1 \advance\count1 by1 \repeat}}
\DeclareSymbolFont{italic}{OT1}{\rmdefault}{m}{it}
\let\mathit\undefined
\DeclareSymbolFontAlphabet{\mathit}{italic}
\edef\@tempa{\hexnumber@\symitalic}
\@setmcodes{`A}{`Z}{"7\@tempa41}
\@setmcodes{`a}{`z}{"7\@tempa61}
\makeatother
\newdimen\asmindent     
\asmindent=\parindent
\newcount\asmi
\def\inc{\global\advance\asmi by 1}
\def\dec{\global\advance\asmi by-1}
\def\nl{{}$\par\hangindent\asmi em
  \noindent\kern\asmi em\ignorespaces$} 
\def\asmskip{{}$\par\smallskip\hangindent\asmi em
  \noindent\kern\asmi em\ignorespaces$} 
\def\asm{\global\asmi=0 
\parskip=0pt
 \def\+{\inc\nl}
 \def\-{\dec\nl}
 \def\\{\nl}
 \begin{trivlist}\item[]\leftskip=\asmindent\relax$}
\def\endasm{$\end{trivlist}}
\def\asmarray{\begin{array}[t]{@{}l@{\;}l@{\;}l@{}}}
\def\endasmarray{\end{array}}
\newcount\asmii
\def\subasm{\vtop\bgroup\asmii=0\normalbaselines
 \def\nl##1{$\egroup\advance\asmii by##1\relax\hbox\bgroup\hskip\asmii em$}
 \def\\{\nl{0}}
 \def\+{\nl{1}}
 \def\-{\nl{-1}}
 \hbox\bgroup\hskip\asmii em$}
\def\endsubasm{$\egroup\egroup}
\def\ASM#1{\hbox{\sc#1}}        

\def\CHOOSE  {\mathrel{\mathbf{choose}}}

\def\DO      {\mathrel{\mathbf{do}}}
\def\ELSE    {\mathrel{\mathbf{else}}}

\def\FORALL  {\mathrel{\mathbf{forall}}}

\def\IF      {\mathrel{\mathbf{if}}}

\def\IN      {\mathrel{\mathbf{in}}}
\def\LET     {\mathrel{\mathbf{let}}}

\def\NOT     {\mathrel{\mathbf{not}}}

\def\SEQ     {\mathrel{\mathbf{seq}}}

\def\THEN    {\mathrel{\mathbf{then}}}

\def\WITH    {\mathrel{\mathbf{with}}}

\def\SEQ    {\mathrel{\mathbf{seq}}}

\makeatletter
\def\enumerate{%
  \ifnum \@enumdepth >\thr@@\@toodeep\else
    \advance\@enumdepth\@ne
    \edef\@enumctr{enum\romannumeral\the\@enumdepth}%
      \expandafter
      \list
        \csname label\@enumctr\endcsname
        {\usecounter\@enumctr\def\makelabel##1{\hss\llap{##1}}
         \itemsep 0pt\parskip 0pt\parsep 0pt\topsep\smallskipamount}%
  \fi}
\def\itemize{%
  \ifnum \@itemdepth >\thr@@\@toodeep\else
    \advance\@itemdepth\@ne
    \edef\@itemitem{labelitem\romannumeral\the\@itemdepth}%
    \expandafter
    \list
      \csname\@itemitem\endcsname
      {\def\makelabel##1{\hss\llap{##1}}
       \itemsep 0pt\parskip 0pt\parsep 0pt\topsep\smallskipamount}%
  \fi}
\makeatother
\usepackage{tlatex_asm}
\usepackage{xcolor}
\definecolor{boxshade}{gray}{.9}\setboolean{shading}{true}

\usepackage{multirow}
\usepackage{graphicx}
\usepackage{acronym}
\usepackage{ifthen}
\usepackage{substr}
\usepackage[left=2cm, top=2cm, right=2cm, bottom=2cm]{geometry}
\usepackage{array}  
\usepackage[figuresright]{rotating}          
\usepackage{url}
\usepackage[shortlabels]{enumitem}
\usepackage{epsfig}
\usepackage{colordvi}
\usepackage{makeidx}
\usepackage[absolute]{textpos}
\textblockorigin{-6mm}{0mm}

\usepackage{booktabs}
\usepackage{dsfont}
\usepackage{wallpaper}
\usepackage{mathtools,cancel}
\usepackage{soul}
\usepackage{float}
\usepackage{setspace}
\usepackage{amsmath,amssymb,amscd,bm,amsfonts}

\usepackage{amsthm,bm}
\usepackage{tabularx, graphics, longtable,makecell}
\usepackage[toc,page]{appendix}
\usepackage{tikz}
\usepackage{tikz-cd}
\usepackage[tikz]{bclogo}
\usetikzlibrary{positioning}
\usepackage{setspace}
\usetikzlibrary{arrows}
\usetikzlibrary{automata}
\usetikzlibrary{calc,shapes,decorations.pathreplacing}
\usetikzlibrary{backgrounds,decorations.markings}
\usepackage{verbatim} 
\usepackage{authblk}
\definecolor{green}{rgb}{0,0.6,0} 
\definecolor{red}{rgb}{1,0,0} 
\definecolor{blue}{rgb}{0,0,1}
\definecolor{LINK_COLOR}{rgb}{0,0,0.7}
\definecolor{CITE_COLOR}{rgb}{0,0.5,0}
\definecolor{light-gray}{gray}{0.95}

\usepackage{listings}
\lstdefinelanguage{casm}
{ basicstyle         = \ttfamily\scriptsize
, keywordstyle       = \color{blue}\bfseries
, commentstyle       = \color{black!50}\emph
, stringstyle        = \color{green}\bfseries
, numberstyle        = \color{black}\tiny
, backgroundcolor    = \color{black!7.5}
, aboveskip          = 5mm
, belowskip          = 5mm
, framexleftmargin   = 1mm
, framextopmargin    = 1mm
, framexbottommargin = 1mm
, framerule          = 0mm
, frame              = tb
, captionpos         = b
, numbers            = left
, numbersep          = 5pt
, stepnumber         = 1
, firstnumber        = auto
, morecomment        = [l]{//}
, morecomment        = [l]{///}
, morecomment        = [s]{/*}{*/}
, morestring         = [b]"
, sensitive          = true
, escapeinside       = {@(}{)@}
, morekeywords       =
{ CASM, init
, function, derived, rule, enumeration, controlled, symbolic, static
, implement, behavior, for, domain, structure, using, new, feature
, undef, true, false, self, this
, Agent, Boolean, Integer, Binary, String, Rational, Decimal
, Range, Tuple, Record, List, Set, Void, Object
, par, endpar, seq, endseq
, skip, call, let, if, then, else, case, of, forall, in, iterate, do
, choose, exists, with, while, match
, not, and, or, xor, implies, pow
}
}

\usepackage{listings}
\usepackage{lstautogobble}

\renewcommand*\lstlistingname{Specification}
\renewcommand\lstlistlistingname{Specifications}
\providecommand*{\lstlistingautorefname}{spec.}
\providecommand*{\lstnumberautorefname}{line}

\makeatletter
\newcounter{asmcounter}[section]
\renewcommand{\theasmcounter}{\arabic{asmcounter}} 

\lstnewenvironment{lstasm}[2][]{
	\renewcommand*\lstlistingname{Specification}
	\renewcommand\lstlistlistingname{Specs.}
	\renewcommand*{\lstlistingautorefname}{spec.}
	\renewcommand*{\lstnumberautorefname}{line}
	\let\c@lstlisting=\c@asmcounter
	\let\thelstlisting=\theasmcounter
	\lstset{language=ASM, #1}
} {}

\newcommand\includelst[1][]{
	\renewcommand*\lstlistingname{Specification}
	\renewcommand\lstlistlistingname{specs.}
	\renewcommand*{\lstlistingautorefname}{spec.}
	\renewcommand*{\lstnumberautorefname}{line}
	\let\c@lstlisting=\c@asmcounter
	\let\thelstlisting=\theasmcounter
	\expandafter\lstinputlisting\expandafter[language=ASM, #1]
}
\makeatother

\definecolor{lightlightgray}{gray}{0.9}
\definecolor{primary}{HTML}{F7542D} 
\definecolor{secondary}{HTML}{F7932D} 
\definecolor{tertiary}{HTML}{207F9A} 
\definecolor{fourth}{HTML}{008000} 
\definecolor{fifth}{HTML}{800000} 

\newcommand{\asmlstkeyword}{\color{fifth}\bfseries}
\newcommand{\asmlstrule}{\ttfamily}
\newcommand{\asmlstdomain}{\scshape\normalsize}
\newcommand{\asmlstfunction}{\rmfamily}
\newcommand{\asmlstpredicate}{\asmlstfunction \itshape}
\newcommand{\asmlstderived}{\asmlstfunction \slshape}
\newcommand{\asmlstvariable}{\ttfamily}
\newcommand{\asmlstconstant}{\em}
\newcommand{\asmstring}{\color{fourth}\ttfamily}
\newcommand{\asmlstcommentstyle}{\rmfamily\color{gray}}

\usepackage{pbox}
\allowdisplaybreaks
\def \constzeroindent {0cm}

\newenvironment{mycustomindent}[1]
{\setlength{\parindent}{#1}}
{\setlength{\parindent}{\constzeroindent}}

\newenvironment{packed_item1}{
\begin{itemize}[topsep=0pt, partopsep=0pt]
  \setlength{\itemsep}{5pt}
  \setlength{\parskip}{0pt}
  \setlength{\parsep}{0pt}
}{\end{itemize}}

\newenvironment{packed_enumerate}{
\begin{enumerate}[topsep=0pt, partopsep=0pt]
  \setlength{\itemsep}{5pt}
  \setlength{\parskip}{0pt}
  \setlength{\parsep}{0pt}
}{\end{enumerate}}

\let\subparagraph\paragraph
\usepackage[compact]{titlesec}
\usepackage{caption,subcaption}
\titlespacing{\section}{0pt}{12pt}{*0}
\titlespacing{\subsection}{0pt}{6pt}{0pt}
\titlespacing{\subsubsection}{0pt}{6pt}{0pt}
\usepackage{breakcites}
\usepackage{tocloft}
\usepackage{abstract}

\renewcommand{\abstractname}{Abstract}
\title{\textbf{Tutorial on the Executable ASM Specification of\\ the AB Protocol and Comparison with TLA$^+$}}
\author[1]{Paolo Dini}
\author[1]{Manuel Bravo}
\author[2]{Philipp Paulweber}
\author[3]{Alexander Raschke}
\author[1]{Gabriela Moreira}
\affil[1]{Informal Systems
    \ \texttt{\small https://informal.systems/, 
            [paolo,manuel,gabriela]@informal.systems}}
\affil[2]{fiskaly GmbH
    \ \texttt{\small https://fiskaly.com,
            ppaulweber@fiskaly.com}}
\affil[3]{Universit\"at Ulm
    \ \texttt{\small https://www.uni-ulm.de/in/sp/,
            alexander.raschke@uni-ulm.de}
\vspace{-0.8cm}}

\date{\today\vspace{-0.5cm}}

\begin{document}
\maketitle

\begin{abstract}
The main aim of this report is to provide an introductory tutorial on the Abstract State Machines (ASM) specification method for software engineering to an audience already familiar with the Temporal Logic of Actions (TLA$^+$) method. The report asks to what extent the ASM and TLA$^+$ methods are complementary in checking specifications against stated requirements and proposes some answers. A second aim is to provide a comparison between different executable frameworks that have been developed for the same specification languages. Thus, the ASM discussion is complemented by executable Corinthian ASM (CASM) and CoreASM models. Similarly, the two TLA$^+$ specifications presented, which rely on the TLC and Apalache model checkers, respectively, are complemented by a Quint specification, a new language developed by Informal Systems to serve as a user-friendly syntax layer for TLA$^+$. For the basis of comparison we use the specification of the Alternating Bit (AB) protocol because it is a simple and well-understood protocol already extensively analysed in the literature. While the models reported here and developed with the two methods are semantically equivalent, ASMs and Quint are better suited for top-down specification from abstract requirements by iterative refinement. TLA$^+$ seems to be more easily used bottom-up, to build abstractions on top of verified components in spite of the fact that it, too, emphasizes iterative refinement. In the final section, the report begins to scope out the possibility of a homomorphism between the specification of the AB protocol and its finite-state machine (FSM) through state space visualizations, motivated by a search for a formal decomposition method.
\end{abstract}

\newpage
\tableofcontents

\newpage
\section{Introduction}
\label{sec:introduction}

The original purpose of this document was to serve as a tutorial for the Abstract State Machines (ASM) specification and modelling method for software engineering \cite{BoeSta03,BoeRas18}, and for how such ASM specifications can be turned into executable models using the Corinthian Abstract State Machine\footnote{\url{https://casm-lang.org}} (CASM) language and framework \cite{paulweber2022phd}. The scope then grew to write a report that could present a comparison of the ASM and TLA$^+$ specification perspectives and of their tooling and executable frameworks. The tutorial assumes that the reader is already familiar with TLA$^+$.

The focus of the report and the basis for the comparisons is the specification of the half-duplex\footnote{The three basic types of communication protocols are: (1) simplex, in which messages are sent in only one direction and an ack or alternation bit is sent in the other; (2) half-duplex, in which the two terminals take turns at sending messages in each direction, with the ack bit for each message travelling in the opposite direction; and full-duplex, in which both terminals send messages in both directions simultaneously and independently.} Alternating Bit (AB) protocol, first published by Bartlett et al.\ in 1969 \cite{Bartlettetal1969} as an improvement on a protocol proposed by Lynch in 1968 \cite{Lynch1968}. Although the ASM specification of the AB protocol is already available in Section 6.3 of the main reference text on ASMs \cite{BoeSta03}, the tutorial part of the report provides a stand-alone introduction to the basic ASM and CASM concepts and practices that will hopefully make the learning ramp easier for newcomers. A chapter on a different executable specification framework, CoreASM,\footnote{\url{https://slideplayer.com/slide/17819082/}} is also included.

After a high-level introduction to ASMs in Chapter \ref{asms}, Chapter \ref{protocols} presents and analyses the Lynch and AB protocols in detail. Chapter \ref{asmspec} derives the ASM specification from a list of requirements, Chapter \ref{casm} presents a basic and a refined version of a CASM model of the ASM rules, and Chapter \ref{coreasm} presents the CoreASM model. Chapter \ref{tla} introduces the TLA$^+$ specification in two versions, a simpler one that emulates the single-thread execution of the CASM code and a more sophisticated one meant for more general behaviour and for the Apalache model checker.\footnote{\url{https://apalache.informal.systems/}} The audience is assumed to be already familiar with TLA$^+$, whose basics can be learned through Leslie Lamport's video course.\footnote{\url{http://lamport.azurewebsites.net/tla/tla.html?from=https://research.microsoft.com/users/lamport/tla/tla.html&type=path}} Chapter \ref{quint} casts the TLA$^+$ model in the newly developed Quint language\footnote{\url{https://github.com/informalsystems/quint}} for expressing TLA$^+$ specifications in a more user-friendly way, and explains how it improves the engineer's Ux while retaining the full power of TLA$^+$. Chapter \ref{discussion} presents a discussion of the similarities and differences between the ASM and TLA$^+$ perspectives, at both theoretical level and at the level of the tools, and Chapter \ref{conclusion} offers some conclusions and hints on possible directions for future work.

Regarding scope, this report addresses only a very abstract version of AB protocol. We originally thought of developing also a first refinement of the specification in order to show how the ASM method handles more concrete implementation details, but lack of time has pushed us instead towards a more extensive ``horizontal'' comparison between different specification languages, methodologies, and tools.

The motivation for this methodological exploration arises from implementation engineers' reluctance to develop specifications for their software applications -- especially in TLA$^+$ -- before they start coding, a problem that is universally recognized. The diagnosis is that specification languages and formal methods tend to be too mathematical, requiring of the engineers and developers a very different kind of thinking from what they rely on when coding. The ASM methodology and the Quint language were developed with a full awareness of this challenge and as a way to address it. The report, therefore, aims to compare the different specification perspectives and methodologies in order to understand their complementarities and the opportunities for integration that could offer more user-friendly tooling to developers while retaining the full generality and rigour of ASM and TLA$^+$ specifications.

\newpage
\section{ASMs}
\label{asms}
\subsection{Conceptual Overview}
The ASM formal specification and modelling method is used for the design of complex, reactive, concurrent, distributed, non-deterministic, multi-agent software systems based on rules for how a given system transitions between different states in response to external stimuli or an internal clock. States are composed of sets of elements together with the (dynamic) functions that operate on them. The elements, the functions, and the rules are all expressed with terminology that reflects the domain in which the application will run. The precise mathematical definition of state transition rules makes ASM models executable, given suitable tooling such as the CASM language. Therefore, ASM models can be thought of as executable pseudo-code that is understandable to the customer or domain expert.

The ASM methodology starts with the definition of a ground model based on the high-level requirements and proceeds by iterative refinement for each implementation decision (vertical refinement) or as new requirements are added (horizontal refinement). At each refinement step the corresponding CASM model can be run to check whether it still satisfies the requirements. The iterative refinement process terminates when the specification has reached a level of detail sufficient for the implementation in the desired target language.

Conceptually, ASMs can be thought of as generalised finite-state machines. Mathematically, they are composed of sets of states and of (dynamic) functions that operate on those states, i.e. they are algebras.
They were in fact first introduced by Yuri Gurevich as \emph{evolving algebras\footnote{\url{https://www.researchgate.net/profile/Yuri-Gurevich/publication/221329427\_Evolving\_Algebras\_and\_Linear\_Time\_Hierarchy/links/0fcfd5100a3f36d80d000000/Evolving-Algebras-and-Linear-Time-Hierarchy.pdf##page=46}}} \cite{Gurevich94b}.
In the most general and abstract terms, ASMs can be thought of as a method to develop a customised programming language for a specific problem. However, since they require the close collaboration of at least four roles/people (software implementer, ASM expert, customer, testing expert), the method also requires the production of a body of detailed documentation that everyone understands and can refer back to in case of problems, change requests, or new version releases. Therefore, ASMs are as important a shared documentation method and central repository of application knowledge as they are a rigorous specification, mathematical verification, and validation (through simulation) framework.

\subsection{Definitions and Basic Concepts}
ASMs were initially developed as single-agent state machines \cite{BoeSta03}. They were then generalised to multi-agent synchronous or asynchronous ASMs \cite{BoeSta03}. More recently, the concept of communicating ASMs was introduced to give more flexibility to the specification of complex distributed systems \cite{BoeSch16,BoeSch17}. Here we start with defining and understanding a single-agent state machine, known as a `Basic ASM'.

This section is a very short summary of parts of Chapter 2 in \cite{BoeSta03}, but this brief summary cannot replace the original book, which the reader is strongly encouraged to consult since it provides a comprehensive discussion of all the theoretical and many practical aspects of the ASM and related concepts.

Basic ASMs are composed of finite sets of transition rules of the form
\begin{asm}
\IF Condition \THEN Updates,
\vspace{-0.4cm}
\end{asm}
where the updates transform abstract ASM states. \emph{Abstract ASM states} are mathematical structures composed of data as elements of sets which are equipped with partial functions and predicates. Predicates are Boolean functions, while constants are treated as 0-ary (static) functions. Partial functions are turned into total functions by adding $f(x) = undef$ for values of the domain where $f$ is not defined.

Following the usual ASM convention where a capitalised variable name indicates a set, \emph{Updates} is a finite set of assignments of the form
\begin{asm}
f(t_1, \cdots, t_n) := t.
\vspace{-0.4cm}
\end{asm}
The values of the functions $f$ in this set change to the values $t$ when these assignments are executed in parallel at the arguments indicated. More precisely, when entering a new state, first all the parameters $t_i$ are evaluated to their values $v_i$, then the value of $f(v_1, \cdots, v_n)$ is changed to (or defined as, if it was \emph{undef}) $v$, which is the value of $f(v_1, \cdots, v_n)$ in the new state. A function name $f$ and the ordered sequence of its arguments $(v_1, \cdots, v_n)$ formed by a list of parameters is called a \emph{location}. `Location-value pairs $(loc, v)$ are called \emph{updates} and represent the basic units of state change' (\cite{BoeSta03}: 29).

If functions are interpreted as `function tables', a location-value pair is a row of the table with the left column holding, for each row, the value of the function and the remaining columns holding the values of the arguments upon which the function depends. A \emph{static} function corresponds to a table that is never changed, whereas \emph{dynamic} functions correspond to tables whose left or value columns are updated as described above.

An ASM computation step in a given state consists in executing simultaneously all updates of all transition rules whose guards are true in that state. A \emph{condition} or \emph{guard} is an arbitrary predicate logic formula without
 \emph{free variables\footnote{\url{https://en.wikipedia.org/wiki/Free_variables_and_bound_variables}}} that evalutes to \emph{true} or \emph{false}. The result of their execution, if it is consistent, yields the next state. A set of updates is \emph{consistent} if it contains no pair of updates with the same location, i.e.\ no two location-value pairs $(loc, v)$, $(loc, v')$ with $v \ne v'$.

When analysing runs $S_0, S_1, S_2, \cdots$ of an ASM, $S_n$ is the $n^{\text{th}}$ state. If $n < m$, we say that $S_n$ is before $S_m$, written $S_n < S_m$.

Simultaneous execution of updates enables the local description of a global state change which, in turn, implies that the next state differs from the previous state only at locations appearing in the update set. The advantage is that, unlike the case of TLA+, the frame problem \cite{Boerger2022} is avoided, i.e.\ only what changes needs to be specified; what is not mentioned does not change by definition.

The simultaneous executions of a rule $R$ for all the values of a free variable $x$ satisfying a given condition $\phi$ is expressed as follows:
\begin{asm}
\FORALL x \WITH \phi \DO R
\vspace{-0.4cm}
\end{asm}
A choice operation to describe non-deterministic behaviour \cite{Boerger2022} is expressed as
\begin{asm}
\CHOOSE x \WITH \phi \DO R
\vspace{-0.4cm}
\end{asm}
Other common constructs such as $\in$ to indicate belonging to a set or $\IF ... \THEN$ are used freely as needed to express various kinds of conditions.

Constraints on an ASM's runs can be imposed to restrict the class of models satisfying a given specification. The constraint mechanism allows the designer to combine in the specification declarative and axiomatic features with operational ones without incurring the cost of the frame problem mentioned above.

The abstract nature of ASMs makes it possible to relate the state evolution of a given `abstract' machine to the state evolution of a more `refined' machine  with a more detailed state set in terms of a notion of equivalence of corresponding run segments of the two ASMs under precisely stated boundary conditions:
\begin{quote}
\small
 The focus is not on generic notions of refinements which can be proved to work in every context and to provide only effects which can never be detected by any user of the new program. Instead the concern is to support a disciplined use of refinements which correctly reflect and explicitly document an intended design decision, adding more details to a more abstract design description, e.g. for making an abstract program executable, for improving a program by additional features or by restricting it through precise boundary conditions which exclude certain undesired behaviors. (\cite{BoeSta03}: 22)
\end{quote}

In summary, an ASM $M$ is defined by its \emph{signature}, i.e.\ the set of declarations of functions and rules, the set of its initial states, and the unique variable-free \emph{main rule} which is often identified with the machine $M$. However, in more recent languages like CASM, the main rule is invisible and has been reduced to the entry point that launches the machine, thereby leaving the programmer more freedom to call the control centre of the ASM rule execution something other than `main'. Such execution control ASM is declared with the \texttt{init} command. The function and rule declarations include the constraints on signature and runs in order to determine the set of possible states of the machine. We now explain how functions are classified.

\subsection{Classification of Functions and Locations}
\begin{figure}[H]
\centering
\includegraphics[width=15cm]{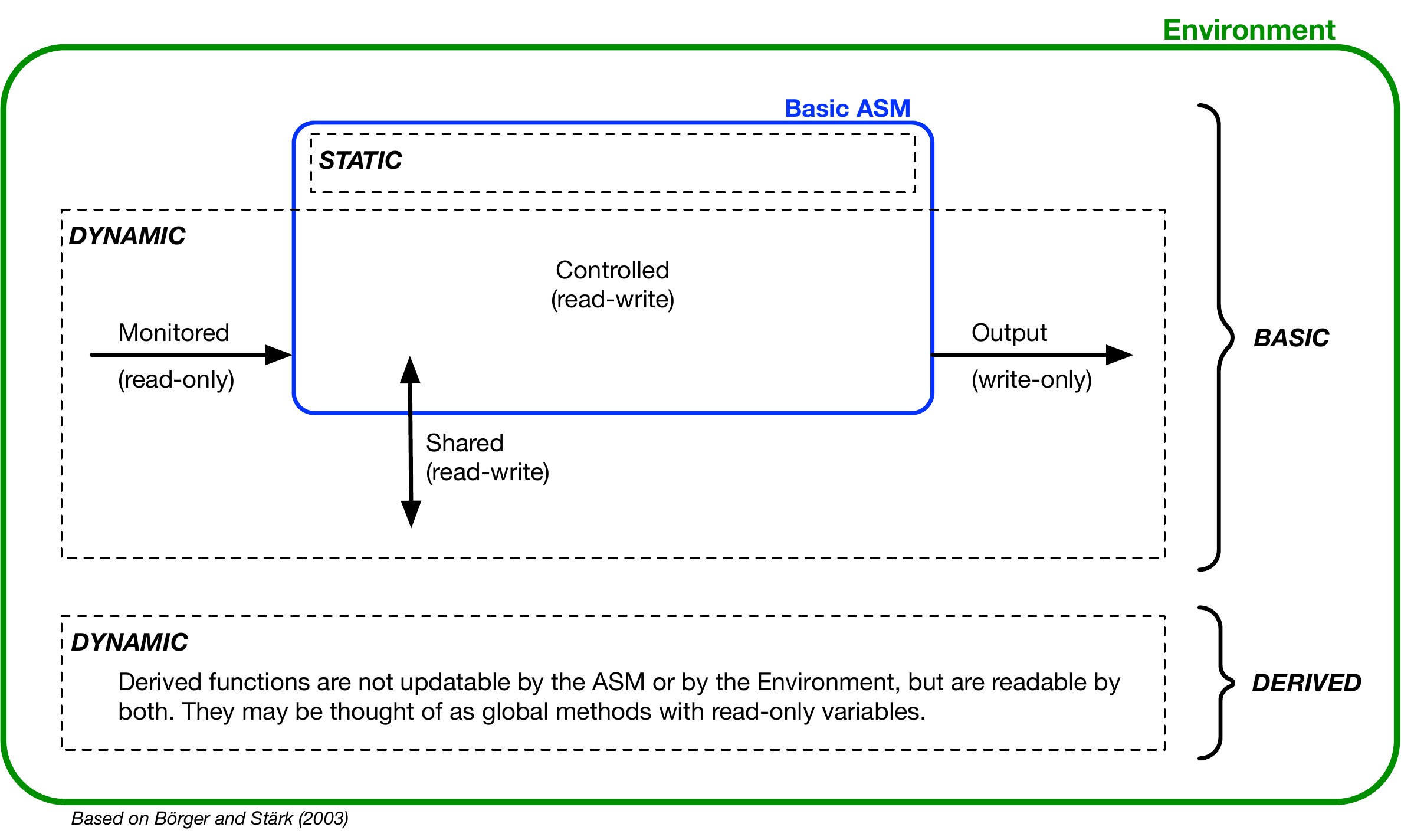}
\caption{\small\textbf{Classification of ASM functions and locations}}
\label{fig:asm_functions}
\vspace{-0.4cm}
\end{figure}

The main distinction for a given ASM $M$ is between its static functions, whose values never change, i.e.\ do not depend on the states of $M$, and its dynamic functions, whose values may change due to updates by $M$ or by the environment, i.e.\ may depend on the states of $M$. As shown in Figure \ref{fig:asm_functions}, ASM functions can alternatively be classified as `basic' or `derived'. Derived functions are not directly updatable by the ASM or the environment. However, they are often expressed in terms of other functions that belong to the ASM signature and may be dynamic. The role of a derived function $f$ is that in different states it allows to produce a different value $f(x)$ for the same argument $x$. For this reason, in the ASM literature derived functions are usually regarded as dynamic. The same classification applies to locations or updates.

With this minimalist set of concepts and definitions we will develop an ASM model of the AB protocol. Chapter \ref{protocols} presents the protocol while Chapter \ref{asmspec} the ASM model. Chapter \ref{casm} presents the corresponding CASM model, Chapter \ref{tla} the TLA$^+$ model of AB, and the final chapter a comparison between the ASM and TLA$^+$ models.

\newpage

\section{Protocol Comparison}
\label{protocols}
The half-duplex Alternating Bit (AB) protocol forms the core of the Kermit\footnote{\url{http://www.columbia.edu/kermit/kermit.html}} file transfer protocol, and was itself an improvement by Bartlett et al.\ \cite{Bartlettetal1969} on a half-duplex protocol developed by Lynch \cite{Lynch1968}. In both cases, the protocol assumes that transmission errors can always be detected.

\subsection{Lynch's 2-Bit Protocol}
\begin{figure}[H]
\centering
\includegraphics[width=16cm]{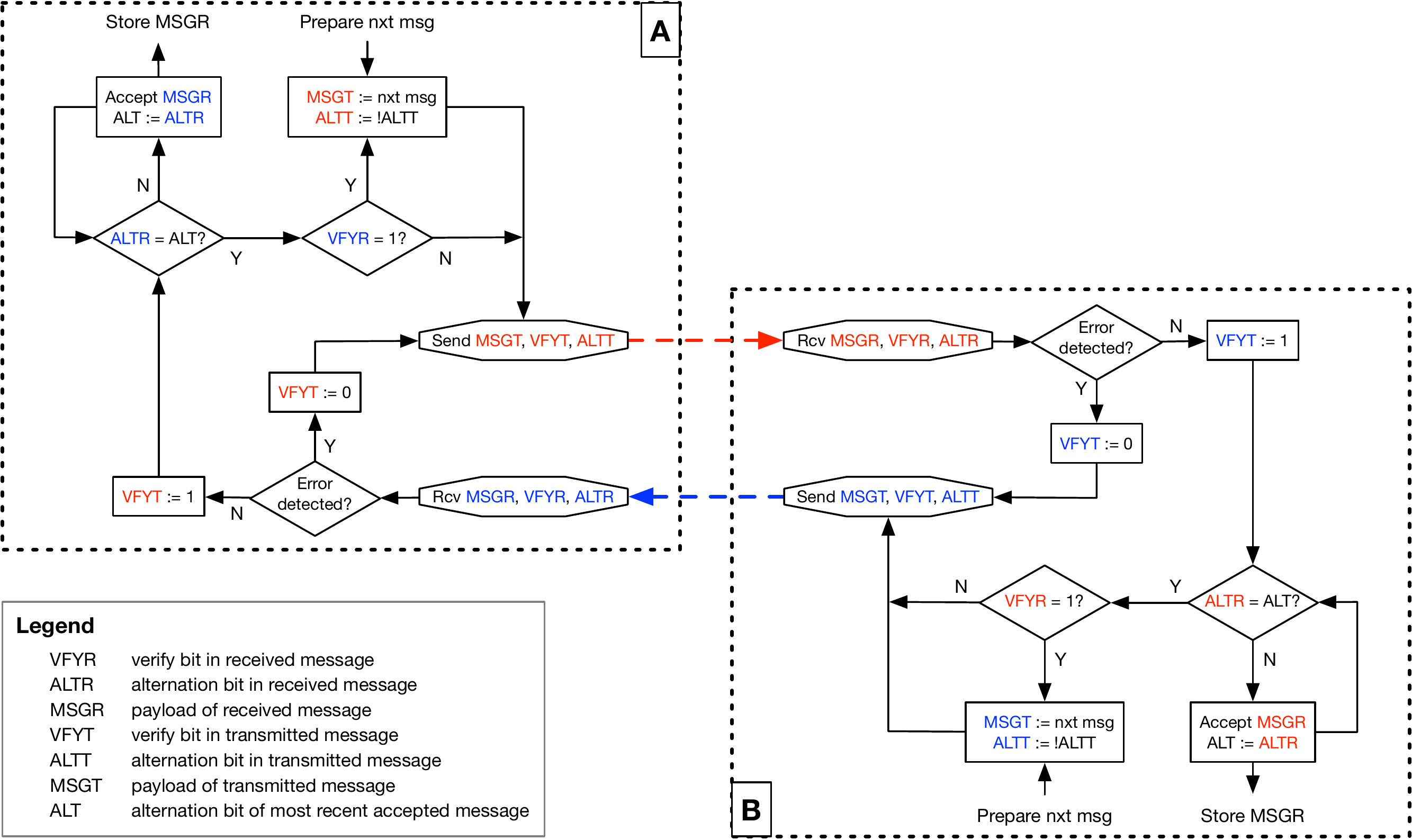}
\caption{\small\textbf{Flowchart of Lynch's reliable 2-bit half-duplex protocol}}
\label{fig:lynch_flowchart}
\vspace{-.3cm}
\end{figure}

Lynch uses an alternation bit to indicate when the next file should be accepted by the receiver. Each terminal stores a local version of its alternation bit, which is compared to the alternation bit sent by the other terminal as an attachment to the file: if they are equal, the in-coming file is rejected even if there are no errors; if they are different, it is accepted. The verify bit is a second attachment and indicates whether the previous file transfer (in the \emph{same} direction) was successful or not. If $VFY = 1$, the next message is loaded and sent; if $VFY = 0$, the previous message is re-sent. Fig.\ \ref{fig:lynch_flowchart} shows identical flowcharts for each terminal, indicating that the protocol is symmetrical. Red and blue colours are used to help distinguish between the two separate data flows in the two opposite directions.

\begin{figure}[H]
\centering
\includegraphics[width=16cm]{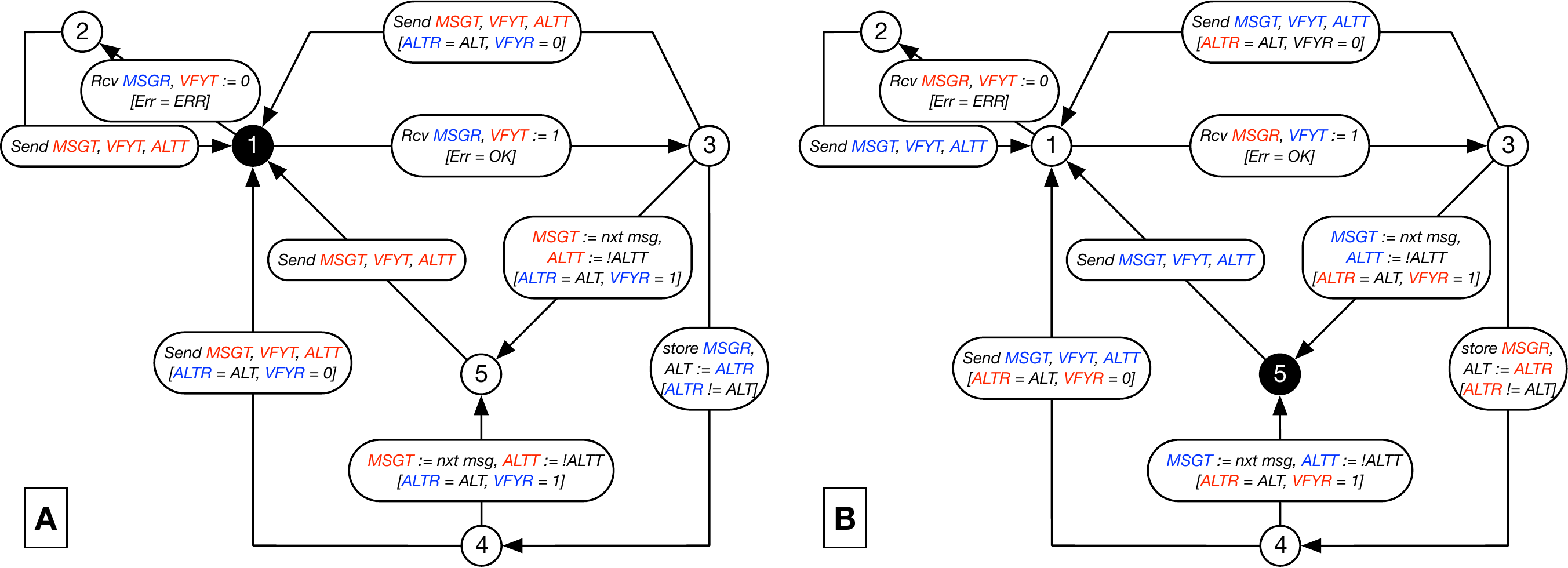}
\caption{\small\textbf{Automata of the Lynch protocol}}
\label{fig:lynch_automata}
\vspace{-0.3cm}
\end{figure}

Fig.\ \ref{fig:lynch_automata} shows the automata for the two terminals, `A' and `B', corresponding to the Lynch protocol of Fig.\ \ref{fig:lynch_flowchart}. The automata's starting states are different because we assume that the first message will be sent from B and will be received by A. The automata are otherwise identical since the protocol is symmetrical. The same red and blue colours are used to highlight the two directions of data flow. Transitions between states are labelled by bubbles that contain a guard in square brackets. If the guard evaluates to True, the rest of the text in the bubble is executed. Where there is no guard, it is assumed that some other independent trigger causes the transition, such as a timeout or user input.

\begin{figure}[H]
\centering
\includegraphics[width=16cm]{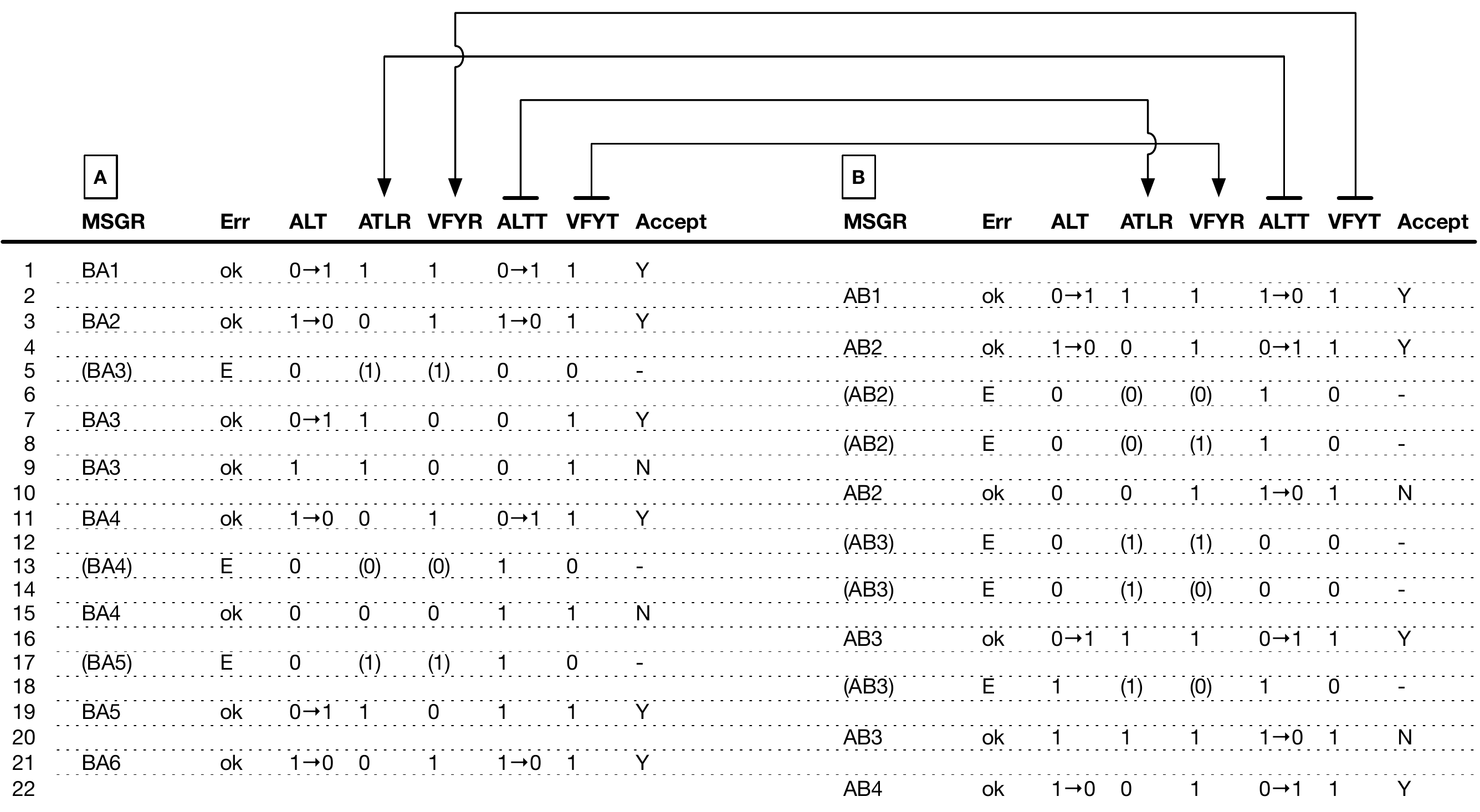}
\caption{\small\textbf{Sequence of messages and errors (based on Fig.\ 2 in \cite{Lynch1968})}}
\label{fig:lynch_sequence}
\vspace{-0.3cm}
\end{figure}

Fig.\ \ref{fig:lynch_sequence} shows a breakdown of the key variables and actions for each terminal, driven by a sequence of message transmission attempts and constrained by a sequence of errors in each direction. These sequences of messages and errors are identical to those used by Lynch in his presentation of the protocol, in Fig.\ 2 of \cite{Lynch1968}, but we provide more details for each step in the transmission in order to make it easier to follow the state changes in the automata and to relate the automata to the flowchart. Entries involving a change indicate the current value on the left and the next-state value on the right.

There are in all 22 attempts at message transmission in both directions. The left-side of the table shows what happens at Terminal A when it receives a message, which may have picked up errors in transmission. The right-side of the table does the same for Terminal B. The circuit diagram-like arrows above the table show the dependency of some of the variable at one terminal to the variables at the other. Parentheses indicate the values that were sent but that may have arrived corrupted at the receiver. Messages in each direction are numbered, such that, given the pattern of errors shown, we can see that B manages to send six files to A successfully, whereas A can only manage four.

Each local $ALT$ bit is updated to equal the $ALTR$ bit just received if and only if the message arrived without error AND it had not already been accepted (stored locally). When the previous message (in the \emph{same} direction) was successful, the $ALTT$ bit is updated to its inverse. When messages in both directions do not incur any errors, the sequence of states in Fig.\ \ref{fig:lynch_automata} is 1-3-4-5, meaning that the arriving file is stored and the next file in the opposite direction is prepared for sending. However, since the two data flows are decoupled and independent, it is possible for an A-to-B message to arrive successfully and be stored even if there was an error in the B-to-A direction, such that a new file in the latter direction is not loaded and the previous file is resent (1-3-4 trace). Equally, it is also possible for an A-to-B file that has already been stored to arrive without error, such that in this case only a new file for the B-to-A transmission is readied (1-3-5 state trace).

\begin{figure}[H]
\centering
\includegraphics[width=17cm]{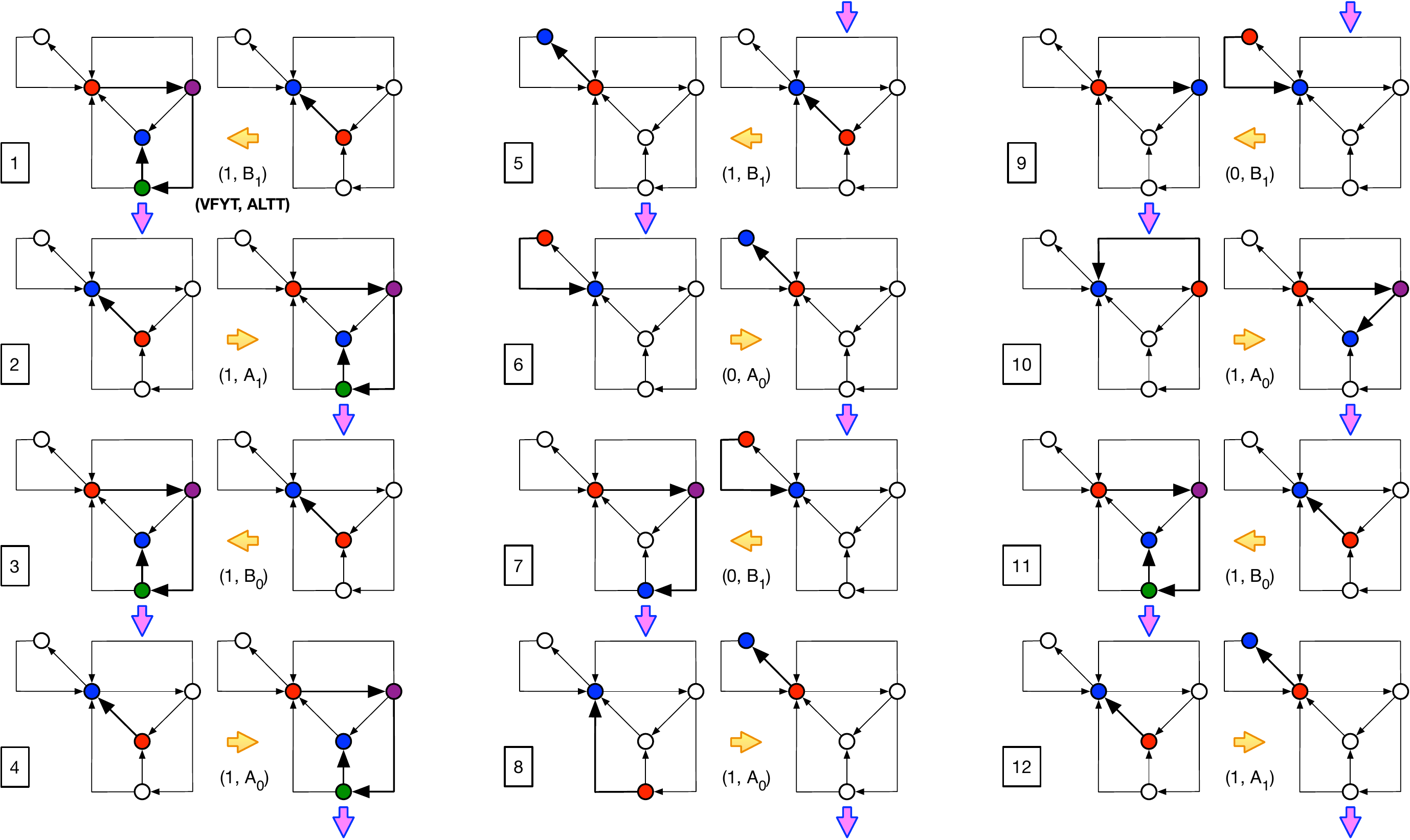}
\caption{\small\textbf{Sequence of automata diagrams following Fig.\ \ref{fig:lynch_sequence}, Lines 1-12 (for Legend see Fig.\ \ref{fig:lynch_automata_diagrams2})}}
\label{fig:lynch_automata_diagrams1}
\end{figure}

\begin{figure}[H]
\centering
\includegraphics[width=17cm]{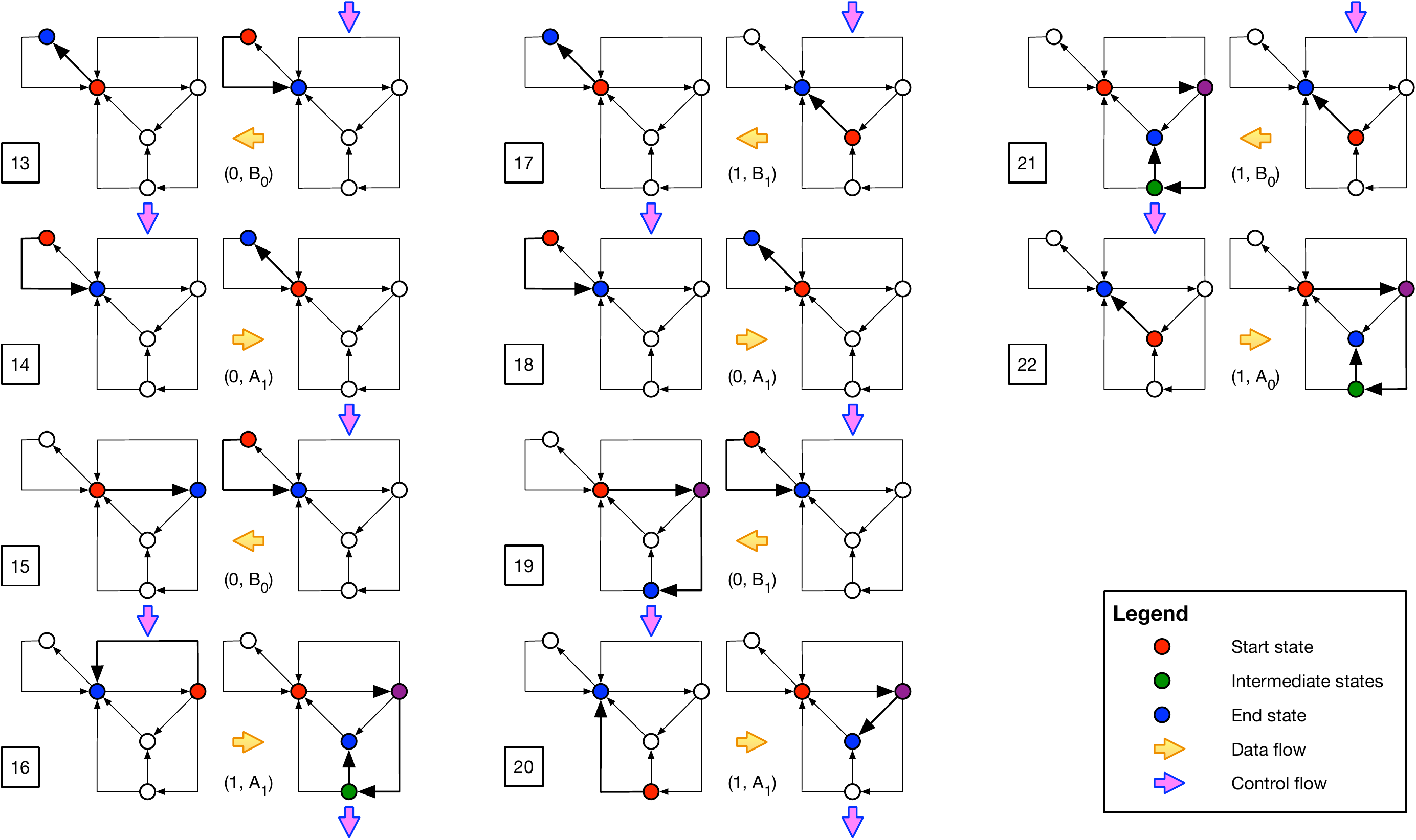}
\caption{\small\textbf{Sequence of automata diagrams following Fig.\ \ref{fig:lynch_sequence}, Lines 13-22}}
\label{fig:lynch_automata_diagrams2}
\end{figure}

Figs.\ \ref{fig:lynch_automata_diagrams1} and \ref{fig:lynch_automata_diagrams2} show the same information as Fig.\ \ref{fig:lynch_sequence} with a simpler graphical rendition of the automata of Fig.\ \ref{fig:lynch_automata}, but making the states and state transitions explicit at \emph{both} terminals. Transitions are highlighted with slightly thicker arrows. Red states are always starting states and blue states are always ending states. The other colours are intermediate states. The subscript indicates the alternation bit being sent, $ALTT$.

\newpage
\subsection{The AB 1-Bit Protocol}
\begin{figure}[H]
\centering
\includegraphics[width=16cm]{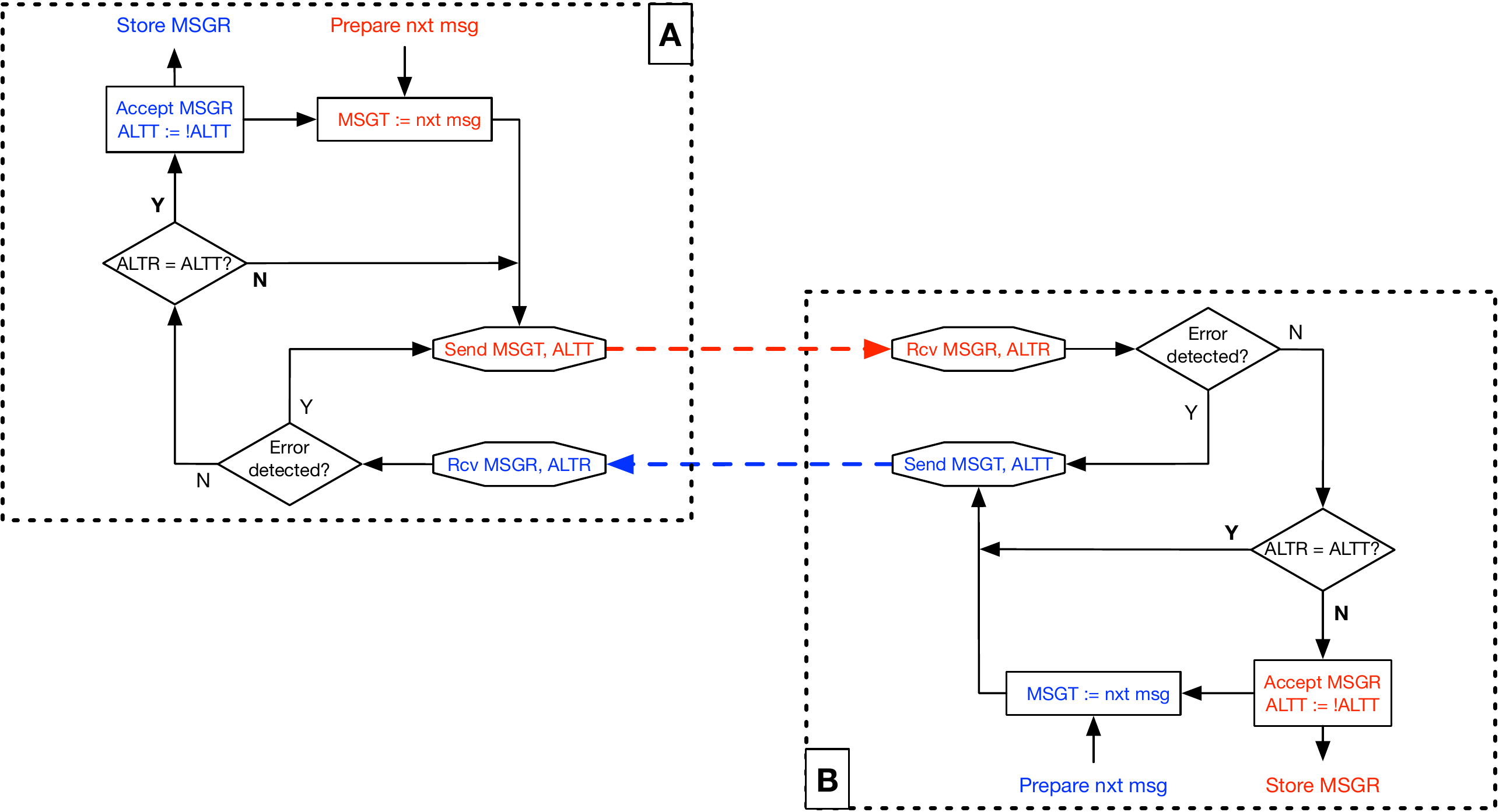}
\caption{\small\textbf{Flowchart of reliable AB 1-bit protocol}}
\label{fig:abp_flowchart}
\end{figure}

AB combines both alternation and verification functions in a single bit. The consequence is that whereas in the Lynch case the protocol is symmetrical, AB is not. Figure \ref{fig:abp_flowchart} shows the flowchart for both terminals, where the asymmetry is highlighted by the opposite handling of the branch point where the alternation bit just received ($ALTR$) is compared to the bit to be sent in the other direction ($ALTT$).

More precisely, where Lynch uses VFY = 0 or 1 to indicate that the previous message was unsuccessful or successful, respectively, Bartlett et al.\ use a \emph{change} in the control bit to indicate success in the previous transfer and no change to indicate failure. However, this rule is reversed for the other terminal. As shown in Fig. \ref{fig:abp_flowchart}, Terminal B follows this rule whereas Terminal A follows the opposite.

\begin{figure}[H]
\centering
\includegraphics[width=16cm]{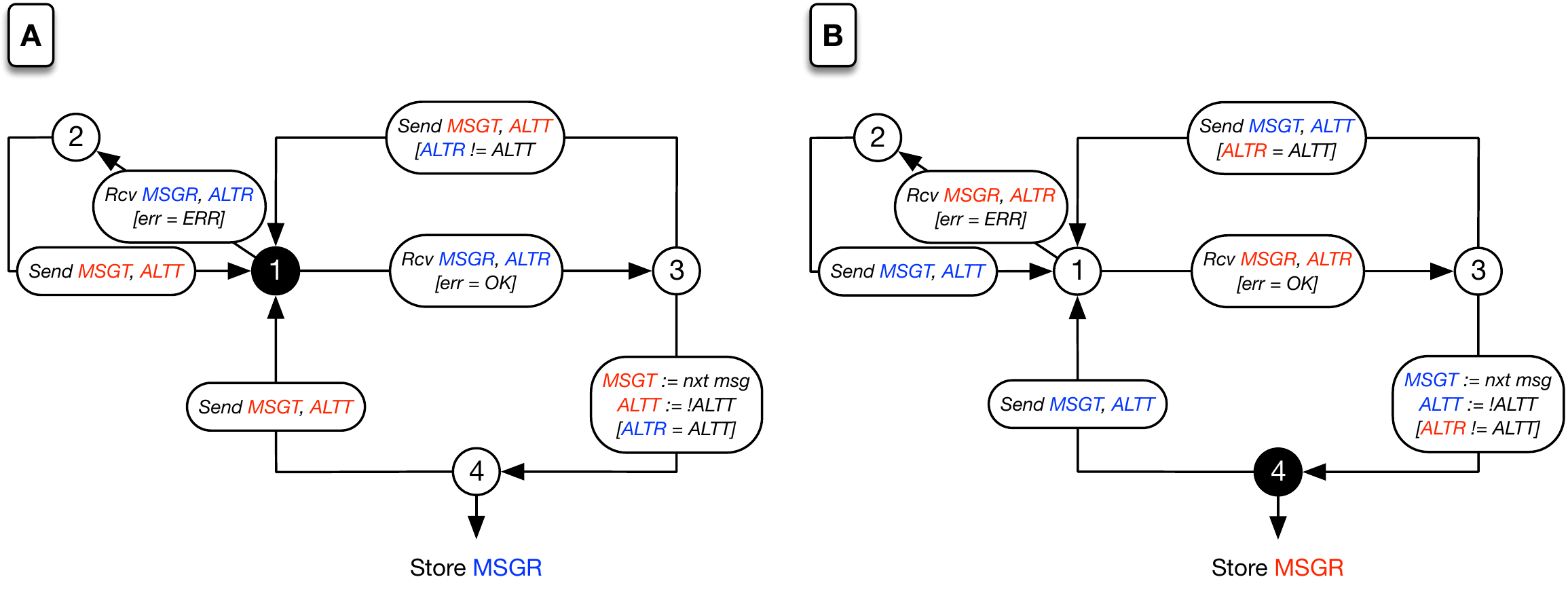}
\caption{\small\textbf{Automata of AB 1-bit protocol}}
\label{fig:abp_automata}
\end{figure}

\begin{figure}[H]
\centering
\includegraphics[width=17cm]{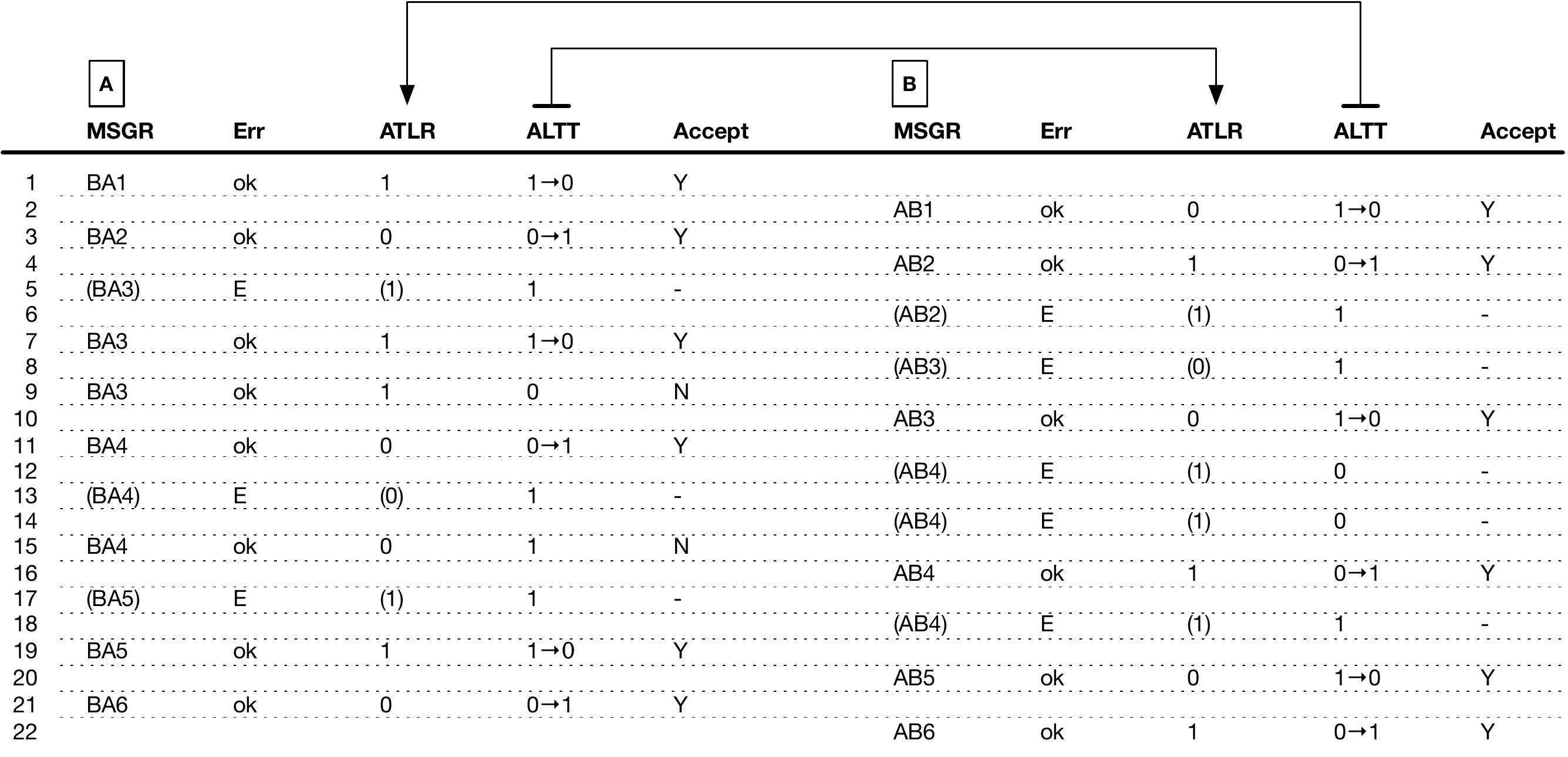}
\caption{\small\textbf{Sequence of messages and errors (based on Fig.\ 2 in \cite{Lynch1968})}}
\label{fig:abp_sequence}
\vspace{-0.3cm}
\end{figure}

Fig.\ \ref{fig:abp_automata} shows the corresponding automata. These are smaller than the automata devised by Bartlett et al.\ but behave the same way. Fig\ \ref{fig:abp_sequence} shows the same sequence of message transfer attempts and errors as Fig.\ \ref{fig:lynch_sequence}. While the B-to-A transmission matches the number of files sent with the Lynch protocol, the A-to-B transmission achieves two additional transfers, suggesting that the AB protocol may be more efficient.

\begin{figure}[H]
\centering
\includegraphics[width=17cm]{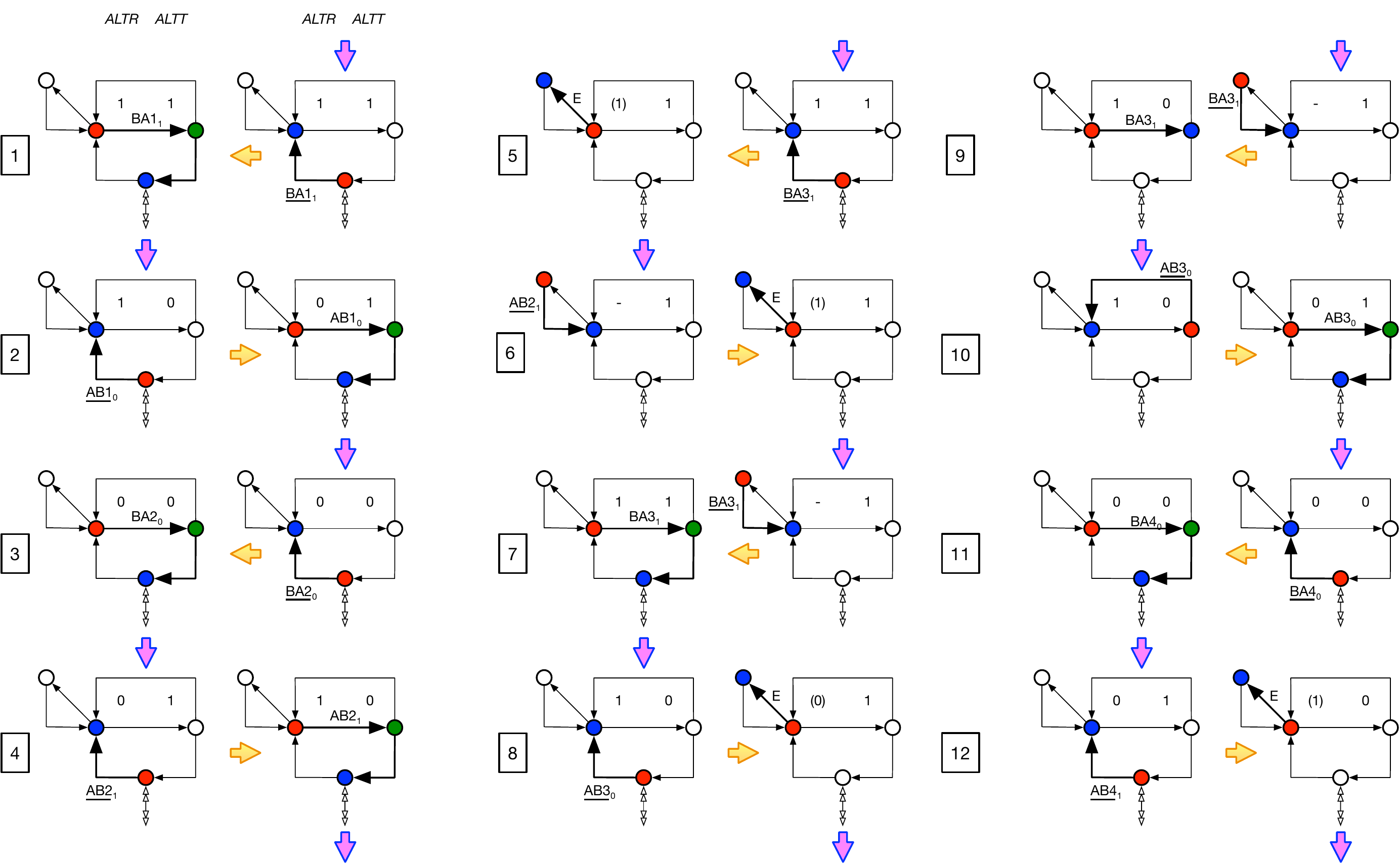}
\caption{\small\textbf{Sequence of automata diagrams corresponding to Fig.\ \ref{fig:abp_sequence}, Lines 1-12}}
\label{fig:abp_automata_diagrams1}
\end{figure}

\begin{figure}[H]
\centering
\includegraphics[width=17cm]{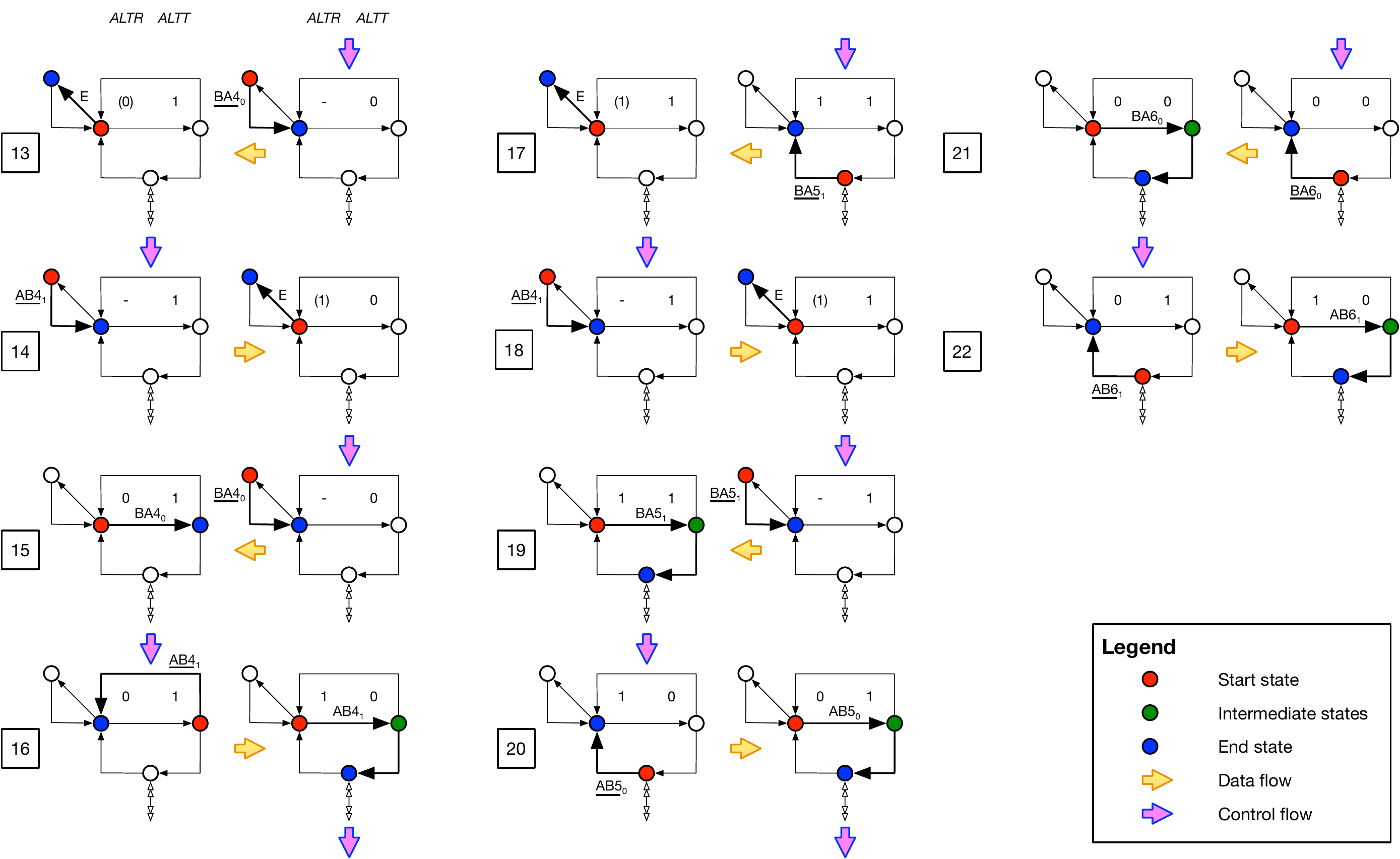}
\caption{\small\textbf{Sequence of automata diagrams corresponding to Fig.\ \ref{fig:abp_sequence}, Lines 13-22}}
\label{fig:abp_automata_diagrams2}
\end{figure}

Figs.~\ref{fig:abp_automata_diagrams1} and \ref{fig:abp_automata_diagrams2} show the detailed automata diagrams for the same sequence. In this case the $ALTR$ and $ALTT$ bits are drawn within each automaton to make it easier to verify that the correct sequence of files is sent in the presence of the given errors. Following the convention used by Bartlett et al., underscores on transition labels indicate the sending transition and absence of underscoring indicates the receiving transition. As previously, the subscript indicates $ALTT$, creating some redundancy since the same information is also provided by the value of the bit written on the right within each sending automaton.

\subsection{Initial Conditions}
We can take the first transmission to be B-to-A without loss of generality. In the asymmetric AB case we indicate what needs to change if A were the sending the first transmission. We use the notation $ALTT_B(0)$ to indicate the value of $ALTT$ of the B terminal before the first transmission. We also assume no errors occur in the first few transmissions.

\subsubsection{Lynch Protocol}
For the Lynch protocol, although we could set $VFYT(0) = 0$ at the beginning, for the starting terminal, there is no loss in generality in pretending that the ``previous'' transmission was successful. So we can set $VFYT_B(0) = 1$. This leaves $(ALT, ALTT)$ as the only variables, which can be set independently for each terminal. $ALTR$ and $VFYR$ are not relevant since their values are overwritten by whatever the other terminal sends them. As shown in Table \ref{tab:lynch_ic}, therefore, there are 16 possible distinct initial conditions (ICs).

Rather than developing a formal proof of which combinations lead to reliable transmission and which do not, here we only lay the groundwork for such a proof and merely suggest the likely trend. The proof can be revisited at a later date if it turns out that it would be helpful to obtain it. To distinguish between ICs that lead to reliable behaviour and those that do not might be difficult in general, meaning with errors in the first transmissions. If instead we focus on the sequence of Fig.\ \ref{fig:lynch_sequence}, it is sufficient to show that the first two files are delivered correctly, since after that the protocol is already known to be reliable and we have already shown all the state transitions for this sequence.

\begin{figure}[H]
\centering
\includegraphics[width=17cm]{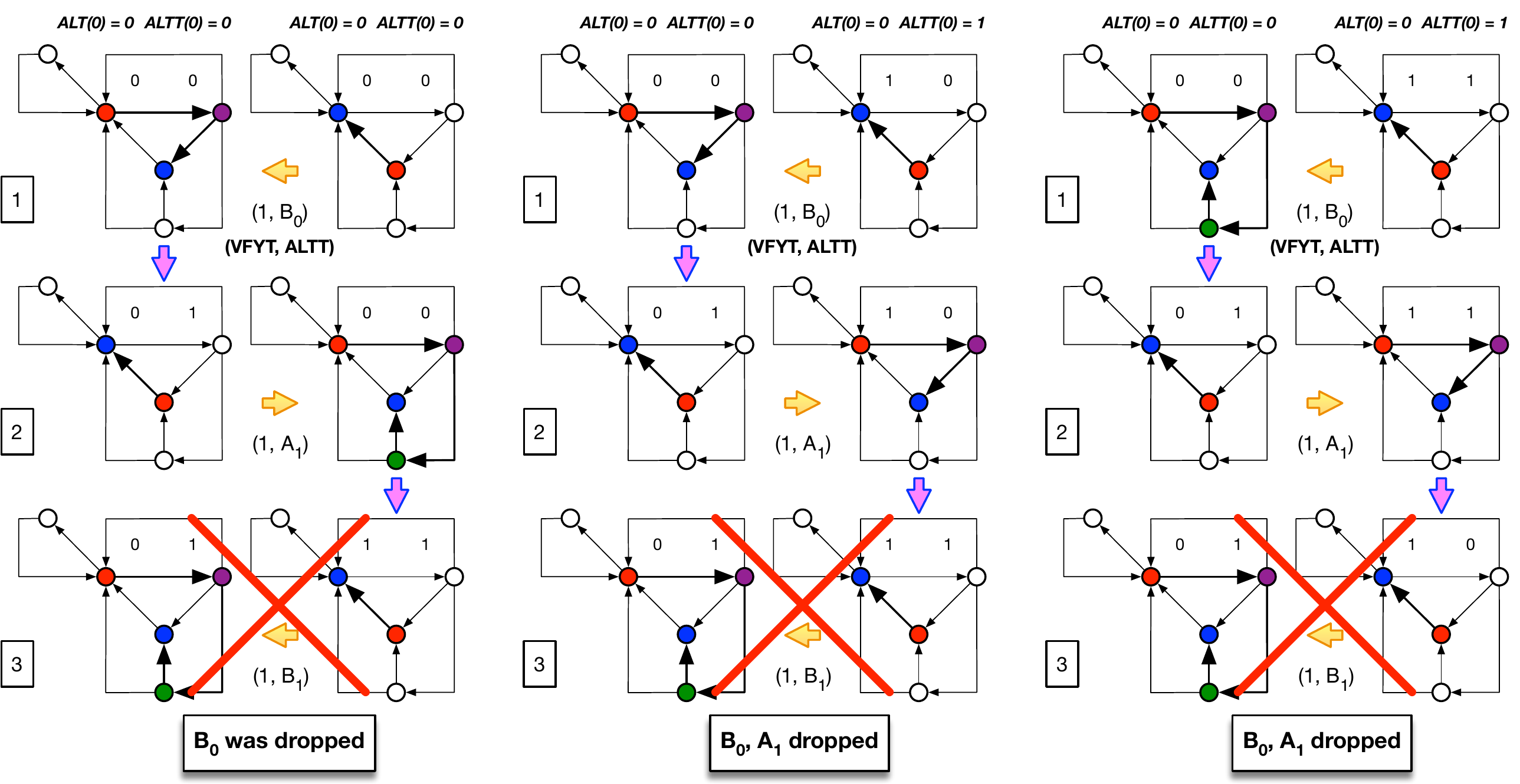}
\caption{\small\textbf{Graphical analysis of ICs 1, 3, and 4 from Table \ref{tab:lynch_ic} for the Lynch protocol}}
\label{fig:lynch_ics_1}
\vspace{-0.3cm}
\end{figure}

\begin{figure}[H]
\centering
\includegraphics[width=17cm]{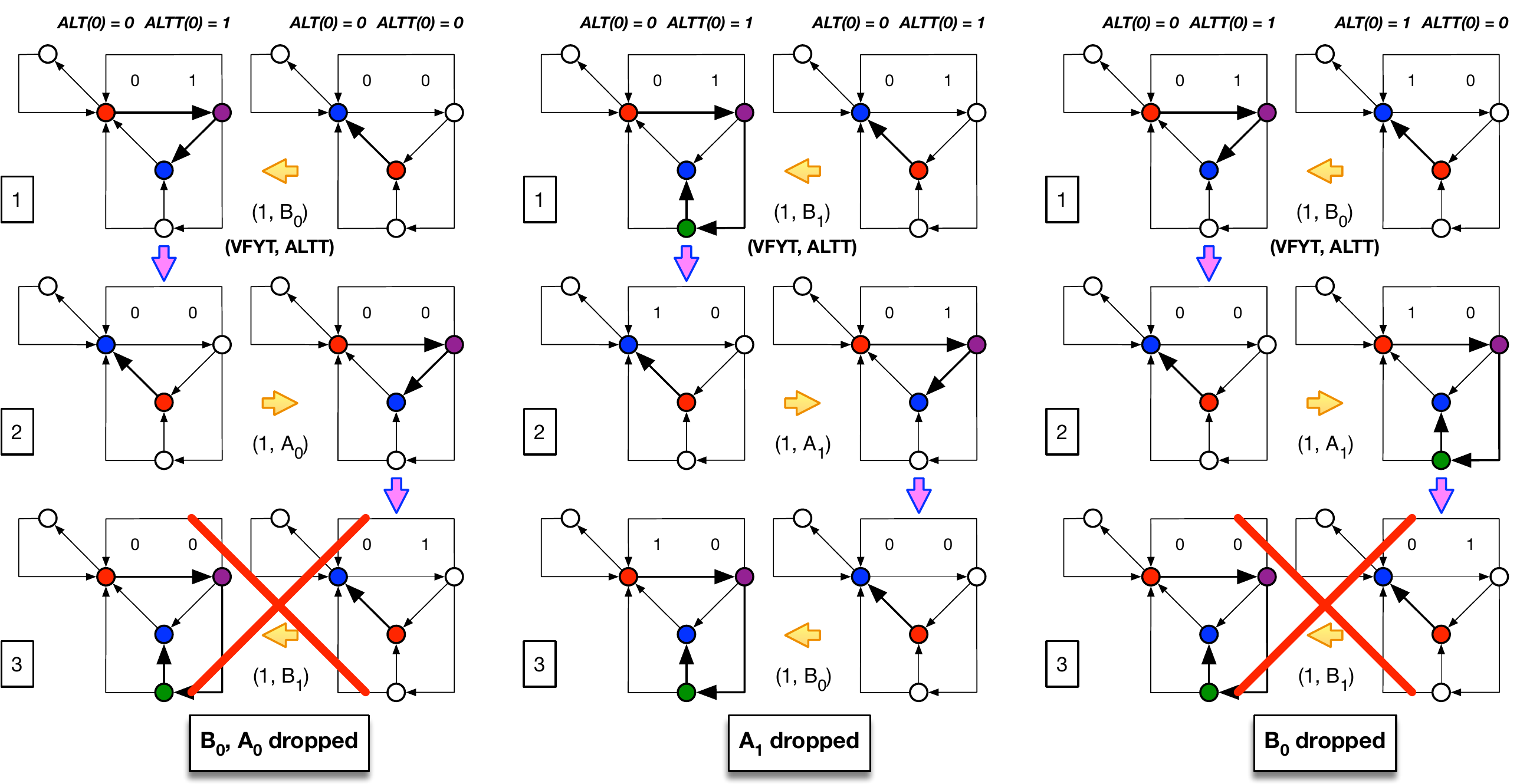}
\caption{\small\textbf{Graphical analysis of ICs 5, 6, and 7 from Table \ref{tab:lynch_ic} for the Lynch protocol}}
\label{fig:lynch_ics_2}
\vspace{-0.3cm}
\end{figure}

\begin{figure}[H]
\centering
\includegraphics[width=17cm]{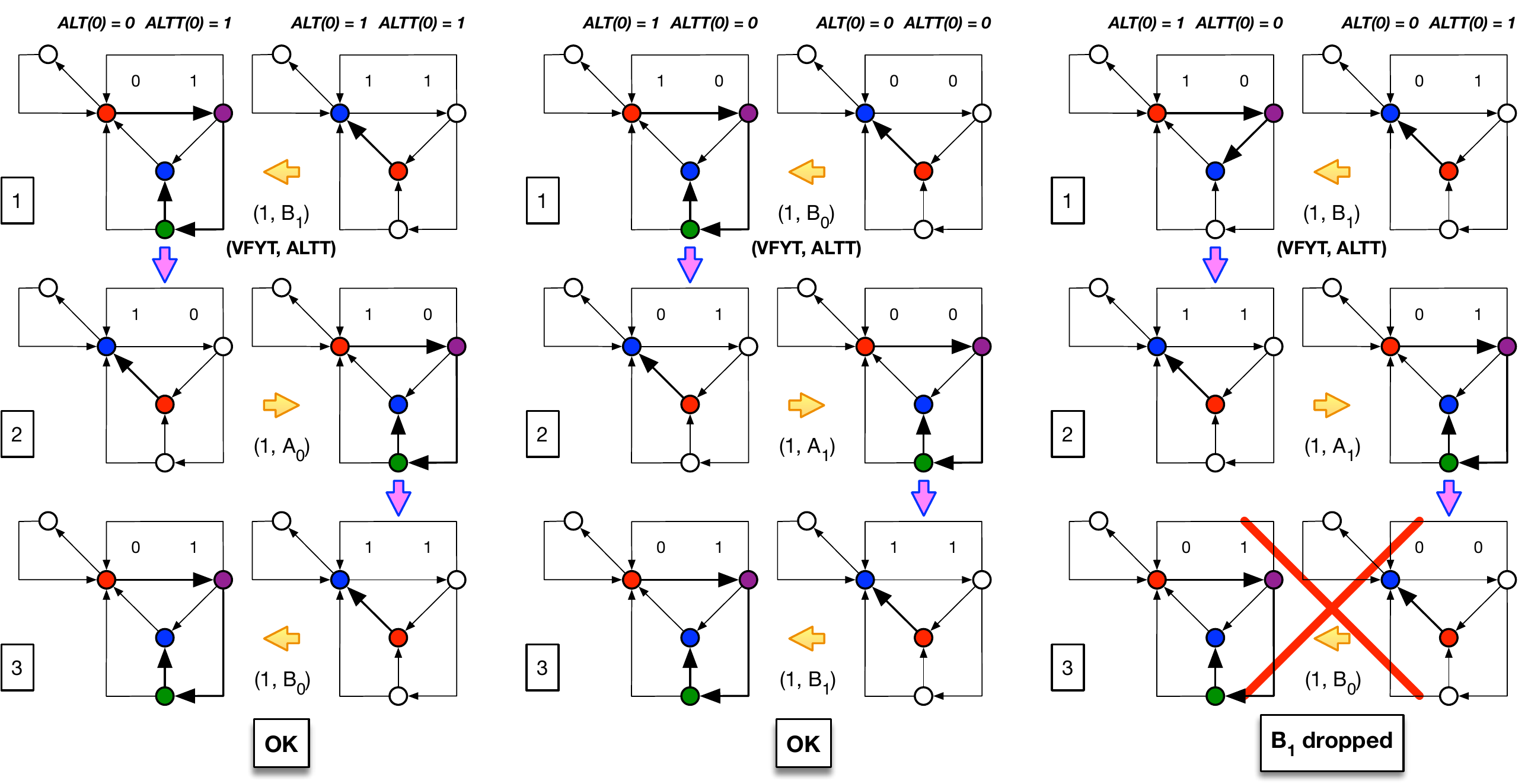}
\caption{\small\textbf{Graphical analysis of ICs 8, 9, and 10 from Table \ref{tab:lynch_ic} for the Lynch protocol}}
\label{fig:lynch_ics_3}
\vspace{-0.3cm}
\end{figure}

The combination shown in red in Table \ref{tab:lynch_ic} corresponds to Figs.\ \ref{fig:lynch_sequence} and \ref{fig:lynch_automata_diagrams1}. The other 9 of the first 10 are shown in Figs.\ \ref{fig:lynch_ics_1}-\ref{fig:lynch_ics_3}. The large red cross indicates that the wrong file is being sent at that stage. Absence of a red cross in the presence of an indication of which file is dropped means that the wrong file is sent at the \emph{next} stage (not shown). ``OK'' means success.

\begin{table}[H]
\begin{centering}
\small
{
\begin{tabular}{r  c  c  c  c  }
IC &$ALT_A(0)$ &$ALTT_A(0)$ &$ALT_B(0)$ &$ALTT_B(0)$ \\
\hline
1 &0&0&0&0 \\
\textbf{2}& \textcolor{red}{\textbf{0}}
    &\textcolor{red}{\textbf{0}}
    &\textcolor{red}{\textbf{0}}
    &\textcolor{red}{\textbf{1}} \\
3 &0&0&1&0 \\
4 &0&0&1&1 \\
5 &0&1&0&0 \\
6 &0&1&0&1 \\
7 &0&1&1&0 \\
\textbf{8}& \textcolor{blue}{\textbf{0}}
    &\textcolor{blue}{\textbf{1}}
    &\textcolor{blue}{\textbf{1}}
    &\textcolor{blue}{\textbf{1}} \\
\textbf{9}& \textcolor{blue}{\textbf{1}}
    &\textcolor{blue}{\textbf{0}}
    &\textcolor{blue}{\textbf{0}}
    &\textcolor{blue}{\textbf{0}} \\
10 &1&0&0&1 \\
11 &1&0&1&0 \\
12 &1&0&1&1 \\
13 &1&1&0&0 \\
14 &1&1&0&1 \\
\textbf{15}& \textcolor{blue}{\textbf{1}}
    &\textcolor{blue}{\textbf{1}}
    &\textcolor{blue}{\textbf{1}}
    &\textcolor{blue}{\textbf{0}} \\
16 &1&1&1&1 \\
\end{tabular}
}
\caption{\small\textbf{Possible combinations of initial conditions for the Lynch protocol}}
\label{tab:lynch_ic}
\end{centering}
\vspace{-0.7cm}
\end{table}

To help with the verification, Table \ref{tab:lynch_alt_bit_traces} shows the traces of alternation bit values at the two terminals for the first 3 steps of the sequence of Fig.\ \ref{fig:lynch_sequence} and for the first 10 ICs in Table \ref{tab:lynch_ic}.


\setlength{\tabcolsep}{8pt}
\begin{table}[H]
\begin{centering}
\small
{
\begin{tabular}{r  l  l  l  c  c  l  l  l }
IC & $ALT_A\ \ $             &$ALTR_A$ &$ALTT_A$     &&&$ALT_B\ \ $           &$ALTR_B$ &$ALTT_B$ \\
\hline
\multirow{3}{*}{1}&
0                       &0  &0$\rightarrow$1    &&&0                 &-  &0 \\
&0                      &0  &1                  &&&0$\rightarrow$1   &0  &0$\rightarrow$1 \\
&0$\rightarrow$1        &1  &1$\rightarrow$0    &&&1                 &1  &1 \\
&&&&&&&&\\
\multirow{3}{*}{3}&
0                       &0  &0$\rightarrow$1    &&&1                &-  &0 \\
&0                      &0  &1                  &&&1                &1  &0$\rightarrow$1 \\
&0$\rightarrow$1        &1  &1$\rightarrow$0    &&&1                &1  &1 \\
&&&&&&&&\\
\multirow{3}{*}{4}&
0$\rightarrow$1         &0  &0$\rightarrow$1    &&&1                &-  &1 \\
&1                      &1  &1                  &&&1                &1  &1$\rightarrow$0 \\
&0$\rightarrow$1        &0  &1$\rightarrow$0    &&&1                &1  &0 \\
&&&&&&&&\\
\multirow{3}{*}{5}&
0$\rightarrow$1         &0  &1$\rightarrow$0    &&&0                &-  &0 \\
&0                      &0  &0                  &&&0                &0  &0$\rightarrow$1 \\
&0$\rightarrow$1        &1  &0$\rightarrow$1    &&&0                &0  &1 \\
&&&&&&&&\\
\multirow{3}{*}{6}&
0$\rightarrow$1         &1  &1$\rightarrow$0    &&&0                &-  &1 \\
&1                      &1  &0                  &&&0                &0  &1$\rightarrow$0 \\
&1$\rightarrow$0        &0  &0$\rightarrow$1    &&&0                &0  &0 \\
&&&&&&&&\\
\multirow{3}{*}{7}&
0                       &0  &1$\rightarrow$0    &&&1                &-  &0 \\
&0                      &0  &0                  &&&1$\rightarrow$0  &0  &0$\rightarrow$1 \\
&0$\rightarrow$1        &1  &0$\rightarrow$1    &&&0                &0  &1 \\
&&&&&&&&\\
\multirow{3}{*}{8}&
0$\rightarrow$1         &1  &1$\rightarrow$0    &&&1                &-  &1 \\
&1                      &1  &0                  &&&1$\rightarrow$0  &0  &1$\rightarrow$0 \\
&1$\rightarrow$0        &0  &0$\rightarrow$1    &&&0                &0  &0 \\
&&&&&&&&\\
\multirow{3}{*}{9}&
1$\rightarrow$0         &0  &0$\rightarrow$1    &&&0                &-  &0 \\
&0                      &0  &1                  &&&0$\rightarrow$1  &1  &0$\rightarrow$1 \\
&0$\rightarrow$1        &1  &1$\rightarrow$0    &&&1                &1  &1 \\
&&&&&&&&\\
\multirow{3}{*}{10}&
1                       &1  &0$\rightarrow$1    &&&0                &-  &1 \\
&1                      &1  &1                  &&&0$\rightarrow$1  &1  &1$\rightarrow$0 \\
&1$\rightarrow$0        &0  &1$\rightarrow$0    &&&1                &1  &0 \\
\end{tabular}
}
\caption{\small\textbf{Record of alternation bit traces for the two terminals and the first 10 cases of Table \ref{tab:lynch_ic}}}
\label{tab:lynch_alt_bit_traces}
\end{centering}
\vspace{-0.7cm}
\end{table}

We stop at the first 10 cases because they are enough for the trend to be recognised, which will be cast as an ASM rule in the next chapter as the following two conditions which must be satisfied simultaneously:
\begin{align}
ALT_A(0) \ne ALTT_B(0) \qquad\qquad ALTT_A(0) = ALT_B(0).
\end{align}
There are only 4 initial conditions that satisfy these conditions: 2, 8, 9, and 15, where the latter three are shown in blue font in the table. We have not shown the 15$^{th}$ IC explicitly and leave it as an exercise for the reader to verify.

\subsubsection{AB Protocol}
\begin{figure}[H]
\centering
\includegraphics[width=17cm]{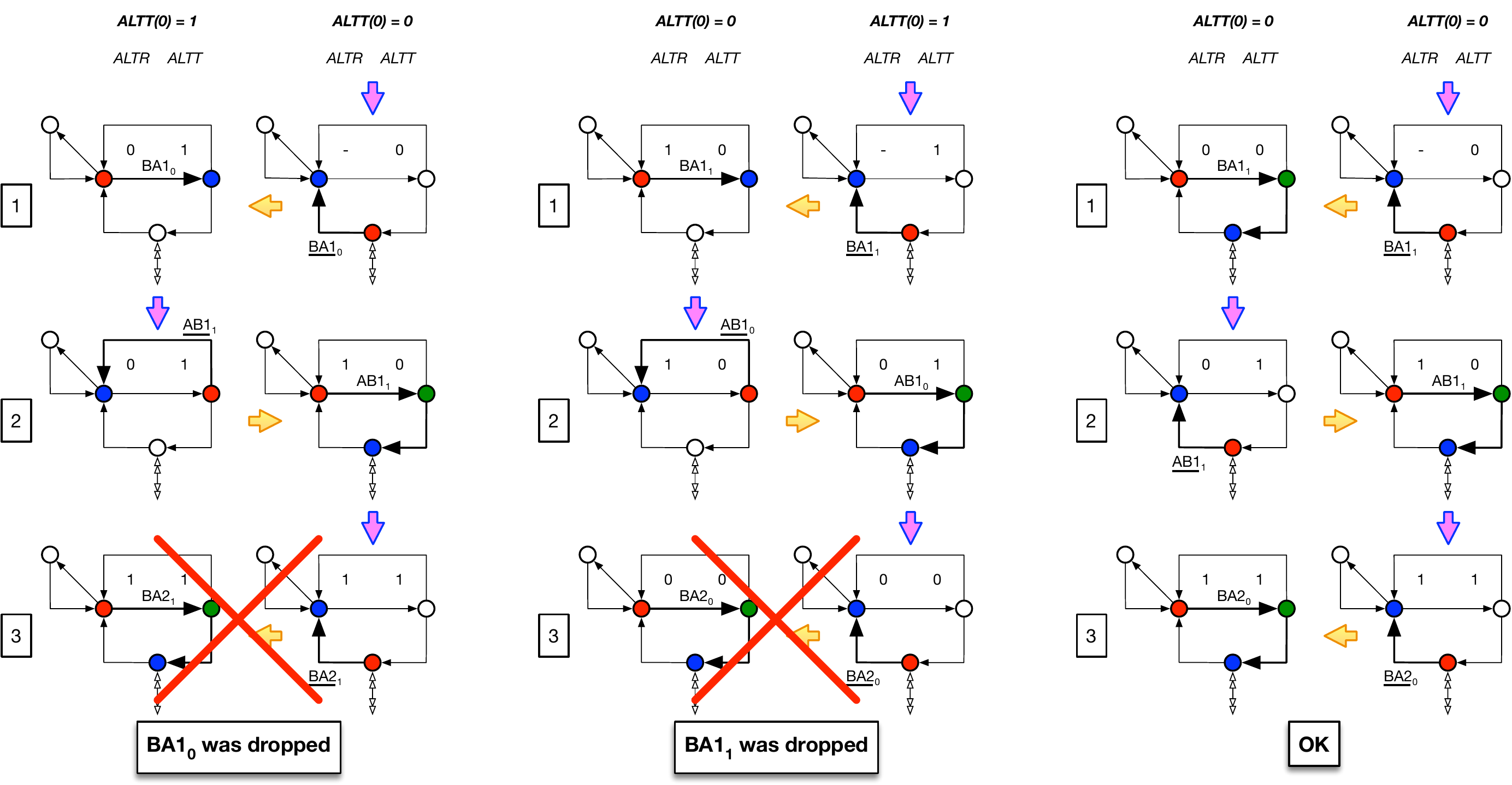}
\caption{\small\textbf{Graphical analysis of remaining initial conditions for the AB protocol}}
\label{fig:abp_ic}
\vspace{-0.3cm}
\end{figure}

For the AB protocol, since $ALTR$ is overwritten by whatever $ALTT$ from the other terminal is sending, we don't need to worry about it. So there are only four cases of interest for the four possible combinations of $(ALTT_A(0), ALTT_B(0))$. The first one, (1, 1) has already been addressed in the figures above. The remaining three possibilities are shown in Fig.\ \ref{fig:abp_ic}, from which we deduce that for this protocol to work reliably the initial conditions when B starts are either (1, 1) or (0, 0). On the other hand, it can easily be verified by inspection that if A starts the initial conditions should be either (0, 1) or (1, 0). Since it may be difficult to synchronise two remote terminals, an easy fix that allows the use of any combination is to add a dummy file at the beginning of the transmission, so that if it is dropped nothing is lost.

The figure assumes that no errors occur in the first few steps. We should examine also the case where one or more errors occur. Let's assume as before that B starts, and that an error occurs. What follows is a series of ``steps'' to help with the logical flow of events, even if in some of the steps no actual event takes place:

\begin{enumerate}
\small
\item B sends msg (BA1, 0) to A, where the 0 is the initial value of $ALTT_B$.
\item BA1 is corrupted en route.
\item A detects the error and goes into its error state.
\item A’s initial message has not been initialized, but its $ALTT$ bit has: $ALTT_A = 0$.
\item A sends ($undef$, 0) back to B. $undef$ here can be anything: 000, or random garbage.
\item Let’s assume that there are no errors and B receives the message. The assumption is that both terminals can tell if an error occurred, so B knows that an error has not occurred. However, its acceptance condition is that $ALTR_B \ne ALTT_B$. In this case they are both 0 so B does not accept the garbage message.
\item B resends (BA1, 0), i.e. without updating the payload or $ALTT_B$.
\item Assuming no error, A receives BA1 and accepts it because its accepting condition is $ALTR_A = ALTT_A$ and they are both 0.
\item A fetches its first message AB1 and flips its ALTT bit, so sends the message (AB1, 1).
\item Assuming no error, B receives AB1 and accepts it since $1 \ne 0$.
\item Etc.
\end{enumerate}

If there is no error at Step 2, then A overwrites the garbage with AB1 at Step 3 and flips $ALTT_A$, so ``go to'' Step 9. If there is an error at Step 6, the following will happen:

\begin{enumerate}
\small
\setcounter{enumi}{5}
\item B detects a transmission error and goes into its error state.
\item B resends (BA1, 0)
\item (The rest is the same)
\end{enumerate}

\newpage
\section{ASM Specification of the AB Protocol}
\label{asmspec}

\subsection{High-Level Requirements}
The Alternating Bit (AB) protocol was formally specified and verified by James Huggins \cite{Huggins1995} using evolving algebras, i.e.\ what later became ASMs. Here we follow the methodology guidelines and start from the requirements, from which the functions and rules and rules are built up step-by-step. In a real implementation there is a notion of timeout that is not present in the automata described by Lynch or Bartlett et al.\ and that can be specified at the next refinement level. We now describe textual descriptions of what each terminal must do. These will become ASM rules in the next section.

\begin{packed_enumerate}
\small
\item For each terminal, the $ALTT$ bit can be thought of as the number of the file being sent mod 2.
\item Each terminal needs to initialise $ALTT$. As shown in the previous chapter, we can use $ALTT_A(0) := 1$ and $ALTT_B(0) := 1$.
\item $ALTR$ of the receiving terminal is always overwritten by the $ALTT$ sent by the sending terminal, so its initial value could be $undef$.
\item During normal operation and consistently with ASM practice, when $ALTR$ and $ALTT$ are not overwritten their local values at each terminal remain unchanged.
\item For each terminal, when a message is received without error and accepted the value of $ALTT$ is inverted: $ALTT := \neg ALTT$. Equivalently, the file number is incremented and $ALTT := N_{file} \mod 2$. In this case, the next file is readied and sent to the other terminal.
\item For each terminal, when a message is received without error but it is not accepted the value of $ALTT$ is left unchanged. In this case, the previous file is resent to the other terminal.
\item The conditions for accepting the files are different and depend on the initial conditions. For the initial conditions used here, the condition for acceptance by Terminal B is $ALTR_B \ne ALTT_B$, whereas for Terminal A the condition is $ALTR_A = ALTT_A$.
\item The protocol assumes that an error detection system is in place, such as a checksum, that allows the receiving terminal to detect reliably the presence of errors generated during transmission.
\item If an error is detected, the receiver resends the current file: it does not store what just arrived and it does not prepare the next file in the other direction. If no error is detected, see Req. 5.
\item The two terminals take turns at sending messages, where each message is composed of one file and the $ALTT$ control bit.
\end{packed_enumerate}

\subsection{ASM Ground Model}
\subsubsection{Ground Model Mapped from Requirements}
\textbf{Req.\ 1}. It turns out that since in the AB protocol the two terminals are not independent this rule is not easy to implement. It is much easier to update $ALTT$ to its complement every time a file is successfully received and the control bit test has been passed. $Terminal$ is a set, while $fileNumber$ and $ALTT$ are dynamic functions of a single variable:
\begin{asm}
Terminal = \{A, B\} \\
fileNumber\colon Terminal \rightarrow \mathbb{N} \\
// ALTT (Terminal \rightarrow \mathbb{B}) = fileNumber(terminal)\ mod\ 2 \\
ALTT\colon Terminal \rightarrow \mathbb{B}
\end{asm}

\textbf{Req.\ 2}. With the preferred way to handle $ALTT$ updates just described, $ALTT$ is no longer a derived function and has to be initialized explicitly. In ASM rule specification, parallel execution is encountered more often than sequential execution. Therefore, the default is parallel execution and, as shown in the following rule, it does not require any special markers. However, where they are deemed necessary for added clarity, single curly brackets are used to indicate a parallel code block.
\clearpage
\begin{asm}
\ASM{Initialize} = \+
    fileNumber(A) := 1 \\
    fileNumber(B) := 1 \\
    ALTT(A) := true \\
    ALTT(B) := true \\
    counter := 1 \\
    initialized := true
\end{asm}

\textbf{Req.\ 3}. Although setting $ALTR(0) = undef$ for both terminals is in principle correct, in the CASM code to be discussed below we set its initial value explicitly. $ALTR$ is a dynamic function of one variable, $\ASM{ReceiveBit}$ is an ASM rule, and $otherTerminal$ is a derived (dynamic) function of one variable:
\begin{asm}
ALTR\colon Terminal \rightarrow \mathbb{B} \\
    \\
\ASM{ReceiveBit}(terminal) = \+
    ALTR(terminal) := ALTT(otherTerminal(terminal)) \-
    \\
otherTerminal(terminal) \rightarrow Terminal = \+
    \IF terminal = A \THEN B \ELSE A
\end{asm}

\textbf{Req.\ 4} is always satisfied by default by an ASM model.

\textbf{Reqs.\ 5-7}. The next rule requires sequential execution of some of its rules and statements. Following CASM syntax, this is indicated by the notation $\{| \quad P \SEQ Q \SEQ R \cdots \quad |\}$. Although in ASM syntax indentation is sufficient to indicate code blocks, the \textbf{if} statement below is an example of code that benefits from delimiters for added clarity. This makes it less likely that the CASM code, for which indentation is \emph{not} sufficient and which requires such delimiters, will be implemented incorrectly.
\begin{asm}
\ASM{ReceiveSuccess}(terminal) = \{| \+
    \ASM{ReceiveBit}(terminal)
        \qquad\qquad\qquad\qquad\qquad \text{//Update the $ALTR$ bit} \\
    \LET condition = \+
        \IF terminal = A \THEN \+
            ALTR(terminal) = ALTT(terminal) \-
        \ELSE \+
            ALTR(terminal) =\ \NOT ALTT(terminal) \-
        \IN \+
            \IF condition \THEN \{ \+
                // \text{Load next file to be sent:} \+
                fileNumber(terminal) :=                 
                    fileNumber(terminal)+1 \-
                // \text{Update the control bit:} \+
                ALTT(terminal) :=\ \NOT ALTT(terminal) \-
                // fileNumber(otherTerminal(terminal)) \text{ would be stored here} \-
            \} \dec\dec\-
|\}
\end{asm}
Remark: Although it is possible to do without a sequential rule here (e.\,g.,\ by replacing each occurence of $ALTR(terminal)$ with its definition $ALTT(otherTerminal(terminal))$), we decided to model it like this in order to emphasize that a bit is \emph{received} before it is processed.

\textbf{Req.\ 8}. In the ground model we are not modelling random error occurrence. Rather, since we wish to validate the ASM model with the CASM executable model we assume the error occurrence shown in Fig.\ \ref{fig:abp_sequence}. 
This is specified with the following static functions, where round parentheses denote the usual mathematical meaning of a fixed-order tuple:
\begin{asm}
errTraceA\colon \mathbb{Z} \rightarrow \mathbb{B} = (false, false, true, false, false, false, true, false, true, false, false) \\
errTraceB\colon \mathbb{Z} \rightarrow \mathbb{B} = (false, false, true, true, false, true, true, false, true, false, false)
\end{asm}

\textbf{Req.\ 9}. The \ASM{SendMsg} rule is just a stub since at the current level of abstraction all the work is done by the \ASM{ReceiveSuccess} rule:
\begin{asm}
\ASM{ReceiveMsg}(terminal, error) = \+
    \IF\ \NOT error \THEN
        \ASM{ReceiveSuccess}(terminal)\-
\\
\ASM{SendMsg} = skip
\end{asm}

\textbf{Req.\ 10}. ASM specifications normally involve a set of rules and a ``main'' rule that contains the control flow of the algorithm. In the CASM language the $\ASM{Main}$ rule is built-in and invisible to the programmer in order to allow more flexibility in the naming of the top rule, called $\ASM{Run}$ in our case:
\begin{asm}
\ASM{Run} = \+
    \IF initialized\ \neq true \THEN \+
        \ASM{Initialize} \-
    \ELSE\+ 
        counter := counter + 1 \\
        \IF counter < 12 \THEN  \{|  \inc \+
                \ASM{SendMsg} \\
                \ASM{ReceiveMsg}(A, errTraceA(counter)) \\
                \ASM{SendMsg} \\
                \ASM{ReceiveMsg}(B, errTraceB(counter)) \-
            |\} \-
        \ELSE \+
            stop
\end{asm}
Remark: Sequential execution can be avoided by introducing a ``phase'' variable that alternates through the different phases $\ASM{SendMsg}$ and $\ASM{ReceiveMsg}$ for both terminals. See the specifications in  Section~\ref{CASMrefinement} or \ref{coreasmSpec} for an example. 

Remark: The variable $counter$ provides an example of the difference in thinking required between sequential and parallel coding. With reference to the rule \ASM{Initialize}, above, in a sequential program $counter$ should be initialized to 0. Because $counter$'s increment occurs in a parallel block in the rule \ASM{Run}, however, the whole rule will be executed with its current value, i.e.\ 1 at the beginning, such that $counter$ will equal 2 at the next state, as desired.

\subsubsection{Ground Model in Compact Form}
Universe(s):
\begin{asm}
Terminal = \{A, B\}
\end{asm}

Static functions:
\begin{asm}
errTraceA\colon \mathbb{Z} \rightarrow \mathbb{B} = (false, false, true, false, false, false, true, false, true, false, false) \\
errTraceB\colon \mathbb{Z} \rightarrow \mathbb{B} = (false, false, true, true, false, true, true, false, true, false, false)
\end{asm}

Dynamic functions:
\begin{asm}
fileNumber\colon Terminal \rightarrow \mathbb{N} \\
ALTT\colon Terminal \rightarrow \mathbb{B} \\
ALTR\colon Terminal \rightarrow \mathbb{B}
\end{asm}

Derived function(s):
\begin{asm}
otherTerminal(terminal) \rightarrow Terminal = \+
    \IF terminal = A \THEN B \ELSE A
\end{asm}

Rules:
\begin{asm}
\ASM{Initialize} = \+
    fileNumber(A) := 1 \\
    fileNumber(B) := 1 \\
    ALTT(A) := true \\
    ALTT(B) := true \\
    counter := 1 \\
    initialized := true \-
\\
\ASM{ReceiveBit}(terminal) = \+
    ALTR(terminal) := ALTT(otherTerminal(terminal)) \-
\\
\ASM{ReceiveSuccess}(terminal) = \{| \+
    \ASM{ReceiveBit}(terminal)
        \qquad\qquad\qquad\qquad\qquad \text{//Update the $ALTR$ bit} \\
    \LET condition = \+
        \IF terminal = A \THEN \+
            ALTR(terminal) = ALTT(terminal) \-
        \ELSE \+
            ALTR(terminal) =\ \NOT ALTT(terminal) \-
        \IN \+
            \IF condition \THEN \{ \+
                // \text{Load next file to be sent:} \+
                fileNumber(terminal) :=                 
                    fileNumber(terminal)+1 \-
                // \text{Update the control bit:} \+
                ALTT(terminal) :=\ \NOT ALTT(terminal) \-
                // fileNumber(otherTerminal(terminal)) \text{ would be stored here} \-
            \} \dec\dec\-
|\} \\
\\
\ASM{ReceiveMsg}(terminal, error) = \+
    \IF\ \NOT error \THEN
        \ASM{ReceiveSuccess}(terminal)\-
\\
\ASM{SendMsg} = skip \\
\\
\ASM{Run} = \+
    \IF initialized\ \neq true \THEN \+
        \ASM{Initialize} \-
    \ELSE\+ 
        counter := counter + 1 \\
        \IF counter < 12 \THEN  \{|  \inc \+
                \ASM{SendMsg} \\
                \ASM{ReceiveMsg}(A, errTraceA(counter)) \\
                \ASM{SendMsg} \\
                \ASM{ReceiveMsg}(B, errTraceB(counter)) \-
            |\} \-
        \ELSE \+
            stop
\end{asm}

\newpage
\section{CASM Model of the AB Protocol}
\label{casm}

\subsection{Introduction to the CASM Language}
\label{casm_intro}

The Corinthian Abstract State Machine (CASM) language, along with its tooling and framework, represents a concrete ASM implementation of the ASM theory defined by Börger and Stärk \cite{BoeSta03} whose purpose is to simulate (execute) ASM specifications.

CASM features a statically strong, inferred, and typed language to aid the specifier in defining only the necessary types for definition elements.
The intermediate types are completely inferred and statically checked by appropriate compiler techniques \cite{paulweber2022phd}.
The language implementation\footnote{\url{https://casm-lang.org/download}} consists currently of three tools -- a numeric and symbolic interpreter \texttt{casmi}, a source code format beautifier \texttt{casmf}, and a Language Server Protocol\footnote{\url{https://langserver.org}} (LSP) daemon \texttt{casmd} for LSP client editor integration.

Historically speaking, the first version of CASM was created during a research project at the Vienna University of Technology (TU Wien) in order to formally describe and simulate computer architectures using ASMs \cite{lezuo2013casm}.
The research effort started out using CoreASM \cite{farahbod2007coreasm} but, due to a strong demand on simulation (execution) performance, the Java-based interpreter implementation of CoreASM could not satisfy the desired goals. Therefore, a specific subset of language features of CoreASM was initially used -- all Basic ASM rules -- with some minor syntax adaptations.
At the time, the project featured a C$^{++}$-based parser, static code analysis, interpreter, and compiler prototype implementation \cite{lezuo2014casm} \cite{paulweber2014msc}.
Sadly this project and its outcome were covered by an NDA. Therefore, since 2014 a completely new CASM implementation written from scratch was created as an open-source project\footnote{\url{https://github.com/casm-lang}} by Paulweber et al. \cite{paulweber2016abz} \cite{paulweber2018abz} \cite{paulweber2021abz}.

In addition to researching core aspects for the (improved) execution of ASM models, Paulweber et al.\ \cite{paulweber2021tosem} \cite{paulweber2021jss} started another investigation to find empirical evidence of how the understandability and usability of ASM languages can be improved using object-oriented language abstractions.
The result of this research led to the integration of a trait-based syntax extension \cite{paulweber2020abz}. The trait-based integration provides the ability to specify even CASM language and run-time features within CASM itself, and makes it possible to move progressively more and ultimately all parts of the language definition and compiler behaviour away from the C$^{++}$-based implementation and to a CASM-based specification \cite{paulweber2020abz}.

\begin{figure}[H]
\centering
\includegraphics[width=17.5cm]{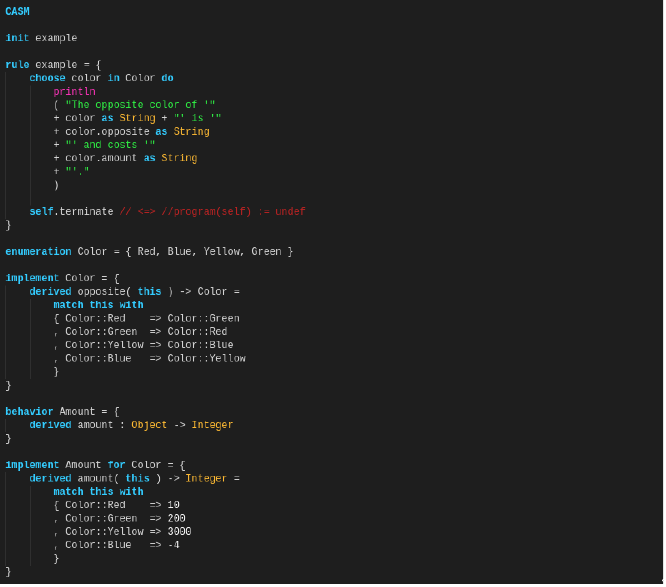}
\includegraphics[width=17.5cm]{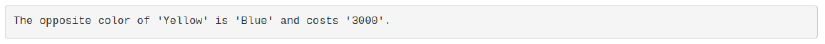}
\caption{\small\textbf{Trait-Based CASM Specification with Example ASM Run Output}}
\label{fig:casm_3}
\end{figure}

For example, Fig.\ \ref{fig:casm_3} demonstrates how the default behaviour for the type \texttt{Color} is defined in a way that makes it possible to derive the colour opposite to the current one.
Furthermore, we can see that a trait \texttt{Amount} is defined and the implementation of that trait for \texttt{Color} specifies the color-to-amount mapping.

\subsection{Paolo's Executable CASM Model}
\label{paolocasm}
\begin{figure}[H]
\raggedright
\includegraphics[width=17cm]{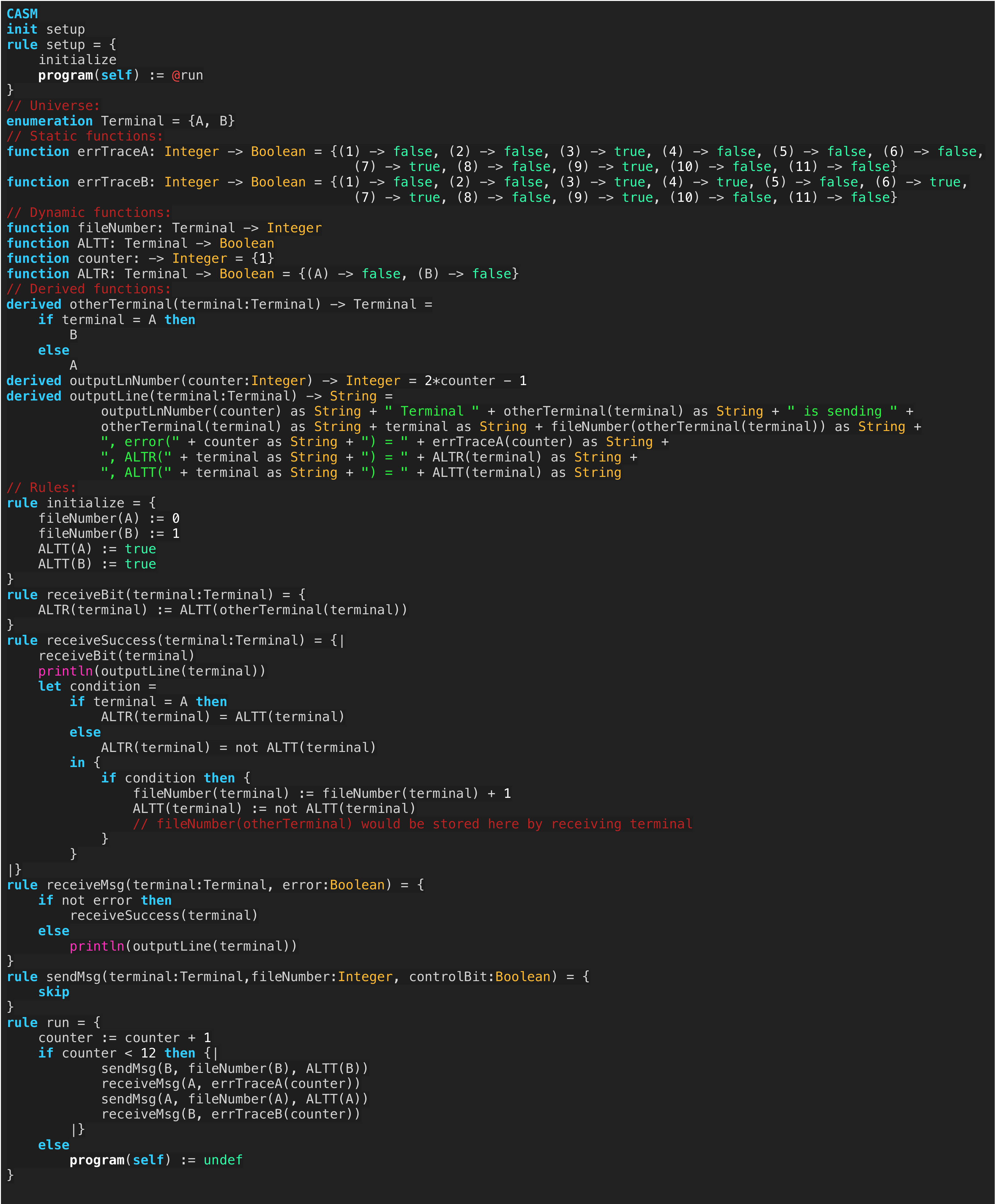}
\caption{\small\textbf{Executable CASM specification of the AB protocol}}
\label{fig:casm_1}
\end{figure}

Fig.\ \ref{fig:casm_1} shows the executable CASM model as a screenshot of the browser-based CASM editor. This model was put together mainly by Paolo but with close guidance from Philipp.

The level of abstraction of this model is very high. Thus, rather than sending actual files a counter for file transmission is incremented when there is no error in the transmission. The errors, in turn, follow the same pattern of the Lynch sequence in Figs.\ \ref{fig:lynch_sequence} and \ref{fig:abp_sequence} in order to be able to check if the model replicates the same behaviour, which in this case is given by file numbers in both directions.

\begin{figure}[H]
\centering
\includegraphics[width=9cm]{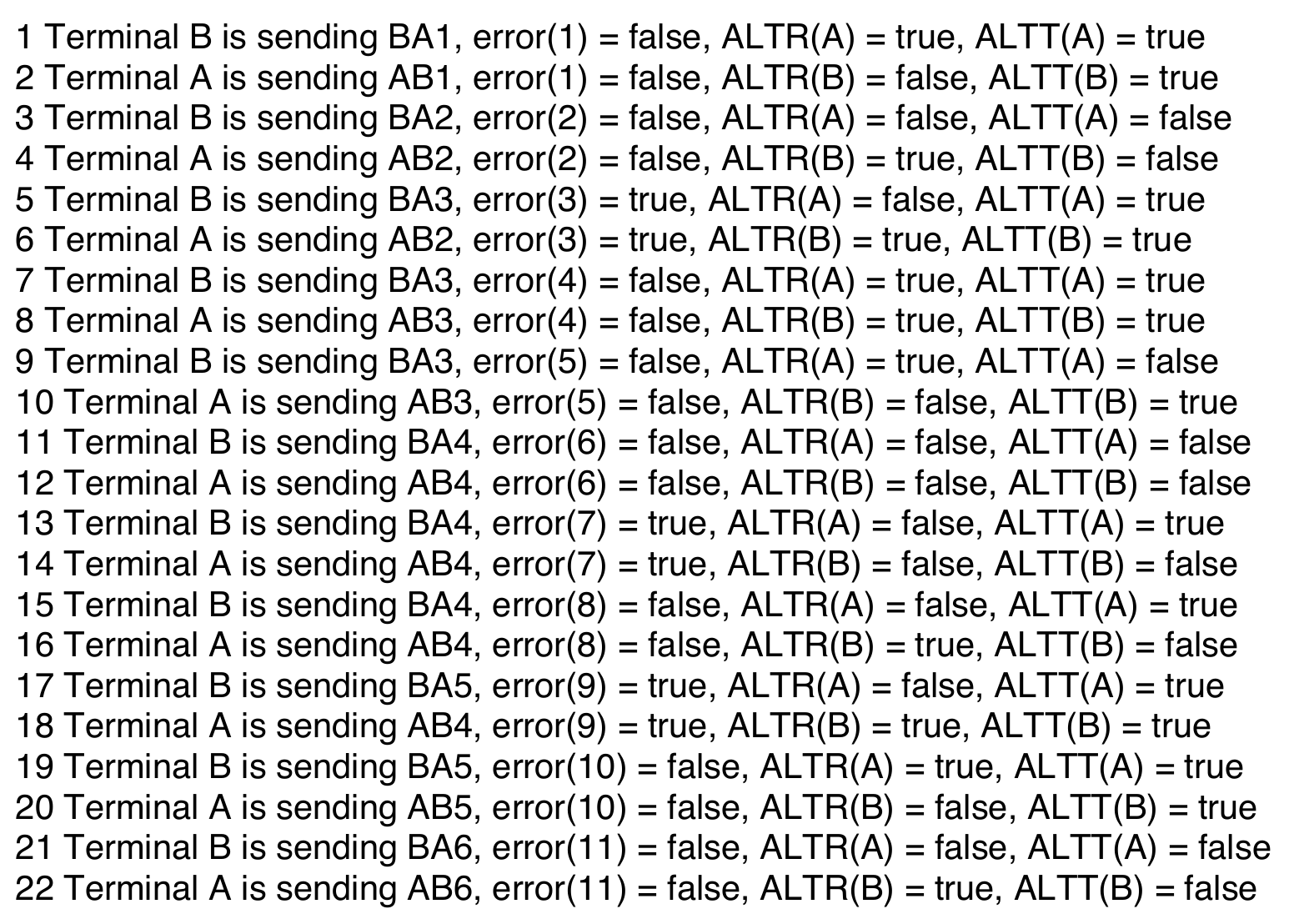}
\caption{\small\textbf{Output of the executable CASM specification of Fig.\ \ref{fig:casm_1}}}
\label{fig:casm_output}
\end{figure}

The output is shown in Fig.\ \ref{fig:casm_output}. This figure is not as easy to read as Fig.\ \ref{fig:abp_sequence} but it contains the same information, thereby validating the model. Line numbers were introduced to make it easier to compare to the output in Fig.\ \ref{fig:abp_sequence}.

\subsection{Philipp's CASM Refinement}
\label{CASMrefinement}

Based on the ground model specification shown in Fig.\ \ref{fig:casm_1}, a lot of refinement steps can be performed.
The first concern is to remove the hidden computation steps implied by the use of the sequential execution semantics block inside the \texttt{run} rule.
The computation is ``hidden'' to the ASM agent, meaning that the intermediate states are not part of the global state set of the ASM.
This has negative verification and computational efficiency implications and, therefore, it is best to avoid it.

\begin{figure}[H]
\centering
\includegraphics[width=17cm]{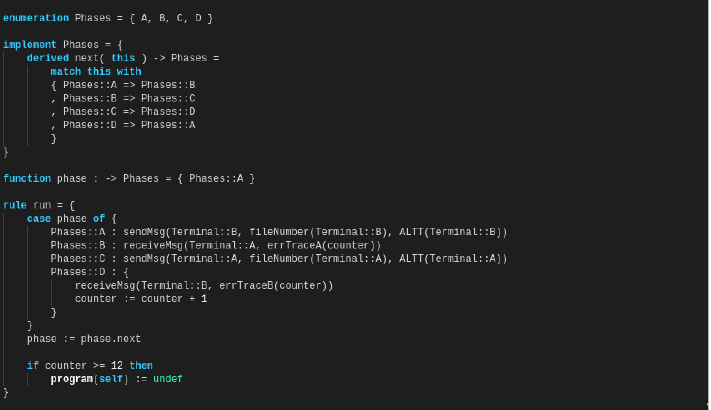}
\caption{\small\textbf{Refinement of Fig.\ \ref{fig:casm_1}}}
\label{fig:casm_2}
\end{figure}

Thus, Fig.\ \ref{fig:casm_2} depicts the \texttt{run} rule's refinement that removes the sequential execution semantics block, which executes the receive and send message rules, and uses a \texttt{Phases} abstraction to represent each phase of the protocol computation within a dedicated ASM step.
Furthermore, to showcase the trait implementation of CASM, a default behaviour for the enumeration \texttt{Phases} was defined to retrieve the \texttt{next} phase value given a current phase value.
This behaviour encapsulation allows us to decouple the specification of the action performed during a phase and the update of the phase to the next phase by using just a single update rule.

\newpage
\section{CoreASM Model of the AB Protocol}
\label{coreasm}
\subsection{Introduction to CoreASM}

As discussed in Chapter \ref{asms}, Abstract State Machines (ASMs)\cite{Boerger03a} are algebraic structures with rules that manipulate them, without a precise language definition. In place of a language, mathematical notation is used in a flexible way and new abbreviations or constructs are introduced in an ad-hoc manner, with the goal to improve the readability and understandability of formal specifications.
The drawback of this flexibility is the difficulty in developing an execution engine that allows simulation of even not-yet-completed abstract specifications.

CoreASM has addressed this challenge with the objective to preserve the specification character in the executable language and avoid slipping into a programming language style. To achieve this goal, CoreASM was designed based on a rigorous plugin architecture and a less strict handling of types. It originated around 2003 as a PhD project by Roozbeh Farahbod at Simon Fraser University in Vancouver, Canada \cite{farahbod2009coreasm}. 

Besides a very small core (hence the name), each language construct is provided by plugins that must be declared at the beginning of a specification. Each plugin consists of a parsing component and an execution component. A primitive bootstrap parser loads all of the named plugins and creates the parser of the currently ``plugged-in'' language by combining the partial parsers of the loaded plugins. Each partial parser adds subtrees to the general abstract syntax tree (AST) of the specification. An abstract interpreter component in CoreASM then traverses this AST and plugin-specific execution functions are called for each node. 

On the one hand, the plugin architecture obviously reduces execution speed, which was one of the main reasons also for the development of CASM. On the other hand, it allows a relatively easy extension of the executable specification language itself and thus more domain-specific specifications. The plugin architecture also enables the developer to provide language constructs that interact with Java classes or even other Java applications (see \cite{gervasi2009jasmine, farahbod2010}). 

Another interesting (and very helpful) aspect of CoreASM is that it is \emph{itself} specified precisely in ASMs. This is a good example of how formal specifications can serve as an abstract yet precise documentation of the architecture and the semantics of an application, greatly improving maintenance \cite{farahbod2009coreasm,farahbod2007coreasm}. 

CoreASM is implemented in Java and published on GitHub\footnote{\url{https://github.com/coreasm}} under the Academic Free License 3.0.\footnote{\url{https://opensource.org/licenses/AFL-3.0}} In addition to the core parsing and execution engine, an Eclipse plugin is provided that includes a language-sensitive editor, a debugger, and various other integrations. The fact that the language of a given specification is compiled based on the loaded plugins complicates the use of existing language modeling tools like Xtext.\footnote{\url{https://www.eclipse.org/Xtext/}}

Besides several improvements to the original source code, the Institute of Software Engineering and Programming Languages at Ulm University has developed the debugger component\footnote{\url{https://github.com/CoreASM/coreasm.core/tree/master/org.coreasm.eclipse/rsc/doc}} and introduced a plugin that allows aspect-oriented specifications \cite{Dausend2014}.

In addition to the formal specification in the PhD thesis \cite{farahbod2009coreasm}, a user manual is provided that shows how to use the language constructs.\footnote{\url{https://github.com/CoreASM/coreasm.core/tree/master/org.coreasm.engine/rsc/doc/user_manual}} CoreASM has been used in several projects to validate specifications, including more complex ones such as \cite{Farahbod2011, Wolski2019}.

\subsection{Alexander's CoreASM specification of AB Protocol}
\label{coreasmSpec}

As mentioned in Sect.~\ref{casm_intro}, CASM is based on CoreASM and therefore both languages are very similar. Therefore, only few adaptations were required to convert the CASM specification into an executable CoreASM specification that produces the same results. 

In detail these are (see Figs.~\ref{fig:coreasm_1},\ref{fig:coreasm_2}):
\begin{itemize}
\item Definition of used plugins at the beginning of the specification (line 4).
\item Removal of many type annotations (functions must be declared with types in order to know the arity) and replacement of some operators/keywords (e.\,g., \texttt{`='} after a function definition in CASM is equivalent to the keyword \texttt{initially} in CoreASM).
\item Due to the absence of traits in CoreASM, the current phase and its changes must be expressed more explicitly than in CASM (lines 13 -- 39).
\end{itemize}

\begin{figure}[bt]
\centering
\lstset{numbers=left, stepnumber=5, firstnumber=1, numberfirstline=true}
\begin{lstlisting}[backgroundcolor=\color{gray!5!white},frame=none]
CoreASM abp

//use standard plugins
use Standard

//set program for the executing agent
init setup
rule setup = {
    initializeALTT
    program(self) := @run
}

enum Phases = { sendAtoB, receiveB, sendBtoA, receiveA }

function phase : -> Phases initially sendAtoB

rule run = {
    case phase of 
        sendAtoB : {
        	sendMsg(B, fileNumber(B), ALTT(B))
        	phase := receiveB
        }
        receiveB : {
        	receiveMsg(A, errTraceA(counter))
        	phase := sendBtoA
        }
        sendBtoA : {
        	sendMsg(A, fileNumber(A), ALTT(A))
        	phase := receiveA
        }
        receiveA : {
		    receiveMsg(B, errTraceB(counter))
	    	counter := counter + 1
	    	phase := sendAtoB
		}
	endcase
    if counter $>$= 12 then
        program(self) := undef
}

// Req. 1:
enum Terminal = { A, B }
function fileNumber: Terminal -> Integer
function ALTT: Terminal -> Boolean

derived otherTerminal(terminal) = if terminal = A then B else A
            
function counter: -> Integer initially 1
function errTraceA: Integer -> Boolean 
    initially {1 -> false, 2 -> false, 3 -> true,  4 -> false,  5 -> false, 6 -> false,
               7 -> true,  8 -> false, 9 -> true, 10 -> false, 11 -> false}
function errTraceB: Integer -> Boolean 
    initially {1 -> false, 2 -> false, 3 -> true,  4 -> true,   5 -> false, 6 -> true,
               7 -> true,  8 -> false, 9 -> true, 10 -> false, 11 -> false}
                   
\end{lstlisting}
\caption{\small\textbf{Executable CoreASM specification }}
\label{fig:coreasm_1}
\end{figure}

\begin{figure}[bt]
\centering
\lstset{numbers=left, stepnumber=5, firstnumber=55, numberfirstline=true}
\begin{lstlisting}[backgroundcolor=\color{gray!5!white},frame=none]
// Req 1. (cont'd)
derived outputLnNumber(terminal) = 2*counter - 1 + if terminal = B then 1 else 0
derived outputLine(terminal) =
            outputLnNumber(terminal) + " Terminal " + otherTerminal(terminal)  + 
            " is sending " + otherTerminal(terminal) + terminal  + 
            fileNumber(otherTerminal(terminal))  + 
            ", error(" + counter  + ") = " + errTraceA(counter)  +
            ", ALTR(" + terminal  + ") = " + ALTR(terminal)  +
            ", ALTT(" + terminal  + ") = " + ALTT(terminal) 

// Req. 2:
rule initializeALTT = {
    fileNumber(A) := 0
    fileNumber(B) := 1
    ALTT(A) := true
    ALTT(B) := true
}

// Req. 3:
function ALTR: Terminal -> Boolean initially {A -> false, B -> false}
rule receiveBit(terminal) = {
    ALTR(terminal) := ALTT(otherTerminal(terminal))
}

// Reqs. 5 and 6 (several actions):
rule receiveSuccess(terminal) = seq
    receiveBit(terminal)
    print(outputLine(terminal))
    let condition = if terminal = A then
            ALTR(terminal) = ALTT(terminal)
        else
            ALTR(terminal) = not ALTT(terminal)
        in {
            if condition then {
                fileNumber(terminal) := fileNumber(terminal) + 1
                ALTT(terminal) := not ALTT(terminal)
            }
        }
endseq

// Req. 10:
rule receiveMsg(terminal, error) = {
    if error = false then
        receiveSuccess(terminal)
    else
        print(outputLine(terminal))
}

rule sendMsg(terminal, fileNumber, controlBit) = {
    skip
}
\end{lstlisting}
\caption{\small\textbf{Executable CoreASM specification }}
\label{fig:coreasm_2}
\end{figure}

\begin{figure}
\centering
\begin{minipage}{\textwidth-1.5cm}
\small
\begin{verbatim}
1 Terminal B is sending BA1, error(1) = false, ALTR(A) = true, ALTT(A) = true
2 Terminal A is sending AB1, error(1) = false, ALTR(B) = false, ALTT(B) = true
3 Terminal B is sending BA2, error(2) = false, ALTR(A) = false, ALTT(A) = false
4 Terminal A is sending AB2, error(2) = false, ALTR(B) = true, ALTT(B) = false
5 Terminal B is sending BA3, error(3) = true, ALTR(A) = false, ALTT(A) = true
6 Terminal A is sending AB2, error(3) = true, ALTR(B) = true, ALTT(B) = true
7 Terminal B is sending BA3, error(4) = false, ALTR(A) = true, ALTT(A) = true
8 Terminal A is sending AB3, error(4) = false, ALTR(B) = true, ALTT(B) = true
9 Terminal B is sending BA3, error(5) = false, ALTR(A) = true, ALTT(A) = false
10 Terminal A is sending AB3, error(5) = false, ALTR(B) = false, ALTT(B) = true
11 Terminal B is sending BA4, error(6) = false, ALTR(A) = false, ALTT(A) = false
12 Terminal A is sending AB4, error(6) = false, ALTR(B) = false, ALTT(B) = false
13 Terminal B is sending BA4, error(7) = true, ALTR(A) = false, ALTT(A) = true
14 Terminal A is sending AB4, error(7) = true, ALTR(B) = false, ALTT(B) = false
15 Terminal B is sending BA4, error(8) = false, ALTR(A) = false, ALTT(A) = true
16 Terminal A is sending AB4, error(8) = false, ALTR(B) = true, ALTT(B) = false
17 Terminal B is sending BA5, error(9) = true, ALTR(A) = false, ALTT(A) = true
18 Terminal A is sending AB4, error(9) = true, ALTR(B) = true, ALTT(B) = true
19 Terminal B is sending BA5, error(10) = false, ALTR(A) = true, ALTT(A) = true
20 Terminal A is sending AB5, error(10) = false, ALTR(B) = false, ALTT(B) = true
21 Terminal B is sending BA6, error(11) = false, ALTR(A) = false, ALTT(A) = false
22 Terminal A is sending AB6, error(11) = false, ALTR(B) = true, ALTT(B) = false
\end{verbatim}
\end{minipage}
\vspace{0.5cm}
\caption{\small\textbf{Output of the executable CoreASM specification of Figs.\ \ref{fig:coreasm_1} and \ref{fig:coreasm_2}, which is identical to the output of CASM in Fig.~\ref{fig:casm_output}}}
\label{fig:coreasm_output}
\end{figure}

\clearpage

\section{TLA$^+$ Model of the AB Protocol}
\label{tla}
There are several ways to write a TLA$^+$ spec for a given application. In this chapter we discuss a simple-minded version, developed by Paolo, and a more sophisticated version, developed by Manuel.

\subsection{Paolo's Spec for TLC}
\begin{figure}[H]
\raggedright
\includegraphics[width=18cm]{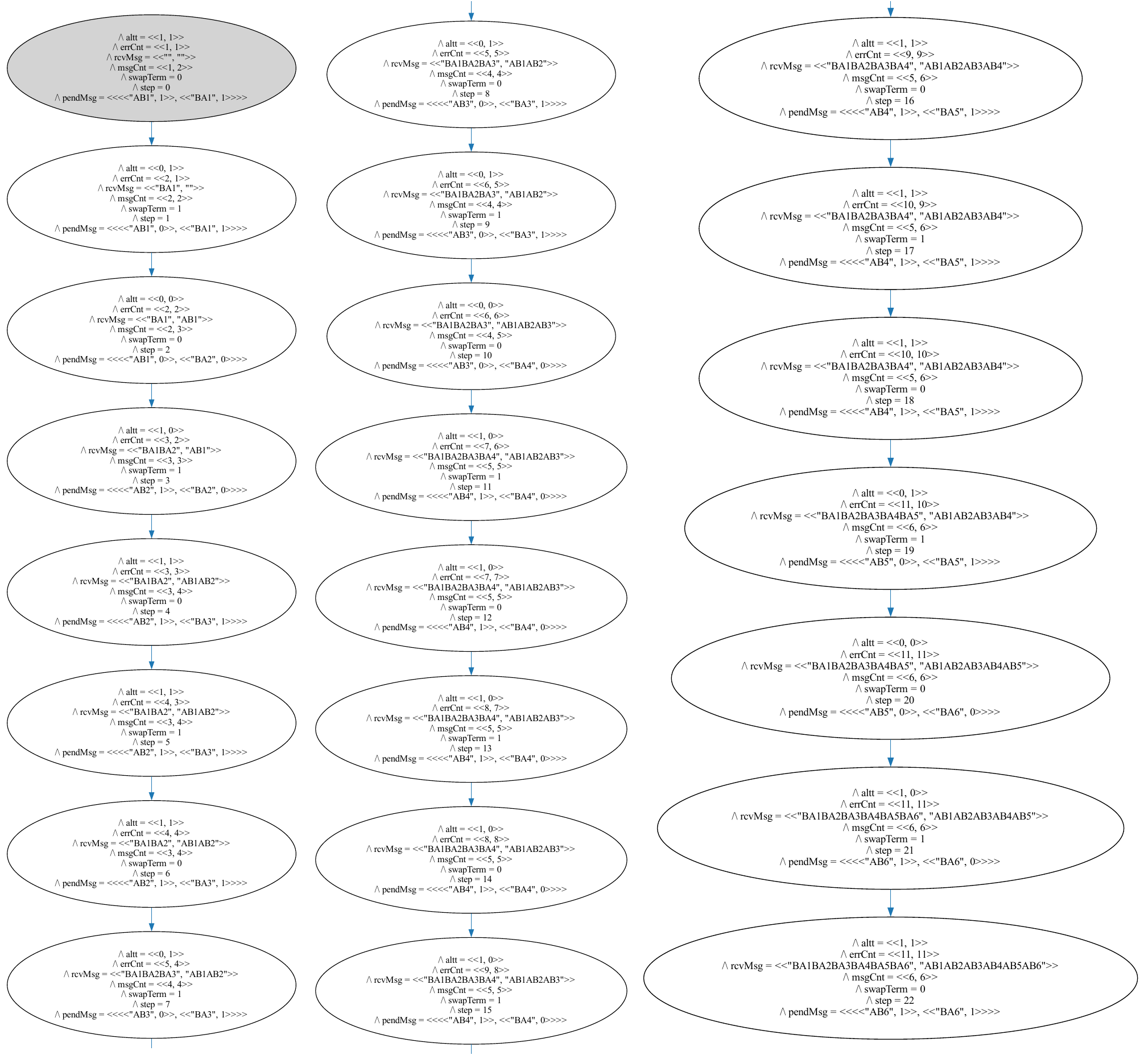}
\caption{\small\textbf{TLC state trace for simple TLA$^+$ spec}}
\label{fig:tlc_output}
\end{figure}

In this first spec the aim was to make the comparison with the ASM and CASM versions as easy as possible. Thus, the spec could be regarded as an ``emulation'' of the single-thread execution of the CASM code. The result is shown in Module ABPaolo2, below, whose constants are initialized as follows:
\begin{align*}
Term        &\triangleq \{ 1, 2 \} \\
errTrace    &\triangleq \langle\langle
                FALSE, FALSE, TRUE, FALSE,
                FALSE, FALSE, TRUE, FALSE,
                TRUE, \\
            &\qquad FALSE, FALSE \rangle, \\
            &\quad\ \ \langle FALSE, FALSE,
                TRUE, TRUE, FALSE, TRUE, TRUE,
                FALSE, TRUE, \\
            &\qquad FALSE, FALSE \rangle\rangle \\
msgs        &\triangleq \langle\langle
                ``AB1", ``AB2", ``AB3", ``AB4",
                ``AB5", ``AB6"\rangle \\
            &\quad\ \ \langle ``BA1", ``BA2", ``BA3",
                ``BA4", ``BA5", ``BA6" \rangle\rangle
\end{align*}
Similarly to the CASM model, this spec can be validated by treating the $errTrace(terminal)$ as an input for each terminal, and checking whether the behaviour of each terminal matches the behaviour that was derived manually in Figs.\ \ref{fig:abp_automata_diagrams1} and \ref{fig:abp_automata_diagrams2} and verified as CASM output in Fig.\ \ref{fig:casm_output}. The output provided by TLC is shown in Fig.\ \ref{fig:tlc_output} and confirms that this spec reproduces the desired behaviour.

In Chapter \ref{discussion} we will compare the roles and assess the usefulness of the different specification methodologies for and perspectives on the software engineering process that we have examined in this report. For now we can say that although this specification can be considered successful, it is not clear how it is actually ``specifying'' imperative code to be implemented. Rather, developing this spec felt more like an implementation effort. Perhaps, as Lamport says in his video course, the point is to think about the implementation abstractly, and perhaps this is the greatest value of the exercise. This is the same claim made by ASMs. More analysis and discussion later.

\newpage
\setlength{\parskip}{0pt}
\begin{tla}
----------------------- MODULE ABPaolo2 ----------------------------
EXTENDS     Integers,Sequences
CONSTANT    msgs,errTrace,Term
VARIABLES   pendMsg,rcvMsg,altt,msgCnt,errCnt,swapTerm,step
TypeOK ==
    /\ pendMsg \in [Term -> STRING \X {0,1}]
    /\ rcvMsg \in [Term -> STRING ]
    /\ altt \in [Term -> {0,1}]
    /\ msgCnt \in [Term -> Int]
    /\ errCnt \in [Term -> Int]
    /\ swapTerm \in {0,1}
    /\ step \in Int
Init ==
    /\ pendMsg = [term \in Term |-> IF term = 1 THEN <<"AB1",1>> ELSE <<"BA1",1>>]
    /\ rcvMsg  = [term \in Term |-> IF term = 1 THEN "" ELSE ""]
    /\ altt    = [term \in Term |-> IF term = 1 THEN 1 ELSE 1]
    /\ msgCnt  = [term \in Term |-> IF term = 1 THEN 1 ELSE 2]
    /\ errCnt  = [term \in Term |-> IF term = 1 THEN 1 ELSE 1]
    /\ swapTerm = 0
    /\ step = 0
vars == <<pendMsg,rcvMsg,altt,msgCnt,errCnt,swapTerm,step>>
flipBit(bit) == IF bit = 0 THEN 1 ELSE 0
cntCounter(n,max) == IF n < max THEN n + 1 ELSE max
alternationTest(term) == IF term = 1	THEN pendMsg[1][2] = pendMsg[2][2]
                         		ELSE pendMsg[1][2] # pendMsg[2][2]
otherTerm(term) == IF term = 1 THEN 2 ELSE 1
ReceiveMsg(term) ==
  /\ IF errTrace[term][errCnt[term]] THEN
        \* Transmission error detected in incoming msg, so only error and global counters are incremented:
        /\ errCnt' = [errCnt EXCEPT ![term] = cntCounter(errCnt[term],11)]
        /\ step' = cntCounter(step,22)
        /\ UNCHANGED <<pendMsg,rcvMsg,altt,msgCnt>>
     ELSE
        IF alternationTest(term) THEN
            \* No error in incoming msg, so it is stored and outgoing msg is prepared:
            /\ rcvMsg' = [rcvMsg EXCEPT ![term] = Append(rcvMsg[term],pendMsg[otherTerm(term)][1])]
            /\ altt' = [altt EXCEPT ![term] = flipBit(altt[term])]
            /\ pendMsg' = [pendMsg EXCEPT ![term] = <<msgs[term][msgCnt[term]],flipBit(altt[term])>>]
            /\ msgCnt' = [msgCnt EXCEPT ![term] = cntCounter(msgCnt[term],6)]
            /\ errCnt' = [errCnt EXCEPT ![term] = cntCounter(errCnt[term],11)]
            /\ step' = cntCounter(step,22)
        ELSE
            \* No error, but incoming msg has already been stored and next outgoing msg prepared:
            /\ errCnt' = [errCnt EXCEPT ![term] = cntCounter(errCnt[term],11)]
            /\ step' = cntCounter(step,22)
            /\ UNCHANGED <<pendMsg,rcvMsg,altt,msgCnt>>
  /\ swapTerm' = flipBit(swapTerm)
Next == IF swapTerm = 0 THEN ReceiveMsg(1) ELSE ReceiveMsg(2)
Spec == Init /\ [][Next]_vars
=============================================================================
\* Modification History
\* Last modified Sun Dec 04 22:01:49 GMT 2022 by paolo
\* Created Tue Nov 22 11:19:40 GMT 2022 by paolo
\end{tla}
\begin{tlatex}
\@x{}\moduleLeftDash\@xx{ {\MODULE} ABPaolo2}\moduleRightDash\@xx{}%
\@x{ {\EXTENDS}\@s{16.4} Integers ,\, Sequences}%
\@x{ {\CONSTANT}\@s{12.29} msgs ,\, errTrace ,\, Term}%
 \@x{ {\VARIABLES}\@s{8.2} pendMsg ,\, rcvMsg ,\, altt ,\, msgCnt ,\, errCnt
 ,\, swapTerm ,\, step}%
\@x{ TypeOK \.{\defeq}}%
 \@x{\@s{16.4} \.{\land}\@s{12.22} pendMsg \.{\in} [ Term \.{\rightarrow}
 {\STRING} \.{\times} \{ 0 ,\, 1 \} ]}%
 \@x{\@s{16.4} \.{\land}\@s{12.22} rcvMsg \.{\in} [ Term \.{\rightarrow}
 {\STRING} ]}%
 \@x{\@s{16.4} \.{\land}\@s{12.22} altt \.{\in} [ Term \.{\rightarrow} \{ 0
 ,\, 1 \} ]}%
 \@x{\@s{16.4} \.{\land}\@s{12.22} msgCnt \.{\in} [ Term \.{\rightarrow} Int
 ]}%
 \@x{\@s{16.4} \.{\land}\@s{12.22} errCnt\@s{4.19} \.{\in} [ Term
 \.{\rightarrow} Int ]}%
\@x{\@s{16.4} \.{\land}\@s{12.22} swapTerm \.{\in} \{ 0 ,\, 1 \}}%
\@x{\@s{16.4} \.{\land}\@s{12.22} step \.{\in} Int}%
\@x{ Init \.{\defeq}}%
 \@x{\@s{16.4} \.{\land} pendMsg \.{=} [ term \.{\in} Term \.{\mapsto} {\IF}\
 term \.{=} 1 \.{\THEN} {\langle}\@w{AB1} ,\, 1 {\rangle} \.{\ELSE}
 {\langle}\@w{BA1} ,\, 1 {\rangle} ]}%
 \@x{\@s{16.4} \.{\land} rcvMsg\@s{7.69} \.{=} [ term \.{\in} Term \.{\mapsto}
 {\IF}\ term \.{=} 1 \.{\THEN}\@w{} \.{\ELSE}\@w{} ]}%
 \@x{\@s{16.4} \.{\land} altt\@s{25.41} \.{=} [ term \.{\in} Term \.{\mapsto}
 {\IF}\ term \.{=} 1 \.{\THEN} 1 \.{\ELSE} 1 ]}%
 \@x{\@s{16.4} \.{\land} msgCnt\@s{4.99} \.{=} [ term \.{\in} Term \.{\mapsto}
 {\IF}\ term \.{=} 1 \.{\THEN} 1 \.{\ELSE} 2 ]}%
 \@x{\@s{16.4} \.{\land} errCnt\@s{9.18} \.{=} [ term \.{\in} Term \.{\mapsto}
 {\IF}\ term \.{=} 1 \.{\THEN} 1 \.{\ELSE} 1 ]}%
\@x{\@s{16.4} \.{\land} swapTerm \.{=} 0}%
\@x{\@s{16.4} \.{\land} step \.{=} 0}%
 \@x{ vars \.{\defeq} {\langle} pendMsg ,\, rcvMsg ,\, altt ,\, msgCnt ,\,
 errCnt ,\, swapTerm ,\, step {\rangle}}%
\@x{ flipBit ( bit ) \.{\defeq} {\IF}\ bit \.{=} 0 \.{\THEN} 1 \.{\ELSE} 0}%
 \@x{ cntCounter ( n ,\, max ) \.{\defeq} {\IF}\ n \.{<} max \.{\THEN} n \.{+}
 1 \.{\ELSE} max}%
 \@x{ alternationTest ( term ) \.{\defeq} {\IF}\ term \.{=} 1\@s{12.29}
 \.{\THEN} pendMsg [ 1 ] [ 2 ] \.{=} pendMsg [ 2 ] [ 2 ]}%
 \@x{\@s{193.5} \.{\ELSE} pendMsg [ 1 ] [ 2 ]\@s{4.65} \.{\neq} pendMsg [ 2 ]
 [ 2 ]}%
 \@x{ otherTerm ( term ) \.{\defeq} {\IF}\ term \.{=} 1 \.{\THEN} 2 \.{\ELSE}
 1}%
\@x{ ReceiveMsg ( term ) \.{\defeq}}%
\@x{\@s{8.2} \.{\land} {\IF}\ errTrace [ term ] [ errCnt [ term ] ] \.{\THEN}}%
\@x{\@s{28.90}}%
\@y{%
 Transmission error detected in incoming msg, so only error and global
 counters are incremented:
}%
\@xx{}%
 \@x{\@s{28.90} \.{\land} errCnt \.{'} \.{=} [ errCnt {\EXCEPT} {\bang} [ term
 ] \.{=} cntCounter ( errCnt [ term ] ,\, 11 ) ]}%
\@x{\@s{28.90} \.{\land} step \.{'} \.{=} cntCounter ( step ,\, 22 )}%
 \@x{\@s{28.90} \.{\land} {\UNCHANGED} {\langle} pendMsg ,\, rcvMsg ,\, altt
 ,\, msgCnt {\rangle}}%
\@x{\@s{20.36} \.{\ELSE}}%
\@x{\@s{32.66} {\IF}\ alternationTest ( term ) \.{\THEN}}%
\@x{\@s{45.30}}%
\@y{%
  No error in incoming msg, so it is stored and outgoing msg is prepared:
}%
\@xx{}%
 \@x{\@s{45.30} \.{\land} rcvMsg \.{'} \.{=} [ rcvMsg {\EXCEPT} {\bang} [ term
 ] \.{=} Append ( rcvMsg [ term ] ,\, pendMsg [ otherTerm ( term ) ] [ 1 ] )
 ]}%
 \@x{\@s{45.30} \.{\land} altt \.{'} \.{=} [ altt {\EXCEPT} {\bang} [ term ]
 \.{=} flipBit ( altt [ term ] ) ]}%
 \@x{\@s{45.30} \.{\land} pendMsg \.{'} \.{=} [ pendMsg {\EXCEPT} {\bang} [
 term ] \.{=} {\langle} msgs [ term ] [ msgCnt [ term ] ] ,\, flipBit ( altt
 [ term ] ) {\rangle} ]}%
 \@x{\@s{45.30} \.{\land} msgCnt \.{'} \.{=} [ msgCnt {\EXCEPT} {\bang} [ term
 ] \.{=} cntCounter ( msgCnt [ term ] ,\, 6 ) ]}%
 \@x{\@s{45.30} \.{\land} errCnt \.{'}\@s{4.19} \.{=} [ errCnt {\EXCEPT}
 {\bang} [ term ]\@s{4.19} \.{=} cntCounter ( errCnt [ term ] ,\, 11 ) ]}%
\@x{\@s{45.30} \.{\land} step \.{'} \.{=} cntCounter ( step ,\, 22 )}%
\@x{\@s{32.66} \.{\ELSE}}%
\@x{\@s{49.06}}%
\@y{%
 No error, but incoming msg has already been stored and next outgoing msg
 prepared:
}
\@xx{}%
 \@x{\@s{49.06} \.{\land} errCnt \.{'} \.{=} [ errCnt {\EXCEPT} {\bang} [ term
 ] \.{=} cntCounter ( errCnt [ term ] ,\, 11 ) ]}%
\@x{\@s{49.06} \.{\land} step \.{'} \.{=} cntCounter ( step ,\, 22 )}%
 \@x{\@s{49.06} \.{\land} {\UNCHANGED} {\langle} pendMsg ,\, rcvMsg ,\, altt
 ,\, msgCnt {\rangle}}%
\@x{\@s{8.2} \.{\land} swapTerm \.{'} \.{=} flipBit ( swapTerm )}%
 \@x{ Next \.{\defeq} {\IF}\ swapTerm \.{=} 0 \.{\THEN} ReceiveMsg ( 1 )
 \.{\ELSE} ReceiveMsg ( 2 )}%
\@x{ Spec\@s{1.60} \.{\defeq} Init \.{\land} {\Box} [ Next ]_{ vars}}%
\@x{}\bottombar\@xx{}%
\@x{}%
\@y{%
  Modification History
}%
\@xx{}%
\@x{}%
\@y{%
  Last modified Sun Dec 04 22:01:49 GMT 2022 by paolo
}%
\@xx{}%
\@x{}%
\@y{%
  Created Tue Nov 22 11:19:40 GMT 2022 by paolo
}%
\@xx{}%
\end{tlatex}
\setlength{\parskip}{12pt}

\newpage
\subsection{Manuel's Spec for Apalache}
\label{apalache-spec}

Manuel wrote a spec for the Apalache symbolic model checker.\footnote{\url{https://apalache.informal.systems/}} Apalache is slightly different from TLC but the language is TLA$^+$ in both cases. Apalache allows (sometimes requires) annotating the types of variables, constants, and functions. Thus, the modules include type annotations. Also, Apalache uses these annotations to run a type checker, which helps debug TLA$^+$ specifications.

The spec involves three modules: ABP3$\_$typedefs, ABP3, and MC$\_$ABP3:
\begin{itemize}
\item The ABP3$\_$typedefs module defines custom types aliases used by the other modules and Apalache.
\item The ABP3 module specifies the AB protocol.
\item The MC$\_$ABP3 module instantiates the ABP3 specification by giving concrete values to the constants MsgsA and MsgsB. It also includes an invariant (consistentPrefix) that checks the correctness of the protocol and two trace invariant generateTrace and generateCompleteTrace22Steps used for debugging and to generate traces.
\end{itemize}

\subsubsection{ABP3$\_$typedefs: Type aliases}

We define two custom type aliases: MSG and STATE. The type MSG defines a message sent between terminals. It contains three fields: the receiver terminal, the message payload, and the alternate bit of the sender terminal. The type STATE defined the state of the state machine.

\footnotesize
\begin{verbatim}
@typeAlias: MSG = [
    receiver: Str,
    msgr: Str,
    altr: Int];

  @typeAlias: STATE = [
    storedMsgs: Str -> Seq(Str),
    alt: Str -> Int,
    msgt: Str -> Str,
    pendingMsg: MSG,
    counterMsgs: Str -> Int];
\end{verbatim}
\normalsize

\subsubsection{ABP3.tla: The model}

Our model defines 2 constants and 5 variables:

\footnotesize
\begin{verbatim}
    CONSTANTS
    \* @type: Seq(Str);
    MsgsA,
    \* @type: Seq(Str);
    MsgsB

VARIABLES
    \* @type: Str -> Seq(Str);
    storedMsgs,
    \* @type: Str -> Int;
    alt,
    \* @type: Str -> Str;
    msgt,
    \* @type: MSG;
    pendingMsg,
    \* @type: Str -> Int;
    counterMsgs
\end{verbatim}
\normalsize

The specification assumes that terminal B sends the first message and that the alternation bit is initially set to 1 at both terminals. Furthermore, the specification assumes that terminal A has not fetched any message initially. This is important to guarantee that no message sent by terminal A is dropped. Variables are initialized as follows to capture these assumptions:

\footnotesize
\begin{verbatim}
Init ==
    /\ storedMsgs = [ terminal \in Terminals |-> <<>> ]
    \* initialize alternate bits
    /\ alt = [ terminal \in Terminals |-> 1 ]
    /\ msgt = [ terminal \in Terminals |-> IF terminal = "terminalA" THEN "garbage" ELSE MsgsA[1] ]
    /\ pendingMsg = [ receiver |-> "terminalA", msgr |-> MsgsB[1], altr |-> 1 ] 
    /\ counterSend = [ terminal \in Terminals |-> IF terminal = "terminalA" THEN 1 ELSE 2 ]
    /\ counterMsgs = [ terminal \in Terminals |-> IF terminal = "terminalA" THEN 1 ELSE 2 ]
\end{verbatim}
\normalsize

The model includes the following auxiliary operators:
\begin{itemize}
\item Terminals: it returns the set of terminals.
\item OtherTerminal: it returns the other terminal.
\item GetMessages: it fetches a message payload from the corresponding message array by index.
\item AcceptMsg: given a terminal and a message, it returns TRUE is the message should be accepted or FALSE otherwise.
\item AlternateBit: alternates a bit.
\end{itemize}

\footnotesize
\begin{verbatim}

Terminals == {"terminalA", "terminalB"}

OtherTerminal(terminal) == IF terminal = "terminalA" THEN "terminalB" ELSE "terminalA"

GetMessage(terminal, sequence) == IF terminal = "terminalA" THEN MsgsA[sequence] ELSE MsgsB[sequence]
\* @type: (Str, MSG) => Bool;

AcceptMsg(receiver, msg, error) ==
    /\ ~error
    /\ \/ receiver = "terminalA" /\ msg.altr = alt[receiver]
       \/ receiver = "terminalB" /\ msg.altr /= alt[receiver]

AlternateBit(bit) == IF bit = 0 THEN 1 ELSE 0
\end{verbatim}
\normalsize

Finally, the main functionality is defined by operators ReceiveMsg and Next. The Next operator picks the pending message from the pendingMsg variable and calls the ReceiveMsg operator.

\footnotesize
\begin{verbatim}
Next ==
    IF \A terminal \in Terminals : Len(storedMsgs[terminal]) = Len(MsgsA)
    THEN UNCHANGED <<storedMsgs, pendingMsg, alt, counterMsgs, msgt>>
    ELSE
      \E receiver \in Terminals: 
      \E error \in BOOLEAN:
        pendingMsg.receiver = receiver /\ ReceiveMsg(receiver, pendingMsg, error)
\end{verbatim}
\normalsize

The ReceiveMsg operator first checks if the message should be accepted. If not, then the terminal resends its last message. If the message should be accepted according to AcceptMsg, then the terminal stores the message payload, alternates the bit, fetches a new message payload, and sends the new message.

\footnotesize
\begin{verbatim}
\* @type: (Str, MSG, Bool) => Bool;
ReceiveMsg(receiver, msg, error) == 
    IF AcceptMsg(receiver, msg, error)
    THEN 
      \* message accepted
      LET altt == AlternateBit(alt[receiver]) IN 
      LET nextMsg == GetMessage(receiver, counterMsgs[receiver]) IN
      LET sendMsg == [ receiver |-> OtherTerminal(receiver),
                       msgr |-> nextMsg,
                       altr |-> altt] IN
      \* store message
      /\ storedMsgs' = [ storedMsgs EXCEPT ![receiver] = Append(@, msg.msgr) ]
      \* alternate bit
      /\ alt' = [ alt EXCEPT ![receiver] = altt ]
      \* fetch next message
      /\ msgt' = [ msgt EXCEPT ![receiver] = nextMsg]
      \* send message and clean processed
      /\ pendingMsg' = sendMsg
      \* update counters
      /\ counterMsgs' = [ counterMsgs EXCEPT ![receiver] = @ + 1 ]
    ELSE
      \* message not accepted; resending last message
      LET sendMsg == [ receiver |-> OtherTerminal(receiver),
                       msgr |-> msgt[receiver],
                       altr |-> alt[receiver]] IN
      /\ pendingMsg' = sendMsg
      /\ UNCHANGED <<storedMsgs, alt, msgt, counterMsgs>>

\end{verbatim}
\normalsize

\subsubsection{MC\_ABP3.tla: Instantiating the model}

We instantiate the constants as follows:

\footnotesize
\begin{verbatim}
MsgsA == <<"AB1", "AB2", "AB3", "AB4", "AB5", "AB6">>
MsgsB == <<"BA1", "BA2", "BA3", "BA4", "BA5", "BA6">>
\end{verbatim}
\normalsize

The correctness of the protocol is verified via the consistentPrefix invariant. This invariant checks that the sequence of messages received by a terminal is always a prefix of the sequence that the counterparty terminal is supposed to send, e.g., it checks that the sequence storedMsgs of terminal A is a prefix of the sequence defined by the constant MsgsB. The invariant implies that a terminal stores messages in a consistent order (as scheduled) dealing successfully with errors and duplicates, without dropping messages.

The consistentPrefix invariant uses the isPrefix operator internally. This operator is implemented using ApaFoldSet, a built-in fold operator of Apalache. Fold operators are common in functional programming and refer to the iterative application of a binary operator over a collection: F and the set of integers from 1 to the length of storedMsgs[terminal] in our case. ApaFoldSet is used in this case to iterate over all entries of storedMsgs in index order and compare each entry to the entry in the counterparty's message array. The fold function returns TRUE if all comparisons match and FALSE otherwise.

\clearpage

\footnotesize
\begin{verbatim}
isPrefix(terminal) == LET F(result, index) ==
                        IF result = FALSE
                        THEN FALSE
                        ELSE storedMsgs[terminal][index] = GetMessage(OtherTerminal(terminal), index)
                      IN ApaFoldSet(F, TRUE, { i \in 1..20: i <= Len(storedMsgs[terminal]) })

consistentPrefix ==
    \A terminal \in Terminals: isPrefix(terminal)
\end{verbatim}
\normalsize

We have checked the correctness of the model by checking consistentPrefix with Apalache on executions of up to 22 steps. This guarantees that for executions of 22 or less steps, the invariant consistentPrefix is never violated, given any sequence of errors.

MC\_ABP3.tla also includes two trace invariants: generateTrace and generateCompleteTrace22Steps. The trace invariant generateTrace simply generates a trace of 12 steps. We mostly use it for debugging.  The trace invariant generateCompleteTrace22Steps is more interesting. It generates a trace of 22 steps in which both terminals successfully exchange all their messages but only at the last step.

\footnotesize
\begin{verbatim}
\* @type: Seq(STATE) => Bool;
generateTrace(trace) ==
    LET Example == 
        /\ Len(trace) = 12
    IN
    ~Example

\* @type: Seq(STATE) => Bool;
generateCompleteTrace22Steps(trace) ==
    LET Example == 
        /\ Len(trace) = 23
        /\ \E terminal \in Terminals: trace[22].storedMsgs[terminal] /= IF terminal = "terminalA" 
                                                                        THEN MsgsB
                                                                        ELSE MsgsA
        /\ \A terminal \in Terminals: trace[23].storedMsgs[terminal] = IF terminal = "terminalA"
                                                                       THEN MsgsB
                                                                       ELSE MsgsA
    IN
    ~Example
\end{verbatim}
\normalsize

\subsubsection{Output trace}
We have generated an output trace using the generateCompleteTrace22Steps trace invariant. To find this execution for this particular model, Apalache explores increasingly longer executions for all possible permutations of error sequences until an execution satisfying the conditions in generateCompleteTrace22Steps is found.

\footnotesize
\begin{verbatim}
apalache-mc check --inv=generateCompleteTrace22Steps --length=22 MC_ABP3.tla
\end{verbatim}
\normalsize

The trace:

\footnotesize
\begin{verbatim}
1: <Initial predicate>
/\ alt = SetAsFun({ <<"terminalA", 1>>, <<"terminalB", 1>> })
/\ counterMsgs = SetAsFun({ <<"terminalA", 1>>, <<"terminalB", 2>> })
/\ msgt = SetAsFun({ <<"terminalA", "garbage">>, <<"terminalB", "BA1">> })
/\ pendingMsg = [altr |-> 1, msgr |-> "BA1", receiver |-> "terminalA"]
/\ storedMsgs = SetAsFun({ <<"terminalA", <<>>>>,
                           <<"terminalB", <<>>>> })
\end{verbatim}

\begin{verbatim}
2: <Next>
/\ alt = SetAsFun({ <<"terminalA", 0>>, <<"terminalB", 1>> })
/\ counterMsgs = SetAsFun({ <<"terminalA", 2>>, <<"terminalB", 2>> })
/\ msgt = SetAsFun({ <<"terminalA", "AB1">>, <<"terminalB", "BA1">> })
/\ pendingMsg = [altr |-> 0, msgr |-> "AB1", receiver |-> "terminalB"]
/\ storedMsgs = SetAsFun({ <<"terminalA", <<"BA1">>>>,
                           <<"terminalB", <<>>>> })
\end{verbatim}

\begin{verbatim}
3: <Next>
/\ alt = SetAsFun({ <<"terminalA", 0>>, <<"terminalB", 0>> })
/\ counterMsgs = SetAsFun({ <<"terminalA", 2>>, <<"terminalB", 3>> })
/\ msgt = SetAsFun({ <<"terminalA", "AB1">>, <<"terminalB", "BA2">> })
/\ pendingMsg = [altr |-> 0, msgr |-> "BA2", receiver |-> "terminalA"]
/\ storedMsgs = SetAsFun({ <<"terminalA", <<"BA1">>>>,
                           <<"terminalB", <<"AB1">>>> })
\end{verbatim}

\begin{verbatim}
4: <Next>
/\ alt = SetAsFun({ <<"terminalA", 1>>, <<"terminalB", 0>> })
/\ counterMsgs = SetAsFun({ <<"terminalA", 3>>, <<"terminalB", 3>> })
/\ msgt = SetAsFun({ <<"terminalA", "AB2">>, <<"terminalB", "BA2">> })
/\ pendingMsg = [altr |-> 1, msgr |-> "AB2", receiver |-> "terminalB"]
/\ storedMsgs = SetAsFun({ <<"terminalA", <<"BA1", "BA2">>>>,
                           <<"terminalB", <<"AB1">>>> })
\end{verbatim}

\begin{verbatim}
5: <Next>
/\ alt = SetAsFun({ <<"terminalA", 1>>, <<"terminalB", 1>> })
/\ counterMsgs = SetAsFun({ <<"terminalA", 3>>, <<"terminalB", 4>> })
/\ msgt = SetAsFun({ <<"terminalA", "AB2">>, <<"terminalB", "BA3">> })
/\ pendingMsg = [altr |-> 1, msgr |-> "BA3", receiver |-> "terminalA"]
/\ storedMsgs = SetAsFun({ <<"terminalA", <<"BA1", "BA2">>>>,
                           <<"terminalB", <<"AB1", "AB2">>>> })
\end{verbatim}

\begin{verbatim}
6: <Next>
/\ alt = SetAsFun({ <<"terminalA", 0>>, <<"terminalB", 1>> })
/\ counterMsgs = SetAsFun({ <<"terminalA", 4>>, <<"terminalB", 4>> })
/\ msgt = SetAsFun({ <<"terminalA", "AB3">>, <<"terminalB", "BA3">> })
/\ pendingMsg = [altr |-> 0, msgr |-> "AB3", receiver |-> "terminalB"]
/\ storedMsgs = SetAsFun({ <<"terminalA", <<"BA1", "BA2", "BA3">>>>,
                           <<"terminalB", <<"AB1", "AB2">>>> })
\end{verbatim}

\begin{verbatim}
7: <Next>
/\ alt = SetAsFun({ <<"terminalA", 0>>, <<"terminalB", 0>> })
/\ counterMsgs = SetAsFun({ <<"terminalA", 4>>, <<"terminalB", 5>> })
/\ msgt = SetAsFun({ <<"terminalA", "AB3">>, <<"terminalB", "BA4">> })
/\ pendingMsg = [altr |-> 0, msgr |-> "BA4", receiver |-> "terminalA"]
/\ storedMsgs = SetAsFun({ <<"terminalA", <<"BA1", "BA2", "BA3">>>>,
                           <<"terminalB", <<"AB1", "AB2", "AB3">>>> })
\end{verbatim}

\begin{verbatim}
8: <Next>
/\ alt = SetAsFun({ <<"terminalA", 1>>, <<"terminalB", 0>> })
/\ counterMsgs = SetAsFun({ <<"terminalA", 5>>, <<"terminalB", 5>> })
/\ msgt = SetAsFun({ <<"terminalA", "AB4">>, <<"terminalB", "BA4">> })
/\ pendingMsg = [altr |-> 1, msgr |-> "AB4", receiver |-> "terminalB"]
/\ storedMsgs = SetAsFun({ <<"terminalA", <<"BA1", "BA2", "BA3", "BA4">>>>,
                           <<"terminalB", <<"AB1", "AB2", "AB3">>>> })
\end{verbatim}

\begin{verbatim}
9: <Next>
/\ alt = SetAsFun({ <<"terminalA", 1>>, <<"terminalB", 0>> })
/\ counterMsgs = SetAsFun({ <<"terminalA", 5>>, <<"terminalB", 5>> })
/\ msgt = SetAsFun({ <<"terminalA", "AB4">>, <<"terminalB", "BA4">> })
/\ pendingMsg = [altr |-> 0, msgr |-> "BA4", receiver |-> "terminalA"]
/\ storedMsgs = SetAsFun({ <<"terminalA", <<"BA1", "BA2", "BA3", "BA4">>>>,
                           <<"terminalB", <<"AB1", "AB2", "AB3">>>> })
\end{verbatim}

\begin{verbatim}
10: <Next>
/\ alt = SetAsFun({ <<"terminalA", 1>>, <<"terminalB", 0>> })
/\ counterMsgs = SetAsFun({ <<"terminalA", 5>>, <<"terminalB", 5>> })
/\ msgt = SetAsFun({ <<"terminalA", "AB4">>, <<"terminalB", "BA4">> })
/\ pendingMsg = [altr |-> 1, msgr |-> "AB4", receiver |-> "terminalB"]
/\ storedMsgs = SetAsFun({ <<"terminalA", <<"BA1", "BA2", "BA3", "BA4">>>>,
                           <<"terminalB", <<"AB1", "AB2", "AB3">>>> })
\end{verbatim}

\begin{verbatim}
11: <Next>
/\ alt = SetAsFun({ <<"terminalA", 1>>, <<"terminalB", 0>> })
/\ counterMsgs = SetAsFun({ <<"terminalA", 5>>, <<"terminalB", 5>> })
/\ msgt = SetAsFun({ <<"terminalA", "AB4">>, <<"terminalB", "BA4">> })
/\ pendingMsg = [altr |-> 0, msgr |-> "BA4", receiver |-> "terminalA"]
/\ storedMsgs = SetAsFun({ <<"terminalA", <<"BA1", "BA2", "BA3", "BA4">>>>,
                           <<"terminalB", <<"AB1", "AB2", "AB3">>>> })
\end{verbatim}

\begin{verbatim}
12: <Next>
/\ alt = SetAsFun({ <<"terminalA", 1>>, <<"terminalB", 0>> })
/\ counterMsgs = SetAsFun({ <<"terminalA", 5>>, <<"terminalB", 5>> })
/\ msgt = SetAsFun({ <<"terminalA", "AB4">>, <<"terminalB", "BA4">> })
/\ pendingMsg = [altr |-> 1, msgr |-> "AB4", receiver |-> "terminalB"]
/\ storedMsgs = SetAsFun({ <<"terminalA", <<"BA1", "BA2", "BA3", "BA4">>>>,
                           <<"terminalB", <<"AB1", "AB2", "AB3">>>> })
\end{verbatim}

\begin{verbatim}
13: <Next>
/\ alt = SetAsFun({ <<"terminalA", 1>>, <<"terminalB", 1>> })
/\ counterMsgs = SetAsFun({ <<"terminalA", 5>>, <<"terminalB", 6>> })
/\ msgt = SetAsFun({ <<"terminalA", "AB4">>, <<"terminalB", "BA5">> })
/\ pendingMsg = [altr |-> 1, msgr |-> "BA5", receiver |-> "terminalA"]
/\ storedMsgs = SetAsFun({ <<"terminalA", <<"BA1", "BA2", "BA3", "BA4">>>>,
                           <<"terminalB", <<"AB1", "AB2", "AB3", "AB4">>>> })
\end{verbatim}

\begin{verbatim}
14: <Next>
/\ alt = SetAsFun({ <<"terminalA", 0>>, <<"terminalB", 1>> })
/\ counterMsgs = SetAsFun({ <<"terminalA", 6>>, <<"terminalB", 6>> })
/\ msgt = SetAsFun({ <<"terminalA", "AB5">>, <<"terminalB", "BA5">> })
/\ pendingMsg = [altr |-> 0, msgr |-> "AB5", receiver |-> "terminalB"]
/\ storedMsgs = SetAsFun({ <<"terminalA", <<"BA1", "BA2", "BA3", "BA4", "BA5">>>>,
                           <<"terminalB", <<"AB1", "AB2", "AB3", "AB4">>>> })
\end{verbatim}

\begin{verbatim}
15: <Next>
/\ alt = SetAsFun({ <<"terminalA", 0>>, <<"terminalB", 1>> })
/\ counterMsgs = SetAsFun({ <<"terminalA", 6>>, <<"terminalB", 6>> })
/\ msgt = SetAsFun({ <<"terminalA", "AB5">>, <<"terminalB", "BA5">> })
/\ pendingMsg = [altr |-> 1, msgr |-> "BA5", receiver |-> "terminalA"]
/\ storedMsgs = SetAsFun({ <<"terminalA", <<"BA1", "BA2", "BA3", "BA4", "BA5">>>>,
                           <<"terminalB", <<"AB1", "AB2", "AB3", "AB4">>>> })
\end{verbatim}

\begin{verbatim}
16: <Next>
/\ alt = SetAsFun({ <<"terminalA", 0>>, <<"terminalB", 1>> })
/\ counterMsgs = SetAsFun({ <<"terminalA", 6>>, <<"terminalB", 6>> })
/\ msgt = SetAsFun({ <<"terminalA", "AB5">>, <<"terminalB", "BA5">> })
/\ pendingMsg = [altr |-> 0, msgr |-> "AB5", receiver |-> "terminalB"]
/\ storedMsgs = SetAsFun({ <<"terminalA", <<"BA1", "BA2", "BA3", "BA4", "BA5">>>>,
                           <<"terminalB", <<"AB1", "AB2", "AB3", "AB4">>>> })
\end{verbatim}

\begin{verbatim}
17: <Next>
/\ alt = SetAsFun({ <<"terminalA", 0>>, <<"terminalB", 1>> })
/\ counterMsgs = SetAsFun({ <<"terminalA", 6>>, <<"terminalB", 6>> })
/\ msgt = SetAsFun({ <<"terminalA", "AB5">>, <<"terminalB", "BA5">> })
/\ pendingMsg = [altr |-> 1, msgr |-> "BA5", receiver |-> "terminalA"]
/\ storedMsgs = SetAsFun({ <<"terminalA", <<"BA1", "BA2", "BA3", "BA4", "BA5">>>>,
                           <<"terminalB", <<"AB1", "AB2", "AB3", "AB4">>>> })
\end{verbatim}

\begin{verbatim}
18: <Next>
/\ alt = SetAsFun({ <<"terminalA", 0>>, <<"terminalB", 1>> })
/\ counterMsgs = SetAsFun({ <<"terminalA", 6>>, <<"terminalB", 6>> })
/\ msgt = SetAsFun({ <<"terminalA", "AB5">>, <<"terminalB", "BA5">> })
/\ pendingMsg = [altr |-> 0, msgr |-> "AB5", receiver |-> "terminalB"]
/\ storedMsgs = SetAsFun({ <<"terminalA", <<"BA1", "BA2", "BA3", "BA4", "BA5">>>>,
                           <<"terminalB", <<"AB1", "AB2", "AB3", "AB4">>>> })
\end{verbatim}

\begin{verbatim}
19: <Next>
/\ alt = SetAsFun({ <<"terminalA", 0>>, <<"terminalB", 1>> })
/\ counterMsgs = SetAsFun({ <<"terminalA", 6>>, <<"terminalB", 6>> })
/\ msgt = SetAsFun({ <<"terminalA", "AB5">>, <<"terminalB", "BA5">> })
/\ pendingMsg = [altr |-> 1, msgr |-> "BA5", receiver |-> "terminalA"]
/\ storedMsgs = SetAsFun({ <<"terminalA", <<"BA1", "BA2", "BA3", "BA4", "BA5">>>>,
                           <<"terminalB", <<"AB1", "AB2", "AB3", "AB4">>>> })
\end{verbatim}

\begin{verbatim}
20: <Next>
/\ alt = SetAsFun({ <<"terminalA", 0>>, <<"terminalB", 1>> })
/\ counterMsgs = SetAsFun({ <<"terminalA", 6>>, <<"terminalB", 6>> })
/\ msgt = SetAsFun({ <<"terminalA", "AB5">>, <<"terminalB", "BA5">> })
/\ pendingMsg = [altr |-> 0, msgr |-> "AB5", receiver |-> "terminalB"]
/\ storedMsgs = SetAsFun({ <<"terminalA", <<"BA1", "BA2", "BA3", "BA4", "BA5">>>>,
                           <<"terminalB", <<"AB1", "AB2", "AB3", "AB4">>>> })
\end{verbatim}

\begin{verbatim}
21: <Next>
/\ alt = SetAsFun({ <<"terminalA", 0>>, <<"terminalB", 0>> })
/\ counterMsgs = SetAsFun({ <<"terminalA", 6>>, <<"terminalB", 7>> })
/\ msgt = SetAsFun({ <<"terminalA", "AB5">>, <<"terminalB", "BA6">> })
/\ pendingMsg = [altr |-> 0, msgr |-> "BA6", receiver |-> "terminalA"]
/\ storedMsgs = SetAsFun({ <<"terminalA", <<"BA1", "BA2", "BA3", "BA4", "BA5">>>>,
                           <<"terminalB", <<"AB1", "AB2", "AB3", "AB4", "AB5">>>> })
\end{verbatim}

\begin{verbatim}
22: <Next>
/\ alt = SetAsFun({ <<"terminalA", 1>>, <<"terminalB", 0>> })
/\ counterMsgs = SetAsFun({ <<"terminalA", 7>>, <<"terminalB", 7>> })
/\ msgt = SetAsFun({ <<"terminalA", "AB6">>, <<"terminalB", "BA6">> })
/\ pendingMsg = [altr |-> 1, msgr |-> "AB6", receiver |-> "terminalB"]
/\ storedMsgs = SetAsFun({ <<"terminalA", <<"BA1", "BA2", "BA3", "BA4", "BA5", "BA6">>>>,
                           <<"terminalB", <<"AB1", "AB2", "AB3", "AB4", "AB5">>>> })
\end{verbatim}

\begin{verbatim}
23: <Next>
/\ alt = SetAsFun({ <<"terminalA", 1>>, <<"terminalB", 1>> })
/\ counterMsgs = SetAsFun({ <<"terminalA", 7>>, <<"terminalB", 8>> })
/\ msgt = SetAsFun({ <<"terminalA", "AB6">>, <<"terminalB", "BA6">> })
/\ pendingMsg = [altr |-> 1, msgr |-> "BA6", receiver |-> "terminalA"]
/\ storedMsgs = SetAsFun({ <<"terminalA", <<"BA1", "BA2", "BA3", "BA4", "BA5", "BA6">>>>,
                           <<"terminalB", <<"AB1", "AB2", "AB3", "AB4", "AB5", "AB6">>>> })
\end{verbatim}
\normalsize
\newpage
\section{Quint Specification of AB Protocol}
\label{quint}

\subsection{Introduction to Quint}

Quint is a specification language over the same underlying logic of TLA\textsuperscript{+}. Quint has syntax and tooling that aim to resemble programming languages and their environments in many ways. By restricting the syntax in some aspects, such as avoiding operator overloading, Quint can be parsed and statically analyzed with significantly less effort.

As the specification in Section \ref{quint-spec} shows, Quint's syntax has constructs related to static analysis. The most evident example is typing information. Quint also has different qualifiers for its operators, with which specification writers can state their expectations on how an operator can interact with the state.

Quint is still under construction, and it is not fully integrated with a model checker as of this writing. It offers a REPL (Read-Eval-Print Loop) that is able to perform random simulation and obtain traces of execution. The REPL is a useful tool to enable initial inspection and debugging of specifications, and it makes sense to use it before running a model checker because of its fast feedback for errors and accessible interface. However, since in order to verify properties Quint needs a model checker, it is currently being integrated into Apalache \cite{apalache}.

\subsection{Gabriela's Spec of the AB Protocol}
\label{quint-spec}

This Section describes a Quint specification for the AB Protocol written by Gabriela. This specification follows the same level of abstraction as the CoreASM specification in Section \ref{coreasmSpec}, and it is presented in a broken-up manner to include explanations. 

The only custom type defined is a record for representing a message being transmitted:

{\small
\begin{verbatim}
type MSG = { receiver: str, msgr: str, altr: int, error: bool }
\end{verbatim}
}

The state is composed of four main variables for the protocol, and two auxiliary variables (\texttt{counter} and \texttt{output}) that keep track of extra information required for testing executions:

{\small
\begin{verbatim}
// The state variables
var storedMsgs: str -> List[int]
var altt: str -> bool
var altr: str -> bool
var fileNumber: str -> int

// Auxiliary state variables for testing
var counter: int
var output: List[{ terminal: str, sent: int, error: bool, altr: bool, altt: bool }]
\end{verbatim}
}

This spec has a single (\texttt{pure}) operator, that is, an operator that does not interact with the state at all:

{\small
\begin{verbatim}
pure def otherTerminal(terminal) = if (terminal == "A") "B" else "A"
\end{verbatim}
}

The initial condition is defined by an action called \texttt{Init}, assigning a value for each state variable.

{\small
\begin{verbatim}
action Init = all {
  storedMsgs' = Map("A" -> [], "B" -> []),
  altt' = Map("A" -> true, "B" -> true),
  altr' = Map("A" -> false, "B" -> false),
  fileNumber' = Map("A" -> 0, "B" -> 1),
  counter' = 0,
  output' = [],
}
\end{verbatim}
}

The condition of acceptance depends on the \texttt{altt} state variable, and is therefore defined by an state-level operator, which requires the \texttt{def} modifier:

{\small
\begin{verbatim}
def conditionOfAcceptance(terminal: str, newAltr: bool): bool =
  or {
    and { terminal == "A", newAltr == altt.get(terminal) },
    and { terminal == "B", newAltr != altt.get(terminal) },
  }
\end{verbatim}
}

Message passing also has to be defined through the state. One option is to define a \texttt{pendingMsg} state variable with the latest message's payload, as it was done in the TLA\textsuperscript{+} in Section \ref{apalache-spec}. This spec, similarly to the CoreASM spec, does not specify that level of detail on message passing. Instead, the information that would be transmitted in a message is read directly from the other terminal's state variables. The operators responsible for this have their names prefixed with \texttt{receive} with the intention of making it explicit that they relate to message passing.

{\small
\begin{verbatim}
// These operators simulate message reception. They read the ALTT and the
// message (file number) from the state belonging to the other terminal.
def receiveBit(terminal) = altt.get(otherTerminal(terminal))
def receiveFileNumber(terminal) = fileNumber.get(otherTerminal(terminal))
\end{verbatim}
}

The action for accepting a message is defined as \texttt{ReceiveSuccess} and updates all core state variables. It receives the messages using the \texttt{receiveBit} and \texttt{receiveFileNumber} operators, then uses the new received \texttt{altr} value to determine if the condition of acceptance is satisfied. If it is, \texttt{fileNumber} and \texttt{altt} for the receiving terminal are updated, and the message is stored in \texttt{storedMsgs}.

{\small
\begin{verbatim}
action ReceiveSuccess(terminal: str): bool =
  val newAltr = receiveBit(terminal)
  val newFileNumber = receiveFileNumber(terminal)
  all {
    altr' = altr.set(terminal, newAltr),
    output' = output.append({
      terminal: otherTerminal(terminal),
      sent: newFileNumber,
      error: false,
      altr: newAltr,
      altt: altt.get(terminal)
    }),
    if (conditionOfAcceptance(terminal, newAltr))
      all {
        fileNumber' = fileNumber.setBy(terminal, (n) => n + 1),
        altt' = altt.setBy(terminal, (b) => not(b)),
        storedMsgs' = storedMsgs.setBy(terminal, (msgs) => msgs.append(newFileNumber)),
      }
    else
      all {
        fileNumber' = fileNumber,
        altt' = altt,
        storedMsgs' = storedMsgs,
      }
  }
\end{verbatim}
}

The action for re-sending a message is defined as \texttt{SendMsg} and, as in the CoreASM spec, no variables are updated.

{\small
\begin{verbatim}
/// Sending messages doesn't change the state of the system.
action SendMsg(terminal, error) = all {
  storedMsgs' = storedMsgs,
  altt' = altt,
  altr' = altr,
  fileNumber' = fileNumber,
}
\end{verbatim}
}

The action \texttt{ReceiveMsg} defines which action should be taken according to an error flag.

{\small
\begin{verbatim}
action ReceiveMsg(terminal: str, error: bool): bool =
  if (error) all {
    // Resend the last message
    SendMsg(terminal, error),
    output' = output.append({
      terminal: otherTerminal(terminal),
      sent: fileNumber.get(otherTerminal(terminal)),
      error: error,
      altr: altr.get(terminal),
      altt: altt.get(terminal),
    })
  }
  else
    ReceiveSuccess(terminal)
\end{verbatim}
}

In order to simulate this protocol, a \texttt{run} is defined. The concept of a run is not present in TLA\textsuperscript{+}, and was introduced in Quint for cases similar to this one, where the goal is to guide a simulation of the protocol according to some parameters and check the output. Here, the parameters are the order of actions (between sending and receiving on each terminal) and the sequence of errors, that is, for each step, whether an error occurs in that step. The sequence of actions is defined as the same sequence in \texttt{run} of the CoreASM specification, and the map of errors is equivalent to the error trace functions defined in that specification as well. The value of \texttt{expectedOutput} is omitted here to save space, but the actual output and the assertion result are given at the end of this section.

{\small
\begin{verbatim}
  pure val errors: str -> List[bool] = Map(
    "A" -> [false, false, true, false, false, false, true, false, true, false, false],
    "B" -> [false, false, true, true, false, true, true, false, true, false, false]
  )
  Init.then((all{
    SendMsg("B", errors.get("B")[counter]),
    output' = output,
    counter' = counter,
  }).then(all{
    ReceiveMsg("A", errors.get("A")[counter]),
    counter' = counter,
  }).then(all{
    SendMsg("A", errors.get("A")[counter]),
    output' = output,
    counter' = counter,
  }).then(all{
    ReceiveMsg("B", errors.get("B")[counter]),
    counter' = counter + 1,
  }).repeated(11)).then(all {
    assert(output == expectedOutput),
    output' = output,
    counter' = counter,
    altt' = altt,
    altr' = altr,
    storedMsgs' = storedMsgs,
    fileNumber' = fileNumber,
  })
\end{verbatim}
}

To obtain a trace in the Quint REPL, the previous code blocks need to be wrapped inside a module that needs to be loaded and imported. By wrapping the code blocks inside a module called \texttt{ABP} with `module ABP { ... }`, it can be loaded in the REPL by running `quint -r src/quint/ABP.qnt::ABP` in the shell. Then, the run `test` can be invoked to run the simulation and make the assertion. That will raise an error if the assertion fails. The output can be inspected by evaluating the state variable `output` after invoking `test`. This is the obtained result:

{\small
\begin{verbatim}
Quint REPL v0.0.3
Type ".exit" to exit, or ".help" for more information
true

>>> test
true
>>> output
[
  { terminal: "B", sent: 1, error: false, altr: true, altt: true },
  { terminal: "A", sent: 1, error: false, altr: false, altt: true },
  { terminal: "B", sent: 2, error: false, altr: false, altt: false },
  { terminal: "A", sent: 2, error: false, altr: true, altt: false },
  { terminal: "B", sent: 3, error: true, altr: false, altt: true },
  { terminal: "A", sent: 2, error: true, altr: true, altt: true },
  { terminal: "B", sent: 3, error: false, altr: true, altt: true },
  { terminal: "A", sent: 3, error: true, altr: true, altt: true },
  { terminal: "B", sent: 3, error: false, altr: true, altt: false },
  { terminal: "A", sent: 3, error: false, altr: false, altt: true },
  { terminal: "B", sent: 4, error: false, altr: false, altt: false },
  { terminal: "A", sent: 4, error: true, altr: false, altt: false },
  { terminal: "B", sent: 4, error: true, altr: false, altt: true },
  { terminal: "A", sent: 4, error: true, altr: false, altt: false },
  { terminal: "B", sent: 4, error: false, altr: false, altt: true },
  { terminal: "A", sent: 4, error: false, altr: true, altt: false },
  { terminal: "B", sent: 5, error: true, altr: false, altt: true },
  { terminal: "A", sent: 4, error: true, altr: true, altt: true },
  { terminal: "B", sent: 5, error: false, altr: true, altt: true },
  { terminal: "A", sent: 5, error: false, altr: false, altt: true },
  { terminal: "B", sent: 6, error: false, altr: false, altt: false },
  { terminal: "A", sent: 6, error: false, altr: true, altt: false },
]
\end{verbatim}
}

\newpage
\section{Comparison between the ASM and TLA$^+$ Methodologies}
\label{discussion}
In this chapter we discuss what we have learned from the different specification perspectives of the previous chapters, first at the theoretical level and then at the level of executable languages and tools. The considerations presented in this chapter should be seen more as speculative conjectures meant to stimulate further discussion than certain conclusions or proven results. The objective of the discussion, and indeed of the whole report, is to explore the complementarities between the different specification methods and, therefore, the possibility to combine them in some way that will strengthen the software engineering development process.

\subsection{ASMs and TLA$^+$}
Invoking a ``geometrization'' metaphor, each type of models can be seen as the specification of the boundary surface of a state space shaped like an infinite cone, whose vertex is rooted at the INIT state and that fans out in 2D or 3D space. The boundaries and interior of this cone can be explored by model checkers like TLC or Apalache. Similarly stated, TLA$^+$ models appear to be analogous to a set of simultaneous linear inequalities from elementary analytic geometry, which together define a certain region of the plane. An ASM model can be interpreted similarly, whereas a CASM model requires more data to run, the result of which usually yields a single trajectory through that same space and starting from the same vertex. CoreASM, and Quint specifications are also able to specify single trajectories within the cone, while TLC and Apalache state traces are analogous constructions derivable from TLA$^+$ models plus suitable constraints.

Methodologically, both frameworks embrace the concepts of abstraction and refinement. Both methodologies go out of their way to stress the importance of abstraction, i.e.\ of focusing on macro aspects of the application being specified rather than on the implementation details. In addition, both methods encourage the use of simple high-level models to start with, adding granularity in later iterations.

As an example of refinement in TLA$^+$, Lecture 9 of Leslie Lamport's video course\footnote{\url{https://lamport.azurewebsites.net/video/videos.html}} presents the specification of a simple version of the AB protocol,\footnote{In particular, of the Simplex version of that protocol.} where some messages are lost randomly, while Lecture 10 adds the ability to detect if a message was corrupted, as a refinement. In the cone metaphor, for both modelling methods iterative refinement increases the granularity of the state space, i.e.\ the ``density'' of the states within the cone and on its boundary surface, for a given fixed ``cone volume''.

Another important dimension for the comparison is given by global properties or invariants. Although ASMs can also define invariants, they appear to be used more often and more consistently by TLA$^+$. The reason could be that the specification of the behaviour boundary is itself a global property of the model. Hence, to gain greater purchase on the set of possible behaviours, identifying and then checking the invariants is a very useful and effective way to explore the large size of the state space looking for bugs (i.e.\ states where a given invariant is not satisfied). By contrast, since an ASM/CASM model is already much more specific about a particular behaviour, its effectiveness is less dependent on the discovery of invariants, even though they are still relevant and potentially useful.

Another difference between the two methods that has important methodological implications can be attributed to the fact that ASMs are based on operational semantics whereas TLA$^+$ is based on declarative Boolean logic statements. Writing ASM specifications as pseudo-code is cognitively equivalent to writing the implementation code. It is in fact a form of implementation, but more abstract. The most abstract is the ground model, with iterative refinement steps progressively approaching the implementation code, but each step is itself an algorithm that can be ``executed'' mathematically in one's mind (which we called \emph{verification} in Chapter \ref{asms}). Quint was developed to achieve a similar Ux effect.

There are four additional important aspects that derive more from the experience of some of the authors than from the findings of this report:
\begin{packed_item1}
\item First, because an ASM model is derived directly from the stated requirements, if it is paired with an executable language such as CASM at each level of refinement the CASM specification can be executed to check whether the output matches the requirements, as was done in Section \ref{paolocasm}. This provides a fast process of \emph{validation} that increases the speed and confidence of the developer. Quint aims for a similar effect.
\item Second, the names of ASM variables and functions are written in a language that matches the language of the domain expert (or customer). To some extent this is true of TLA$^+$ as well, but the problem is that the semantics of TLA$^+$ are mathematical statements in Boolean logic, which is a type of abstraction that non-technical people and engineers find difficult to relate to. On the other hand, by using operational semantics ASM rules are much closer to the natural language statements with which the requirements themselves are expressed. The combination of operational semantics, understandability, and rapid validation cycle makes it easier for implementation engineers to start dabbling in ASM specifications and to try to use them. Quint's syntax is similarly aimed at programmers, while Apalache also relies on operational semantics \cite{apalache}.
\item Third, when a change request or a new requirement emerges it is easy to modify the ground model and to then apply the modification to the different refinement levels, all the way to the code. In some cases this process can be automated and actually be performed by a compiler.
\item Finally, the fact that ASM specifications are easily understandable by all the stakeholders implies that such a document becomes the central documentation reference for the whole development team, including the customer. CASM and CoreASM are similar in this regard, although somewhat more technical. Since Quint is a new language, it is too early to assess how effective it will be in fulfilling this function.
\end{packed_item1}

Probably the most useful feature of TLA$^+$ is the opportunity it affords to express invariants. Although the often infinite state space of most applications cannot be explored fully, checking the invariant(s) for representative finite state traces can still provide a high degree of confidence that the application will perform as desired.

\subsection{State Space Visualization}
\label{viz}
\begin{figure}[H]
\centering
\includegraphics[width=12cm]{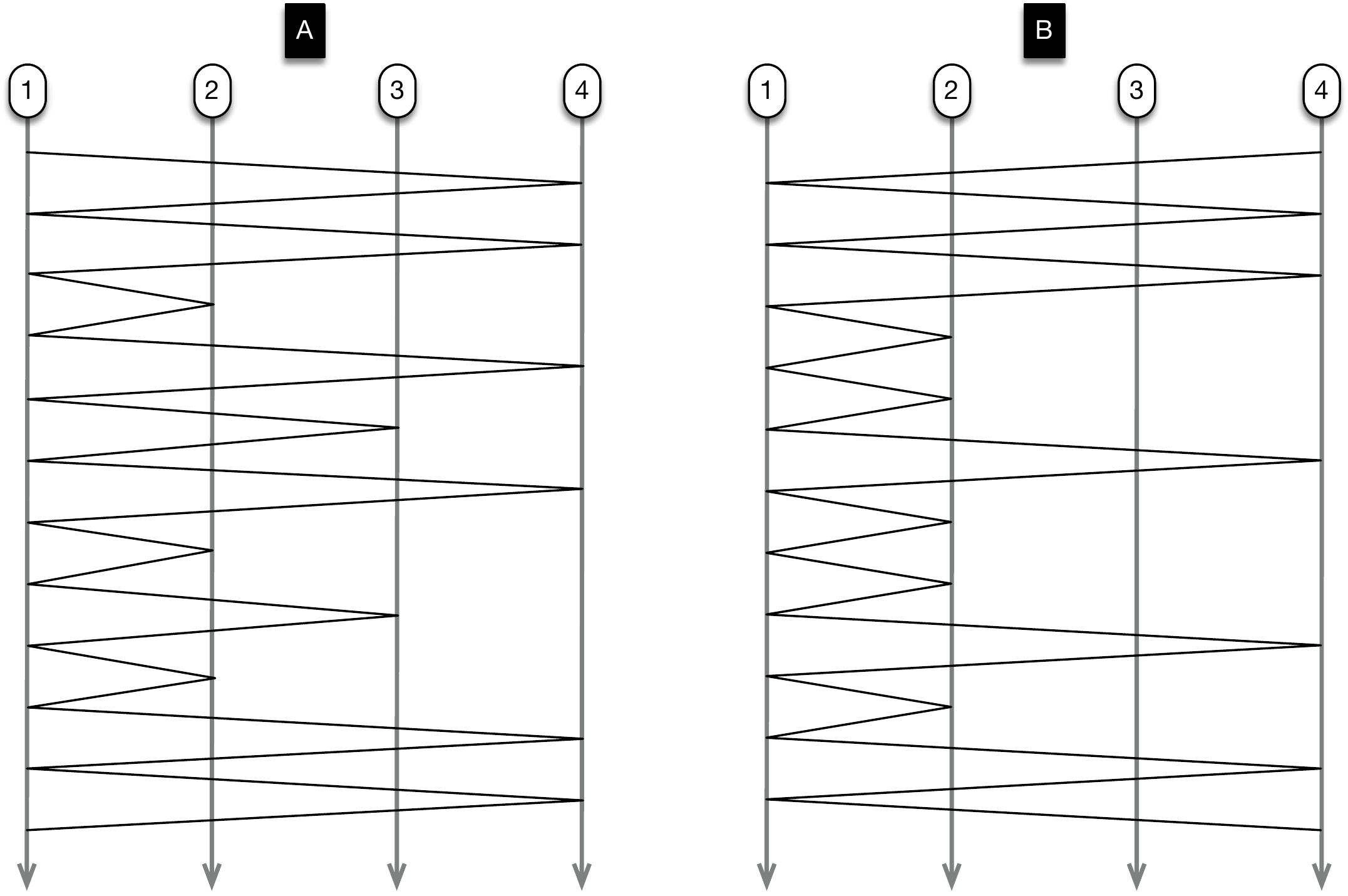}
\caption{\small\textbf{State traces of the two terminals corresponding to the Lynch \cite{Lynch1968} error sequence}}
\label{fig:swimlanes}
\vspace{-0.3cm}
\end{figure}

To begin scoping out a possible formal relationship between ASM and TLA$^+$ models we take advantage of the simplicity of the AB protocol to visualize its state space explicitly. Fig.\ \ref{fig:swimlanes} shows a ``swimlanes'' view of the state trajectories of the two terminals that correspond to the error sequence of Fig.\ \ref{fig:abp_sequence}.

We can develop a more efficient visualization for the system as follows. Since each terminal's automaton can be in one of 4 states, the system of two automata can be in 16 states at most, as shown in Table \ref{tab:sysautomaton}. The reachable states are shown in bold in different colours, where black denotes normal operation, red is an error state, and green is a state that corresponds to the error-free receipt of a message that was previously stored. State 3 (i.e.\ (1,3)) is not visited by the sequence corresponding to Fig.\ \ref{fig:abp_sequence}. Table \ref{tab:statetraces} shows the state traces of the two terminals that are also depicted in the swimlanes figure.

\setlength{\tabcolsep}{6pt}
\begin{table}[H]
\begin{centering}
\small
\begin{tabular}{c | cccccccccccccccc}
Automaton &\multicolumn{3}{l}{States}&&&&&&&&&&&&& \\
\hline
A &1 &\textcolor{red}{\textbf{1}} &\textcolor{green}{\textbf{1}} &\textbf{1}
	&\textcolor{red}{\textbf{2}} &2 &2 &2 &\textcolor{green}{\textbf{3}}
	&3 &3 &3 &\textbf{4} &4 &4 &4  \\
B &1 &\textcolor{red}{\textbf{2}} &\textcolor{green}{\textbf{3}} &\textbf{4}
	&\textcolor{red}{\textbf{1}} &2 &3 &4 &\textcolor{green}{\textbf{1}}
	&2 &3 &4 &\textbf{1} &2 &3 &4 \\
\hline
System &1 &\textcolor{red}{\textbf{2}} &\textcolor{green}{\textbf{3}} &\textbf{4}
	&\textcolor{red}{\textbf{5}} &6 &7 &8 &\textcolor{green}{\textbf{9}}
	&10 &11 &12 &\textbf{13} &14 &15 &16 \\
\hline
\end{tabular}
\caption{\small\textbf{System states, reachable in bold (black: normal op; \textcolor{red}{red: error}; \textcolor{green}{green: already stored})}}
\label{tab:sysautomaton}
\end{centering}
\vspace{-0.7cm}
\end{table}

\medskip

\setlength{\tabcolsep}{6pt}
\begin{table}[H]
\begin{centering}
\small
\begin{tabular}{c | ccccccccccccccccccccccc}
Automaton &\multicolumn{3}{l}{States}&&&&&&&&&&&&&&&&&&& \\
\hline
A &1 &4 &1 &4 &1 &2 &1 &4 &1 &3 &1 &4 &1 &2 &1 &3 &1 &2 &1 &4 &1 &4 &1 \\
B &4 &1 &4 &1 &4 &1 &2 &1 &2 &1 &4 &1 &2 &1 &2 &1 &4 &1 &2 &1 &4 &1 &4 \\
\hline
System &4 &13 &4 &13 &4 &5 &2 &13 &2 &9 &4 &13 &2 &5 &2 &9 &4 &5 &2 &13 &4 &13 &4 \\
\hline
\end{tabular}
\caption{\small\textbf{Individual terminal and system states for the error sequence of Fig.\ \ref{fig:abp_sequence}}}
\label{tab:statetraces}
\end{centering}
\end{table}

\begin{figure}[H]
\centering
\includegraphics[width=10cm]{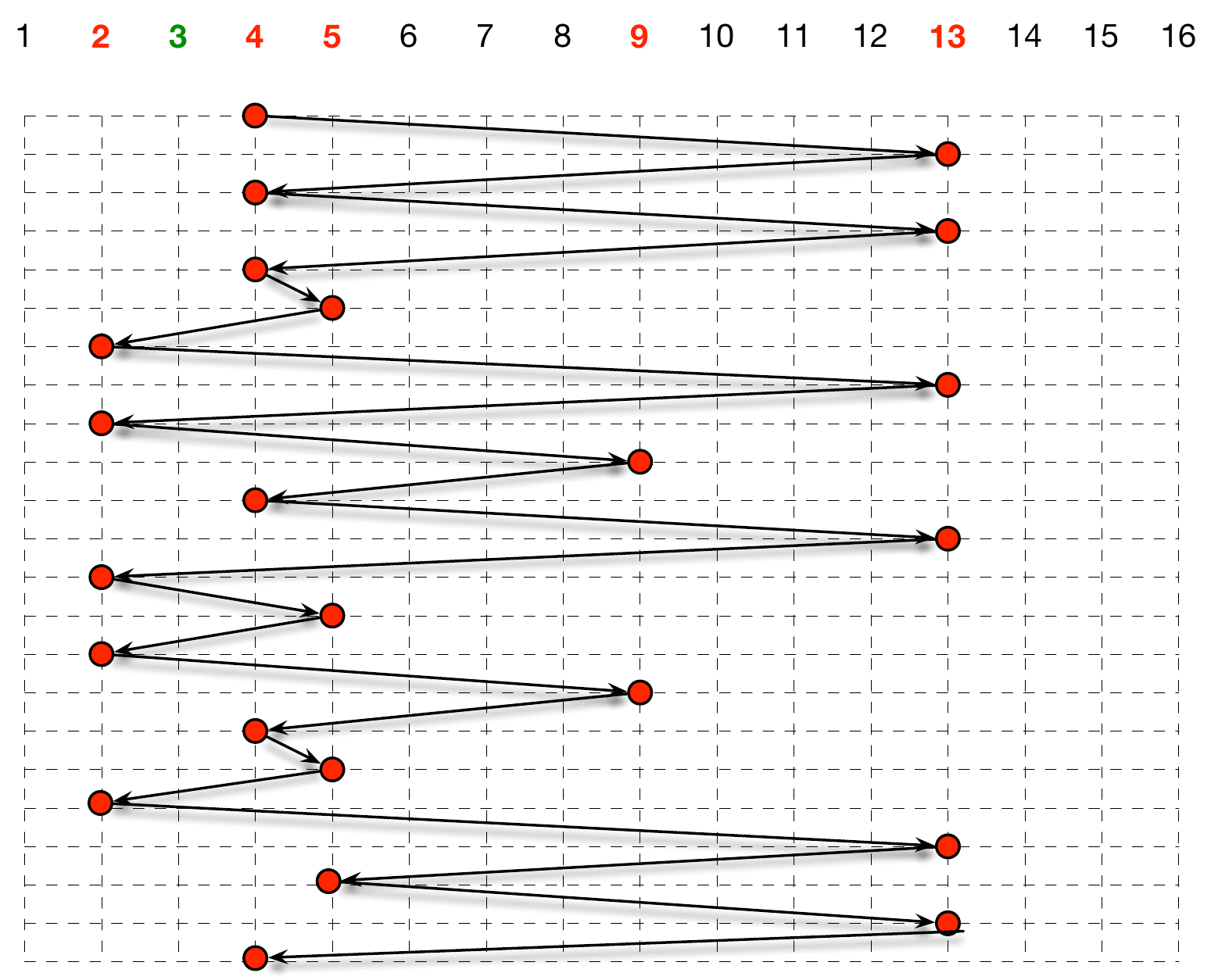}
\caption{\small\textbf{ABP system's state trace corresponding to the Lynch \cite{Lynch1968} error sequence}}
\label{fig:abpstatetrace}
\end{figure}

\begin{figure}[H]
\centering
\includegraphics[width=14cm]{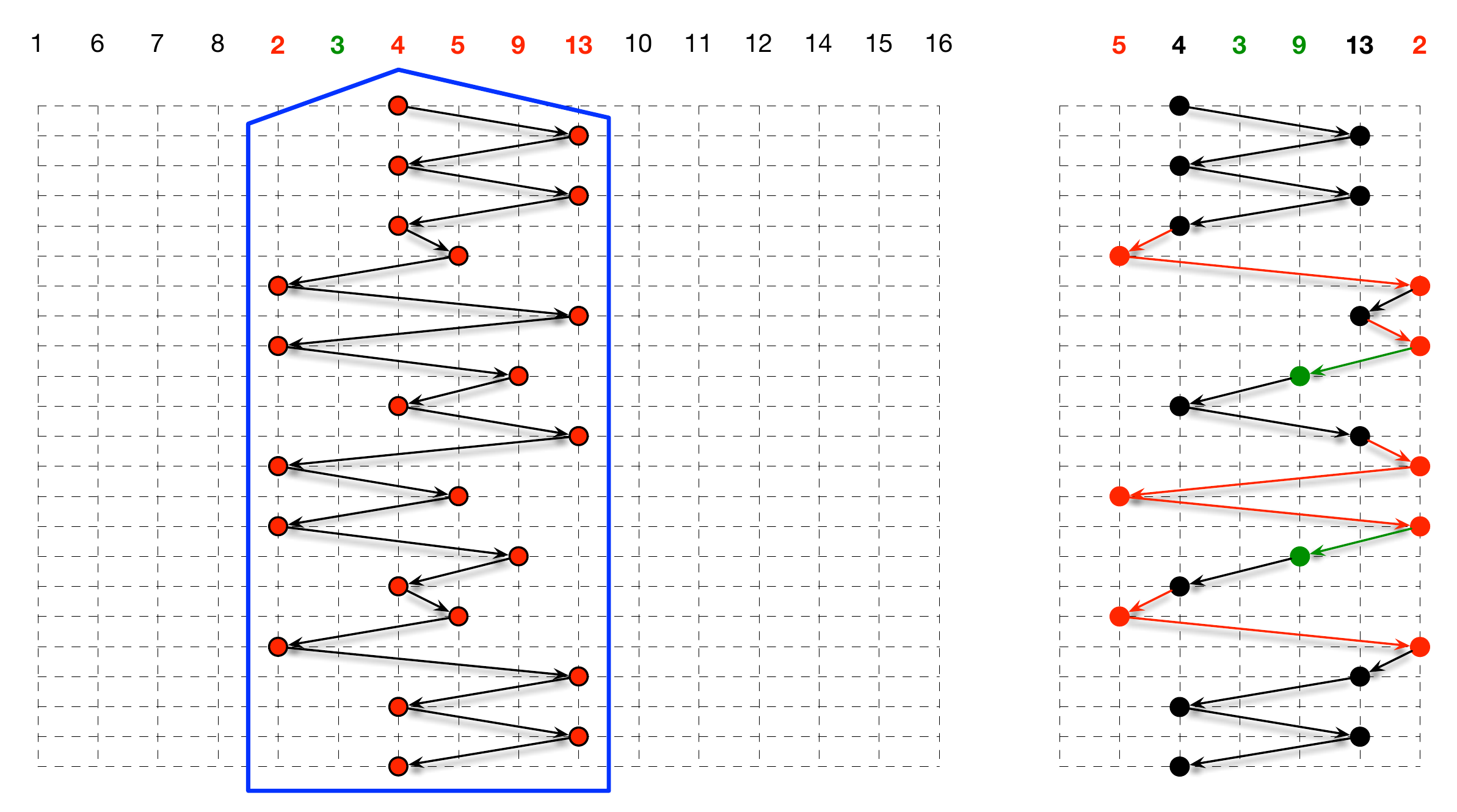}
\caption{\small\textbf{System state trace rearranged for better readability}}
\label{fig:abpstatetrace2}
\end{figure}

\begin{figure}[H]
\centering
\includegraphics[width=12cm]{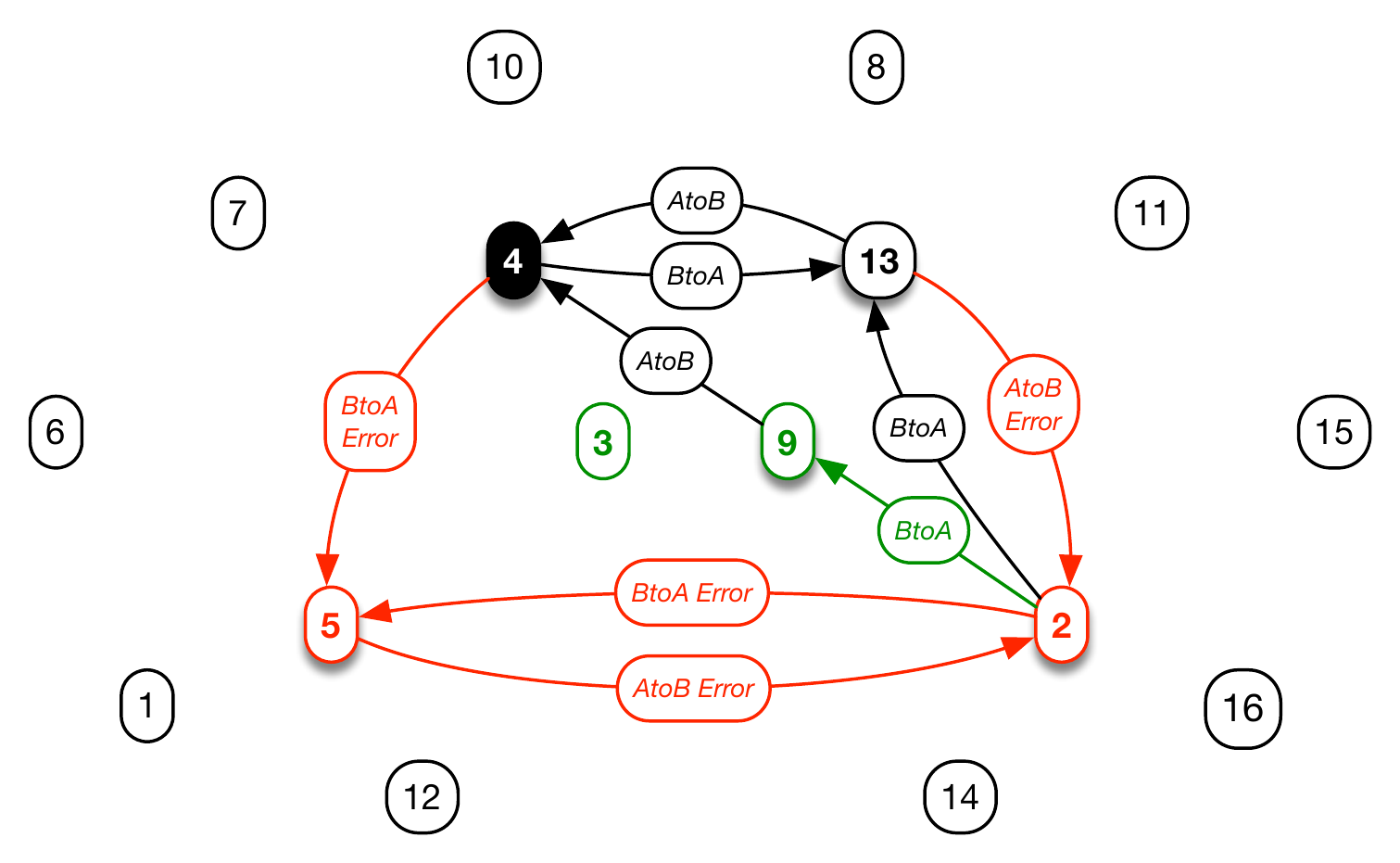}
\caption{\small\textbf{System's finite state machine for Lynch's error sequence}}
\label{fig:abpfsm}
\end{figure}

Fig.\ \ref{fig:abpstatetrace} shows the state trace corresponding to the error sequence of Fig.\ \ref{fig:abp_sequence} in the system's state space. Since this type of state space is not a metric space, how a trace is arranged does not matter as long as its topology is preserved. Thus, Fig.\ \ref{fig:abpstatetrace2} shows the same information in a modified state space where all the reachable states have been bunched together for greater clarity. The ``cone'' is shown by the blue boundary, although this figure shows that when the whole reachable state space is traversed, as in this case, perhaps a rectangle is a better representative shape. On the right of Fig.\ \ref{fig:abpstatetrace2} another view of the state trace is shown that uses the colour coding defined above. Fig.\ \ref{fig:abpfsm} shows the system's finite state machine (FSM) immersed in the global state space.

For a given length of error sequence $n$, the number of possible sequences is $2^n$ of which, for $n = 22$, the Lynch sequence of Fig.\ \ref{fig:abp_sequence} is one. 
Tables \ref{tab:errseqautomata1} and \ref{tab:errseqautomata2} show a set of elementary automata to see whether they could serve as a basis for some kind of automaton decomposition. This is by no means meant to be a representative set of all the possible patterns.

Could the map from the set of error sequences to the set of corresponding automata be a homomorphism? A map $\theta\colon S \rightarrow A$, where $S$ is the set of error sequences and $A$ is the set of corresponding automata, is a homomorphism if both these conditions hold \cite{Cameron2008}:
\begin{align}
\begin{split}
 \theta(s_i + s_j) &=\theta(s_i) + \theta(s_j) \\
\theta(s_i s_j) &=\theta(s_i)  \theta(s_j),
\end{split}
\end{align}
for suitably defined addition and multiplication operations on the set elements $s_i, s_j \in S$, where both indices range from 1 to $2^n$. Glossing over whether or not the $S$ and $A$ sets in question have any algebraic structure (like a ring, group, etc), we can easily see that the simplest possible example of linear superposition for a simple-minded definition of addition operation does not work. In fact, adding $s_2$ and $s_3$ vectorially we get $s_4$, but the automaton corresponding to $s_4$ is very different from the ``addition'' of the automata corresponding to $s_2$ and $s_3$ defined as the union of their edge sets. More seems to be required for this idea to work.

\bigskip

\newcolumntype{i}{ >{\centering\arraybackslash} m{0.5cm} }
\newcolumntype{C}{ >{\centering\arraybackslash} m{5cm} }
\begin{table}[H]
\small
\begin{centering}
\begin{tabular}{| i | m{5cm} | m{2cm} | C |}
\hline
\textbf{No.} & \textbf{Error Sequence} & \textbf{State Trace} & \textbf{Automaton}	\\ 
\hline
	1
	& $(0,0,0,0,0,0,0, \cdots )$
	& \makecell[l]	{$1 4 1 4 1 4 1 4 \cdots$ \\
				 $4 1 4 1 4 1 4 1 \cdots$}
	& \parbox[t]{4cm}{\centering{\includegraphics[width=3 cm]{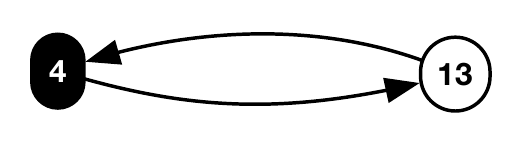}} } \\ 
\hline
	2
	& $(0,0,\textcolor{red}{E},0,0,0,0, \cdots )$
	& \makecell[l]	{$1 4 1 \textcolor{red}{2} \textcolor{green}{1} 4 1 4 \cdots$ \\
				 $4 1 4 \textcolor{red}{1} \textcolor{green}{3} 1 4 1 \cdots$}
	& \parbox[t]{4cm}{\centerline{\includegraphics[width=4 cm]{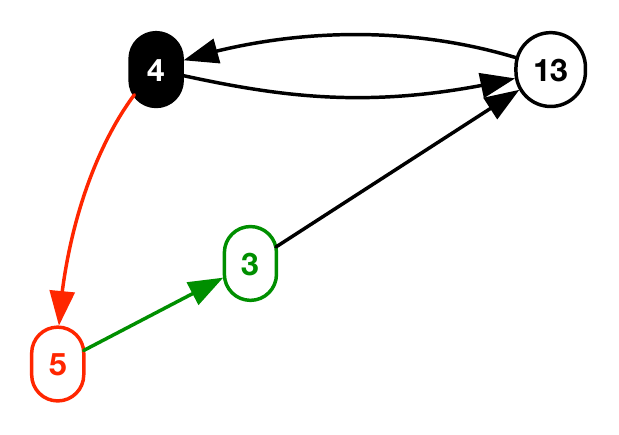}} } \\ 
\hline
	3
	& $(0,0,0,\textcolor{red}{E},0,0,0, \cdots )$
	& \makecell[l]	{$1 4 1 4 \textcolor{red}{1} \textcolor{green}{3} 1 4 \cdots$ \\
				 $4 1 4 1 \textcolor{red}{2} \textcolor{green}{1} 4 1 \cdots$}
	& \parbox[t]{4cm}{\centerline{\includegraphics[width=4 cm]{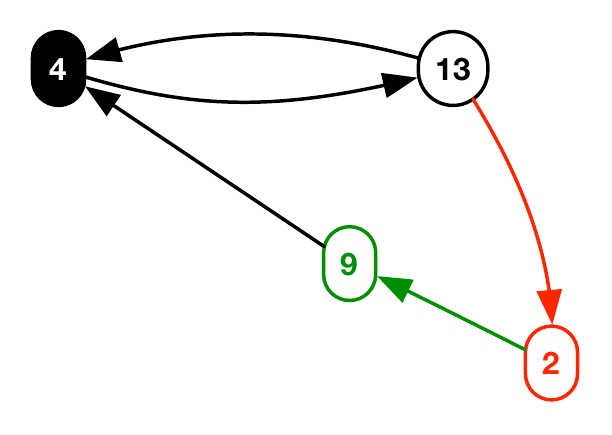}} } \\ 
\hline
	4
	& $(0,0,\textcolor{red}{E},\textcolor{red}{E},0,0,0, \cdots )$
	& \makecell[l]	{$1 4 1 \textcolor{red}{2} \textcolor{red}{1} 4 1 4 \cdots$ \\
				 $4 1 4 \textcolor{red}{1} \textcolor{red}{2} 1 4 1 \cdots$}
	& \parbox[t]{4cm}{\centerline{\includegraphics[width=4 cm]{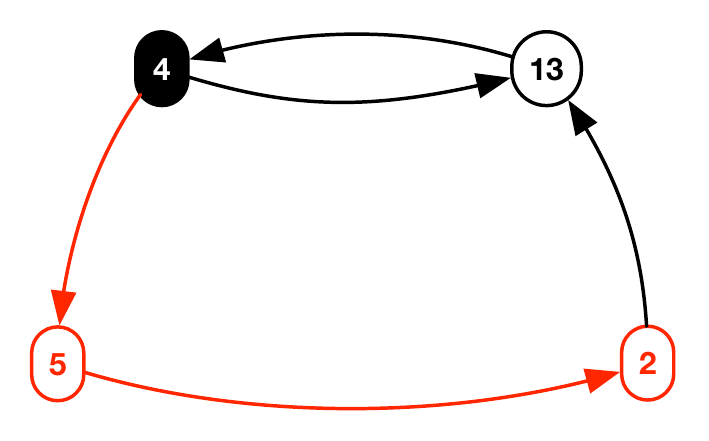}} } \\ 
\hline
	5
	& $(0,0,0,\textcolor{red}{E},\textcolor{red}{E},0,0, \cdots )$
	& \makecell[l]	{$1 4 1 4 \textcolor{red}{1} \textcolor{red}{2} 1 4 \cdots$ \\
				 $4 1 4 1 \textcolor{red}{2} \textcolor{red}{1} 4 1 \cdots$}
	& \parbox[t]{4cm}{\centerline{\includegraphics[width=4 cm]{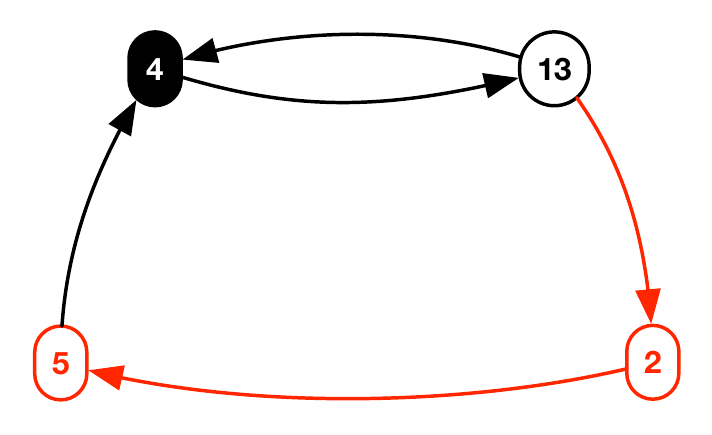}} } \\ 
\hline
	6
	& $(0,0,\textcolor{red}{E},0,\textcolor{red}{E},0,\textcolor{red}{E}, \cdots )$
	& \makecell[l]	{$1 4 1 \textcolor{red}{2} 1 \textcolor{red}{2} 1 \textcolor{red}{2} \cdots$ \\
				 $4 1 4 \textcolor{red}{1} 4 \textcolor{red}{1} 4 \textcolor{red}{1} \cdots$}
	& \parbox[t]{4cm}{\centerline{\includegraphics[width=4 cm]{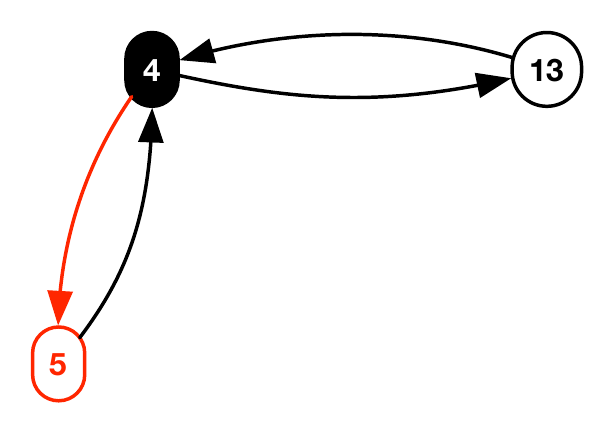}} } \\ 
\hline
\end{tabular}
\caption{\bf \small Automaton motifs for representative error sequences}
\label{tab:errseqautomata1}
\end{centering}
\end{table}

Automata decomposition is studied by algebraic automata theory, an algebra subfield that grew out of semigroup theory about 60 years ago \cite{KrohnRhodes1965,Zeiger1967}. Unfortunately it is so abstract that it is difficult to relate its results and insights to concrete applications. A finite-state automaton can be defined mathematically as a finite set of transformations acting on a finite set of states $Q$. In general, these transformations can be composed by functional composition. The set of all finite sequences of transformations thus satisfies the axioms of a semigroup, meaning that it is closed with respect to a multiplication law (here, functional composition) that satisfies the associative property. However, not all of its members need have an inverse. By convention, the empty sequence yields an identity; although this implies that we have a monoid, the term semigroup is used anyway. Thus, the algebraic version of a finite-state automaton is a ‘transformation semigroup’, or `ts', which is a direct generalization of the permutation group concept to semigroups. As for permutations groups, a potential cause of confusion arises from the fact that one element $s$ of the semigroup $S$ of a given ts acts as an operator on \emph{all} the states
$q \in Q$ \emph{simultaneously}. In other words, the element $s \in S$ should be
seen as the whole function $s\colon Q \rightarrow Q$ that is defined over the whole state set at once. By contrast, the execution of a single step of a
given algorithm implemented by a given automaton, such as we have been discussing here, yields a `state transition' and should be seen as a single value of such a function for a given starting state $q\colon s(q) := q'$. This and similar points about the algebraic structure of automata are explored in more detail in \cite{DiniNehanivEgriNagySchilstra2013}.

\begin{table}[H]
\small
\begin{centering}
\begin{tabular}{| i | m{4cm} | m{2.3cm} | C |}
\hline
\textbf{No.} & \textbf{Error Sequence} & \textbf{State Trace} & \textbf{Automaton}	\\
\hline
	7
	& $(0,0,0,\textcolor{red}{E},0,\textcolor{red}{E},0,\textcolor{red}{E},0, \cdots )$
	& \makecell[l]	{$1 4 1 4 \textcolor{red}{1} 4 \textcolor{red}{1} 4 \textcolor{red}{1} \cdots$ \\
				 $4 1 4 1 \textcolor{red}{2} 1 \textcolor{red}{2} 1 \textcolor{red}{2} \cdots$}
	& \parbox[t]{4cm}{\centerline{\includegraphics[width=4 cm]{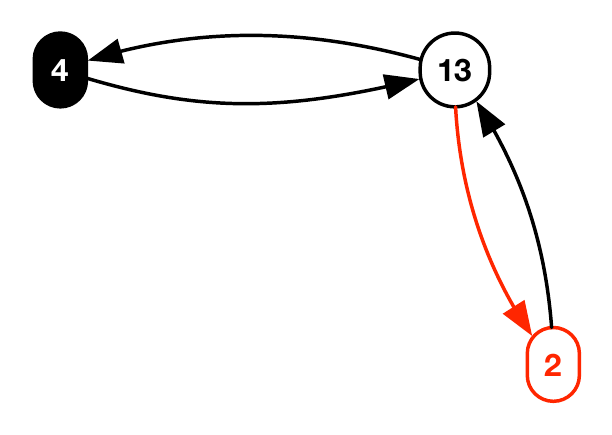}} } \\
\hline
	8
	& $(0,0,\textcolor{red}{E},\textcolor{red}{E},0,\textcolor{red}{E},0,0, \cdots )$
	& \makecell[l]	{$1 4 1 \textcolor{red}{2} \textcolor{red}{1} 4 \textcolor{red}{1} \textcolor{green}{3} 1 \cdots$ \\
				 $4 1 4 \textcolor{red}{1} \textcolor{red}{2} 1 \textcolor{red}{2} \textcolor{green}{1} 4 \cdots$}
	& \parbox[t]{4cm}{\centerline{\includegraphics[width=4 cm]{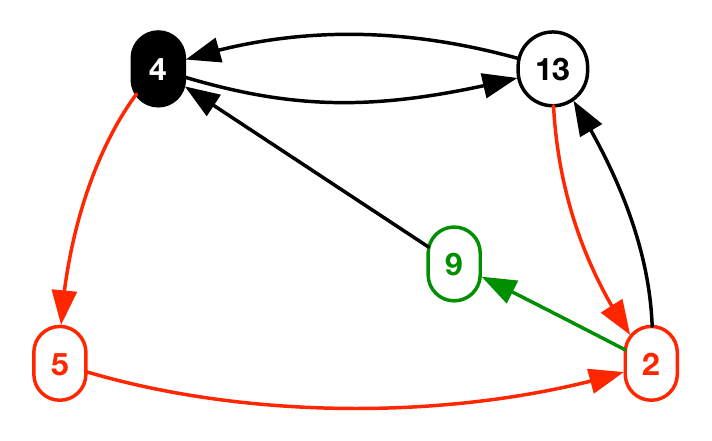}} } \\
\hline
	9
	& $(0,0,0,\textcolor{red}{E},\textcolor{red}{E},0,\textcolor{red}{E},0,0, \cdots )$
	& \makecell[l]	{$1 4 1 4 \textcolor{red}{1} \textcolor{red}{2} 1 \textcolor{red}{2} \textcolor{green}{1} 4 \cdots$ \\
				 $4 1 4 1 \textcolor{red}{2} \textcolor{red}{1} 4 \textcolor{red}{1} \textcolor{green}{3} 1 \cdots$}
	& \parbox[t]{4cm}{\centerline{\includegraphics[width=4 cm]{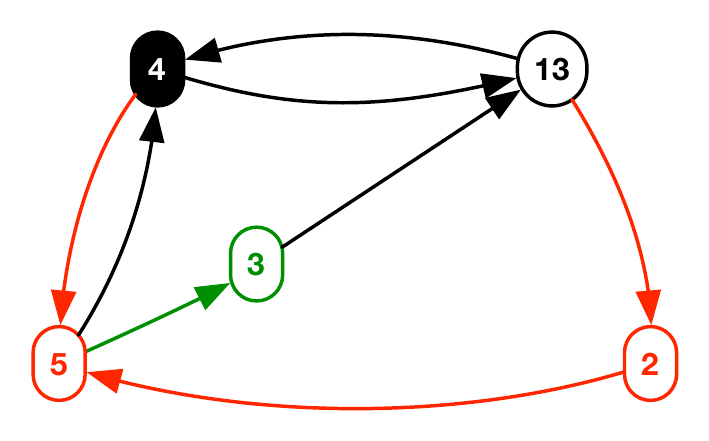}} } \\
\hline
	10
	& $(0,0,\textcolor{red}{E},0,\textcolor{red}{E},\textcolor{red}{E},0,0, \cdots )$
	& \makecell[l]	{$1 4 1 \textcolor{red}{2} \textcolor{green}{1} \textcolor{red}{2} \textcolor{red}{1} 4 1 \cdots$ \\
				 $4 1 4 \textcolor{red}{1} \textcolor{green}{3} \textcolor{red}{1} \textcolor{red}{2} 1 4 \cdots$}
	& \parbox[t]{4cm}{\centerline{\includegraphics[width=4 cm]{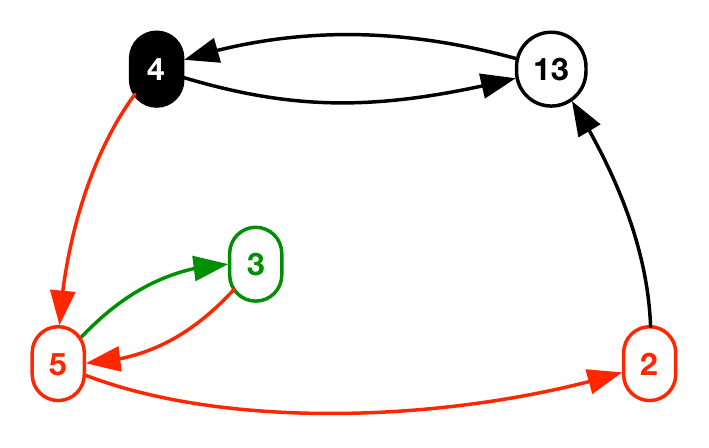}} } \\
\hline
	11
	& $(0,0,0,\textcolor{red}{E},0,\textcolor{red}{E},\textcolor{red}{E},0,0, \cdots )$
	& \makecell[l]	{$1 4 1 4 \textcolor{red}{1} \textcolor{green}{3} \textcolor{red}{1} \textcolor{red}{2} 1 4 \cdots$ \\
				 $4 1 4 1 \textcolor{red}{2} \textcolor{green}{1} \textcolor{red}{2} \textcolor{red}{1} 4 1 \cdots$}
	& \parbox[t]{4cm}{\centerline{\includegraphics[width=4 cm]{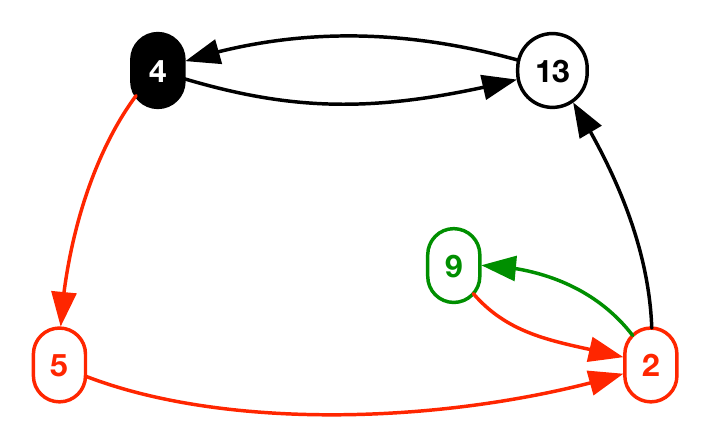}} } \\
\hline
	12
	& $(0,0,\textcolor{red}{E},\textcolor{red}{E},\textcolor{red}{E},\textcolor{red}{E},\textcolor{red}{E}, \cdots )$
	& \makecell[l]	{$1 4 1 \textcolor{red}{2} \textcolor{red}{1} \textcolor{red}{2} \textcolor{red}{1} \textcolor{red}{2} \cdots$ \\
				 $4 1 4 \textcolor{red}{1} \textcolor{red}{2} \textcolor{red}{1} \textcolor{red}{2} \textcolor{red}{1}  \cdots$}
	& \parbox[t]{4cm}{\centerline{\includegraphics[width=4 cm]{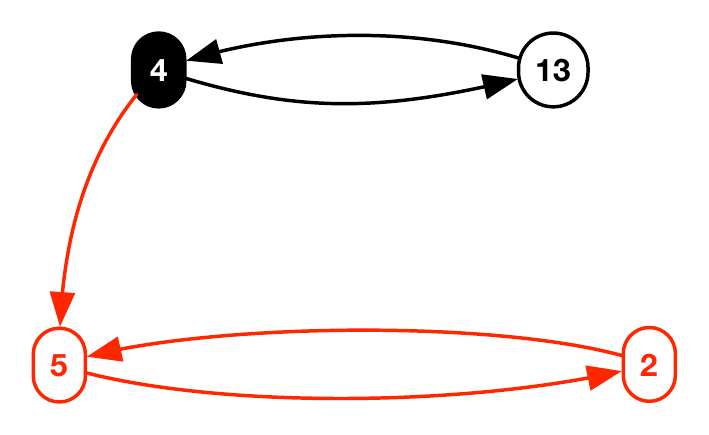}} } \\
\hline
	13
	& $(0,0,0,\textcolor{red}{E},\textcolor{red}{E},\textcolor{red}{E},\textcolor{red}{E},\textcolor{red}{E}, \cdots )$
	& \makecell[l]	{$1 4 1 4 \textcolor{red}{1} \textcolor{red}{2} \textcolor{red}{1} \textcolor{red}{2} \textcolor{red}{1} \cdots$ \\
				 $4 1 4 1 \textcolor{red}{2} \textcolor{red}{1} \textcolor{red}{2} \textcolor{red}{1} \textcolor{red}{2}  \cdots$}
	& \parbox[t]{4cm}{\centerline{\includegraphics[width=4 cm]{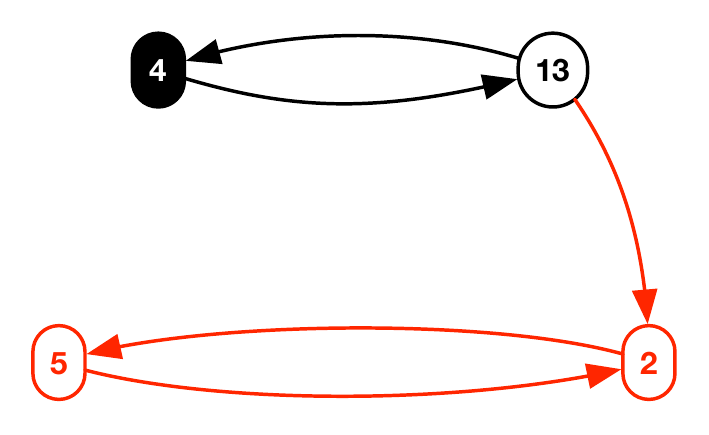}} } \\
\hline
\end{tabular}
\caption{\bf \small Automaton motifs for representative error sequences (Cont'd)}
\label{tab:errseqautomata2}
\end{centering}
\end{table}

\clearpage

\section{Conclusions and Future Work}
\label{conclusion}
The search for a possible algebraic structure in this type of problem is motivated by a different kind of mapping. Namely, if it were possible to ``decompose'' the ASM or TLA$^+$ specification of a given system into elementary sub-specifications in such a way that a homomorphism could be established between the elementary components of the overall specification and the corresponding elementary components of the general automaton, the task of specifying, verifying, and validating complex software systems could be broken down into simpler tasks that could then be composed to achieve the general specification. Such a condition would clearly impose a significant constraint on the formal systems involved, but this does not necessarily imply a constraint on the \emph{computation} being specified. The potential benefits of composability seem significant enough to motivate further exploration in this direction.

Such an algebra-based approach is likely to be more relevant to TLA$^+$ than to ASMs because the latter already rely on a methodology that is fundamentally different from the concept of composability. The ASM methodology begins from a very abstract state-based domain model based on requirements described by the domain expert and adds structure and details by iterative refinement. Although also TLA$^+$ makes extensive use of iterative refinement, its declarative semantics seems to afford it greater at each stage structure, possibly making TLA$^+$ models better suited for decomposition.

\section*{Acknowledgment}
We are very grateful to Prof.\ Egon B\"orger for his feedback on Chapter 2 of this report.
\newpage

\setlength{\parskip}{0.8\baselineskip}
\bibliographystyle{plainurl}
\small
\addcontentsline{toc}{section}{References}

\end{document}